\documentclass[structabstract]{aa}  
\usepackage{longtable,lscape,graphicx,float,amsmath,amssymb,mathrsfs,txfonts}
\usepackage{multirow,natbib,ulem} 

\def\msol{\,\rm{M}$_{\odot}$}
\def\lsol{\,\rm{L}$_{\odot}$}

\def\arcsec{$^{\prime}$$^{\prime}$}

\def\deg{$^{\circ}$}
\def\micron{\,$\mu$m}

\def\kms{\,km~s$^{-1}$}
\def\pccm{\,cm$^{-3}$}

\def\apjs{ApJS}
\def\apjl{ApJL}%
\def\aap{A\&A}%

\newcounter{ppnum4}
\newcounter{ppnum5}
\begin{document}
\title{Waterfalls around protostars}

\subtitle{Infall motions towards Class 0/I envelopes as probed by water \thanks{\textit{Herschel} is an ESA space observatory with science instruments provided by European-led Principal Investigator consortia and with important participation from NASA.}}

\author{J.~C.~Mottram\inst{1}\thanks{E-mail:mottram@strw.leidenuniv.nl} 
          \and
          E.~F.~ van Dishoeck\inst{1,2}
          \and
          M. Schmalzl\inst{1}
          \and
          L.~E.~Kristensen\inst{1,3}
          \and
          R. Visser\inst{4}
          \and
          M.~R.~Hogerheijde\inst{1}
          \and
          S.~Bruderer\inst{2}
}

\institute{Leiden Observatory, Leiden University, PO Box 9513, 2300 RA Leiden, The Netherlands
\and
Max-Planck-Institut f\"{u}r Extraterrestrische Physik, Giessenbachstrasse 1, 85748 Garching, Germany
\and
Harvard-Smithsonian Center for Astrophysics, 60 Garden Street, Cambridge, MA 02138, USA
\and
Department of Astronomy, University of Michigan, 500 Church Street, Ann Arbor, MI 48109-1042, USA
}

   \date{Received 3 May 2013; accepted 21 August 2013}

 
\abstract
{For stars to form, material must fall inwards from core scales through the envelope towards the central protostar. While theories of how this takes place have been around for some time, the velocity profile around protostars is poorly constrained. The combination of observations in multiple transitions of a tracer which is sensitive to kinematics and radiative transfer modelling of those lines has the potential to break this deadlock.}
{Seven protostars observed with the Heterodyne Instrument for the Far-Infrared (HIFI) on board the \textit{Herschel} Space Observatory as part of the ``Water in star-forming regions with \textit{Herschel}'' (WISH) survey show infall signatures in water line observations. We aim to constrain the infall velocity and the radii over which infall is taking place within the protostellar envelopes of these sources. We will also use these data to constrain the chemistry of cold water.}
 {We use 1-D non-LTE \textsc{RATRAN} radiative transfer models of the observed water lines to constrain the infall velocity and chemistry in the protostellar envelopes of 6 Class 0 protostars and one Class I source. We assume a free-fall velocity profile and, having found the best fit, vary the radii over which infall takes place.}
{In the well-studied Class 0 protostar NGC1333-IRAS4A we find that infall takes place over the whole envelope to which our observations are sensitive ($r\gtrsim$1000\,AU). For L1527, L1157, BHR71 and IRAS15398 infall takes place on core to envelope scales (i.e. $\sim$10000$-$3000\,AU). In Serpens-SMM4 and GSS30 the inverse P-Cygni profiles seen in the ground-state lines are more likely due to larger-scale motions or foreground clouds. Models including a simple consideration of the chemistry are consistent with the observations, while using step abundance profiles are not. The non-detection of excited water in the inner envelope in 6 out of 7 protostars is further evidence that water must be heavily depleted from the gas-phase at these radii.}
{Infall in four of the sources is supersonic and infall in all sources must take place at the outer edge of the envelope, which may be evidence that collapse is global or outside-in rather than inside-out. The mass infall rate in NGC1333-IRAS4A is large ($\gtrsim$10$^{-4}$\msol{}\,yr$^{-1}$), higher than the mass outflow rate and expected mass accretion rates onto the star, suggesting that any flattened disk-like structure on small scales will be gravitationally unstable, potentially leading to rotational fragmentation and/or episodic accretion.}


\maketitle

%

\section{Introduction}
\label{S:Intro}

For stars to form from the interstellar medium the density must increase from $\sim$10\pccm{} to $\sim$10$^{24}$\pccm{}. While some of this process is achieved through the formation of molecular clouds ($n$$\sim$10$^{4}$\pccm{}) and then dense filaments and cores ($n$$\sim$10$^{6}$\pccm{}) within these clouds, the remaining 18 orders of magnitude increase must be accomplished through infall of gravitationally bound and unstable envelope material onto the central protostar. This infall may proceed on an inside-out \citep{Shu1977} or outside in basis \citep{Foster1993}. When angular momentum is taken into account \citep[e.g.][]{Ulrich1976,Cassen1981,Terebey1984}, this leads to a flattened inner structure which may eventually form a rotationally supported disk, with infall proceeding from the envelope to the disk, then through the disk onto the star.

Observational attempts to test such models and understand how infall varies both spatially and in time have proven more difficult. The observational signature associated with infalling gas is an asymmetric line profile with more blue than redshifted emission \citep[see e.g.][]{Evans1999,Myers2000}. In some tracers, strong absorption by the outer parts of the envelope results in an inverse P-Cygni line profile where the absorption is red-shifted with respect to the source velocity, usually accompanied by blue-shifted emission. Various authors have used dense gas tracers such as CS, HCO$^{+}$, HCN, N$_{2}$H$^{+}$ and H$_{2}$CO to probe the infall signatures of cores and protostars using both single-dish and interferometric observations \citep[e.g.][]{Zhou1992,Zhou1993,Gregersen1997,Mardones1997,DiFrancesco2001,Fuller2005,Attard2009}. However counter-claims that some profiles can be reproduced with a combination of rotation and foreground layers \citep[][]{Menten1987,Choi2001} have made it difficult to conclusively identify infall without the need for sophisticated modelling \citep[e.g.][]{Brinch2009}. 

In order to trace infall, one needs a good tracer of motion in envelope material. The molecules previously used were chosen because they have critical densities and abundances which make them reasonable probes of the regions where infall is expected to take place. Water, on the other hand, is particularly sensitive to motion: even though sub-millimetre water transitions have higher critical densities of order 10$^{6}$$-$10$^{8}$\pccm{}, they are easily sub-thermally excited due to high Einstein-A coefficients. In addition, the freeze-out of water onto dust grains at temperatures below $\sim$100\,K and the photodesorption or photodissociation of water by both interstellar and cosmic-ray induced UV radiation make the chemistry of water particularly sensitive to temperature and the local radiation field \citep[e.g.][]{Hollenbach2009}. Indeed, the details of water line profile shapes actually allow this chemistry to be constrained on much smaller spatial scales than the observing beam \citep[e.g.][]{Caselli2012}. 

The properties that make H$_{2}$O such a good tracer of motion also make it impossible to observe the vast majority of transitions from the ground due to atmospheric absorption, so space satellites are required to study water comprehensively. The ``Water in star-forming regions with \textit{Herschel}'' \citep[WISH;][]{vanDishoeck2011} survey has obtained velocity-resolved spectra of multiple water lines towards a sample of $\sim$80 targets which cover the luminosity range from 1 to 10$^{6}$\lsol{} and vary in evolutionary stage from pre-stellar cores to gas-rich disks. While most of the water emission detected towards the sub-sample of 29 low-mass Class 0/I protostars is related to outflow motions \citep[e.g.][]{Bergin2003,Kristensen2010,Kristensen2012}, seven sources were identified by \citet{Kristensen2012} which show inverse P-Cygni profiles in the ground-state H$_{2}$O 1$_{10}-$1$_{01}$ line at 557\,GHz. 

The analysis method used by many previous authors, including \citet{Kristensen2012}, assumed two infalling slabs of material, sometimes with the addition of a third central static slab in order to reproduce the `true' inverse P-Cygni profile. While such an approach is a good first step, it only provides information about the characteristic infall velocity at a characteristic radius from the central protostar. The first goal of this paper is to constrain the spatial scales on which infall takes place within these seven WISH sources. This will allow accurate measurement of the mass infall rate ($\dot{M}_{\mathrm{inf}}$) of material from the envelope towards radii where a flattened disk-like structure may be present. This is distinct from the mass accretion rate ($\dot{M}_{\mathrm{acc}}$) of material from any central disk-like structure onto the central protostar. The mass infall and accretion rates need not be directly related, particularly if the disk stability varies as a function of time. Understanding what the mass infall rate is, and over what spatial scales infall takes place, will allow us to predict whether a central flattened disk-like structure is stable during the early phases of star formation. The second goal of this paper is to use the intensity and shape of gas-phase water lines to constrain water chemistry in the cold regions of protostellar envelopes, a topic more extensively discussed in \citet{Schmalzlip}.

Both these goals require more detailed and complex modelling than previously employed. We therefore perform full 1-D spherically symmetric non-LTE radiative transfer modelling of water line profiles, taking into account an envelope with varying density, temperature, velocity and water abundance. The paper is structured as follows. In Section~\ref{S:observations}, we present the observations with which we will constrain our models, as well as removal of outflow-related emission using gaussian fitting. The modelling approach, including the simple water chemistry required to calculate how the water abundance varies within the envelope, is discussed in Section~\ref{S:models}. The results and analysis relating to infall are presented in Section~\ref{S:infall}, while those relating to water chemistry in protostellar envelopes are presented in Section~\ref{S:chemistry}. We then discuss the wider implications of these results in Section~\ref{S:discussion} before reaching our conclusions in Section~\ref{S:conclusions}.

\section{Observations}
\label{S:observations}

The seven sources discussed in this paper are part of the WISH sub-sample of embedded low-mass Class 0/I protostars and were selected because they have inverse P-Cygni line profiles in the H$_{2}$O 1$_{10}-$1$_{01}$ line (E$_{\mathrm{up}}$/k$_{\mathrm{b}}$ = 61\,K), as identified by \citet{Kristensen2012}. Their general properties are given in Table~\ref{T:observations_sources}; six are Class 0 and one (GSS30-IRS1) is a Class I protostar. All sources were observed in two ortho-H$_{2}$O and three para-H$_{2}$O lines with the Heterodyne Instrument for the Far-Infrared \citep[HIFI;][]{deGraauw2010} on the \textit{Herschel} Space Observatory \citep{Pilbratt2010} as part of the WISH survey. These lines have upper-level energies ranging from 50 to 250\,K. In addition, the 3$_{12}-$3$_{03}$ ortho-H$_{2}$O line was observed for two sources (NGC1333-IRAS4A, hereafter referred to as IRAS4A, and Ser-SMM4) and the 2$_{12}-$1$_{01}$ ortho-H$_{2}$O line observed towards IRAS4A only. The observation identification numbers are given in Table~\ref{T:obsids}, which also serves to summarise which lines were observed towards each source. All data were obtained using both the wide-band spectrometer (WBS) and high-resolution spectrometer (HRS) in both horizontal and vertical polarisation. Table~\ref{T:observations_lines} presents details of the observed lines, including the line frequency, main-beam efficiency, spatial and spectral resolutions and the upper level energy of the transition.

\begin{table}
\begin{center}
\caption[]{Source properties}
\begin{tabular}{llcccc}
\hline
\hline \noalign {\smallskip}
Source & \multicolumn{1}{c}{$D$\tablefootmark{a}} & $\varv_{\rm LSR}$\tablefootmark{b} & $L_{\rm bol}$\tablefootmark{c} & $T_{\rm bol}$\tablefootmark{c} & $M_{\rm env}$\tablefootmark{d} \\
 & (pc) & (\kms{}) & \multicolumn{1}{c}{(\lsol{})} & (K) & (\msol{}) \\
\hline\noalign {\smallskip}
IRAS4A  & 235 & $+$7.2 & \phantom{1}9.1 & \phantom{1}33 & 5.2 \\
L1527  & 140 & $+$5.9 & \phantom{1}1.9 & \phantom{1}44 & 0.9\\
BHR71  & 200 & $-$4.4 & 14.8 & \phantom{1}44 & 3.1 \\
IRAS15398  & 130 & $+$5.1 & \phantom{1}1.6 & \phantom{1}52 &	0.5 \\
GSS30-IRS1  & 125 & $+$3.5 & 13.9 & 142 & 0.6 \\
Ser SMM4  & 415 & $+$8.0 & \phantom{1}6.2 & \phantom{1}26 & 6.9 \\
L1157  & 325 & $+$2.6 & \phantom{1}4.7 & \phantom{1}46 & 1.5 \\
\hline\noalign {\smallskip}
\label{T:observations_sources}
\end{tabular}
\tablefoot{\tablefoottext{a}{Taken from \citet{vanDishoeck2011} with the exception of Ser SMM4, where we use the distance determined using VLBA observations by \citet{Dzib2010}.}
        \tablefoottext{b}{Obtained from ground-based C$^{18}$O or C$^{17}$O observations \citep{Yildiz2013a} with the exception of IRAS4A for which the value from \citet{Kristensen2012} is more consistent with our data.}
	\tablefoottext{c}{Measured using \textit{Herschel}-PACS data from the WISH and DIGIT key programmes \citep[][]{Karska2013}.}
	\tablefoottext{d}{Mass within the 10\,K radius, determined by \citet{Kristensen2012} from DUSTY modelling of the sources.}
	}
\end{center}
\end{table}

The data were reduced using \textsc{HIPE} \citep{Ott2010} version 8.2 with all further processing performed using \textsc{python}. A low-order (i.e. $\leq$2) polynomial was used to fit and subtract the baseline from the two WBS polarisations for each line separately, after which the data were co-added. HIFI is a dual-sideband receiver, so each spectrum is a combination of simultaneous observations in both the upper and lower sidebands. As such, where a constant baseline can be fit to a sub-band this gives twice the continuum brightness temperature as it includes continuum photons from both sidebands. Some emission lines are broader than the more limited velocity range covered by the HRS observations (typically 50$-$100\kms{}), so only a constant baseline was used for these data, but otherwise the process was the same. All co-added data were then converted to the main-beam temperature scale using the beam efficiencies from \citet{Roelfsema2012}. A more detailed discussion of the data processing and quality for all WISH low-mass protostars can be found in \citet{Kristensen2012} for the H$_{2}$O 1$_{10}-$1$_{01}$ line, and will be presented in \citet{Mottramip} for all other water observations of low-mass YSOs. 

Due to the higher spectral resolution available with the HRS, and the fact that envelope emission is always within $\varv_{\rm LSR}$$\pm$8\kms{}, only these data will be used in the remainder of this paper with the exception of the 2$_{12}$-1$_{01}$ transition at 1670\,GHz. For this high-frequency line the WBS spectral resolution is comparable to the HRS resolution in the other lines and so the improved noise of the WBS over the HRS (a factor of $\sqrt{2}$) makes these data a better choice. All spectra shown in this paper have been shifted so that the systemic velocity of the source is at 0\kms{}.

\begin{table*}
\begin{center}
\caption[]{Observed water lines}
\begin{tabular}{lrccccr}
\hline
\hline \noalign {\smallskip}
Line & \multicolumn{1}{c}{Rest Freq.} & $\eta_{\mathrm{mb}}$\tablefootmark{a} & $\theta_{\mathrm{mb}}$\tablefootmark{a} & WBS Res. & HRS Res. & \multicolumn{1}{c}{E$_{\mathrm{up}}$/k$_{\mathrm{b}}$} \\
 & \multicolumn{1}{c}{(GHz)} & & (\arcsec{}) & (\kms{}) & (\kms{}) & \multicolumn{1}{c}{(K)} \\
\hline\noalign {\smallskip}
H$_{2}$O 1$_{10}$-1$_{01}$ & 556.93607 & 0.75 & 38.1 & 0.27 & 0.03 & 61.0 \\
H$_{2}$O 2$_{12}$-1$_{01}$\tablefootmark{b} & 1669.90496 & 0.71 & 12.7 & 0.09 & 0.02 & 114.4 \\
H$_{2}$O 3$_{12}$-2$_{21}$ & 1153.12682 & 0.64 & 18.4 & 0.13 & 0.06 & 249.4 \\
H$_{2}$O 3$_{12}$-3$_{03}$\tablefootmark{c} & 1097.36505 & 0.74 & 19.3 & 0.14 & 0.07 & 249.4 \\
\hline\noalign {\smallskip}
H$_{2}$O 1$_{11}$-0$_{00}$ & 1113.34306 & 0.74 & 19.0 & 0.13 & 0.06 & 53.4 \\
H$_{2}$O 2$_{02}$-1$_{11}$ & 987.92670 & 0.74 & 21.5 & 0.15 & 0.07 & 100.8 \\
H$_{2}$O 2$_{11}$-2$_{02}$ & 752.03323 & 0.75 & 28.2 & 0.20 & 0.05 & 136.9 \\
\hline\noalign {\smallskip}
\label{T:observations_lines}
\end{tabular}
\tablefoot{\tablefoottext{a}{Calculated using equations 1 and 3 from \citet{Roelfsema2012}.}\tablefoottext{b}{NGC1333-IRAS4A only.}\tablefoottext{c}{NGC1333-IRAS4A and Ser-SMM4 only}}
\end{center}
\end{table*}

\begin{figure}
\begin{center}
\includegraphics[width=0.45\textwidth]{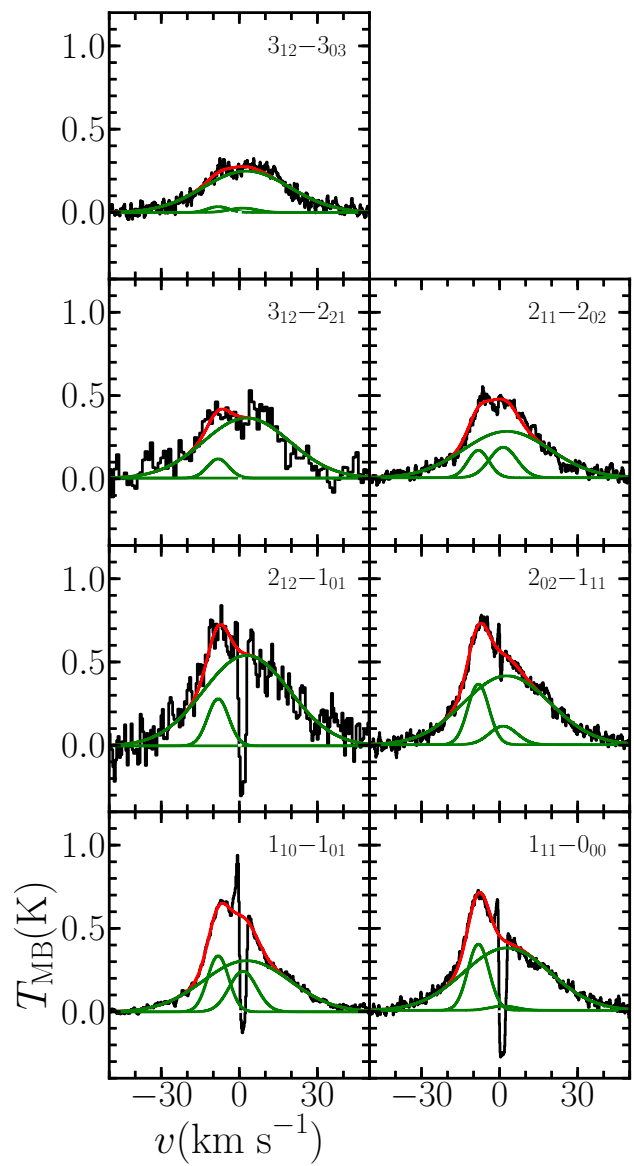}
\caption{Ortho (left) and para (right) continuum subtracted H$_{2}$O WBS spectra for IRAS4A shifted so that the source velocity is at 0\kms{}. The 3$_{12}-$2$_{21}$ line has been smoothed to 1\kms{} resolution and the 2$_{12}-$1$_{01}$ to 0.5\kms{}. All other lines are shown at their native WBS spectral resolution.  The red line indicates the sum of a three-component gaussian fit to the line while the green lines show the individual gaussians.}
\label{F:observations_data}
\end{center}
\end{figure}

Figure~\ref{F:observations_data} provides an example of the WISH H$_{2}$O observations for IRAS4A. The H$_{2}$O emission in many of the lines is complex, with multiple components within the beam. As discussed in \citet{Kristensen2010} for IRAS4A, the broad (FWHM $\gtrsim$ 20\kms{}) gaussian component is associated with emission from a molecular outflow while additional components which are offset from the source velocity and have more moderate linewidths (FWHM $\sim$ 5$-$10\kms{}) likely originate from currently shocked gas \citep{Kristensen2013}. These components are distinct from the narrow envelope emission near the source velocity, and result from very different physical processes which are not directly included in our model.

\begin{figure}
\begin{center}
\includegraphics[width=0.23\textwidth]{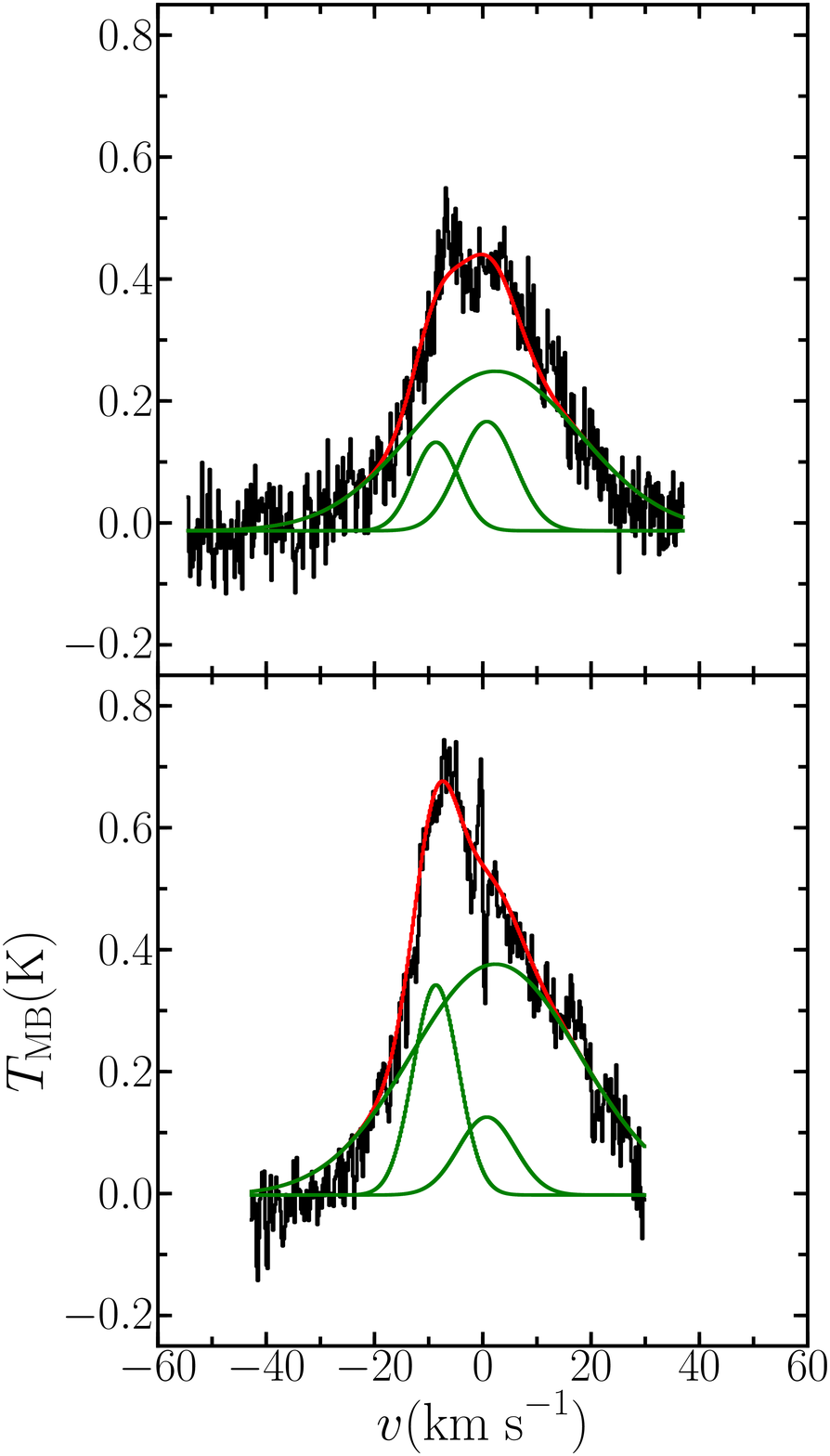}
\includegraphics[width=0.23\textwidth]{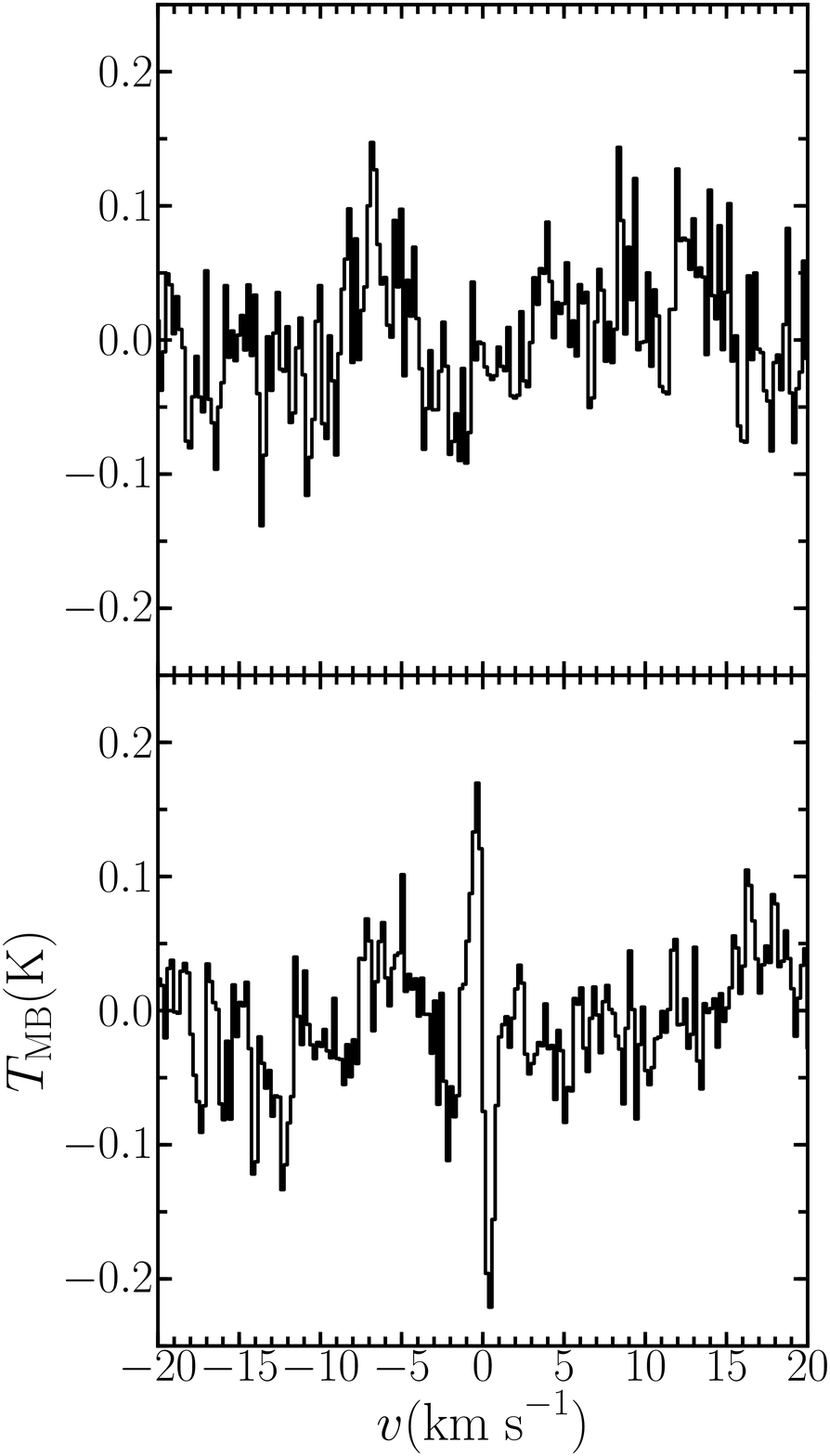}
\caption{Left: HRS spectrum of H$_{2}$O 2$_{11}-$2$_{02}$ (top) and 2$_{02}-$1$_{11}$ (bottom) for IRAS4A rebinned to 0.2\kms{}. The red and green lines have the same meaning as in Fig.~\ref{F:observations_data}. Right: Residual to the fit over a narrower velocity range for emphasis.}
\label{F:observations_decomp}
\end{center}
\end{figure}

In order to isolate only the emission relating to the protostellar envelope, gaussian fits to the line profiles are used to remove the broader and shifted components \citep[see also][]{Kristensen2012}. A combination of up to three gaussian profiles is required to remove the additional distinct physical components. We assume that the line width and central velocity of each component is the same for all lines of a given source. Therefore, after masking emission and absorption associated with the inverse P-Cygni profile, all lines and components are fitted simultaneously with common central velocities and linewidths for each component. An example fit is shown in Figure~\ref{F:observations_data}, with the residuals for two lines presented in Figure~\ref{F:observations_decomp}. All three components identified in this source are observed in multiple lines, though each component is not necessarily detected in all lines. In particular, for IRAS4A the fainter component near the source velocity, which was not included in the fit to the 1$_{10}-$1$_{01}$ line by \citet{Kristensen2012}, is required to fit the higher-excited lines. Removing this component leads to a demonstrably worse fit to the data. This difference in the data and components considered leads to a slight difference in our fit results compared to those presented in \citet{Kristensen2012} but are consistent.

Inverse P-Cygni profiles are observed in the ground-state 1$_{10}-$1$_{01}$ and 1$_{11}$-0$_{00}$ lines for all sources, with only IRAS4A showing such a profile in the excited 2$_{02}$-1$_{11}$ line. There are no detections of narrow envelope emission or absorption in the 2$_{11}$-2$_{02}$, 3$_{12}$-2$_{21}$ or 3$_{12}$-3$_{03}$ lines unlike for high-mass protostars \citep{vanderTak2013}. While some gaussian emission is sometimes observed near the source velocity, this is too broad (FWHM$\sim$10\kms{}) to be consistent with envelope emission. In many sources the absorption in the ground-state water lines due to the envelope is saturated, i.e. the absorption extends below the continuum and is flattened (e.g. see Fig.~\ref{F:observations_data}) indicating that all photons, both from the continuum and the outflow component, have been absorbed by the envelope. The difference between the baseline and the saturated absorption then gives a measure of the continuum level at this frequency. 

\section{Models}
\label{S:models}

In order to constrain and understand the velocity field within the sample of sources discussed above we simulate the water line profiles using 1-dimensional spherically symmetric \textsc{RATRAN} models \citep{Hogerheijde2000b}. \textsc{RATRAN} uses a Monte-Carlo method with accelerated lambda iteration, solving for the level populations including radiative pumping by dust continuum, then performing ray-tracing to create synthetic images. Finally, the synthetic images are convolved to the appropriate \textit{Herschel} beam and the spectrum towards the centre of the model is extracted.

The details and assumptions of these models are discussed in the following subsections, but first it is important to discuss the impact of assuming spherical symmetry. These sources are all known to have prominent outflows which we do not include in the model grid. While we have removed components associated with the outflow from the observations to account for this, the outflow cavity will still lead to the removal of dense material from some sight-lines in the envelope. In addition, the cores around these sources may be flattened, particularly if they reside in filaments \citep[e.g. L1157;][]{Tobin2012a}. Both these effects could lead to 2/3-D variations in the density, and thus temperature, compared to the spherically averaged case. However, using 2-D or 3-D models would require many more free geometrical parameters which the data presented here cannot constrain due to the lack of spatial information, so such models would only add degeneracy. In addition, such models are more computationally expensive, particularly if non-LTE line radiative transfer is included, which for H$_{2}$O is important. A 1-D non-LTE model is therefore the best approach with the data in hand.

On small ($\lesssim$500\,AU) scales, the distribution of material around a protostar is unlikely to be spherical due to rotational flattening, and so the 1-D approximation breaks down \citep{Jorgensen2005}. While we include these regions in our models, most of the observed emission comes from regions well outside such radii because this is inside the $\tau$=1 surface for central velocities and beam dilution minimises the contribution of emission at velocities where the optical depth is lower. Nevertheless, care must be taken not to extrapolate these results to conditions in the inner envelope where 2/3-D modelling is necessary and where the physical conditions are less constrained.

\subsection{Density and Temperature}
\label{S:models_n_and_T}

\begin{figure*}
\begin{center}
\includegraphics[width=0.33\textwidth]{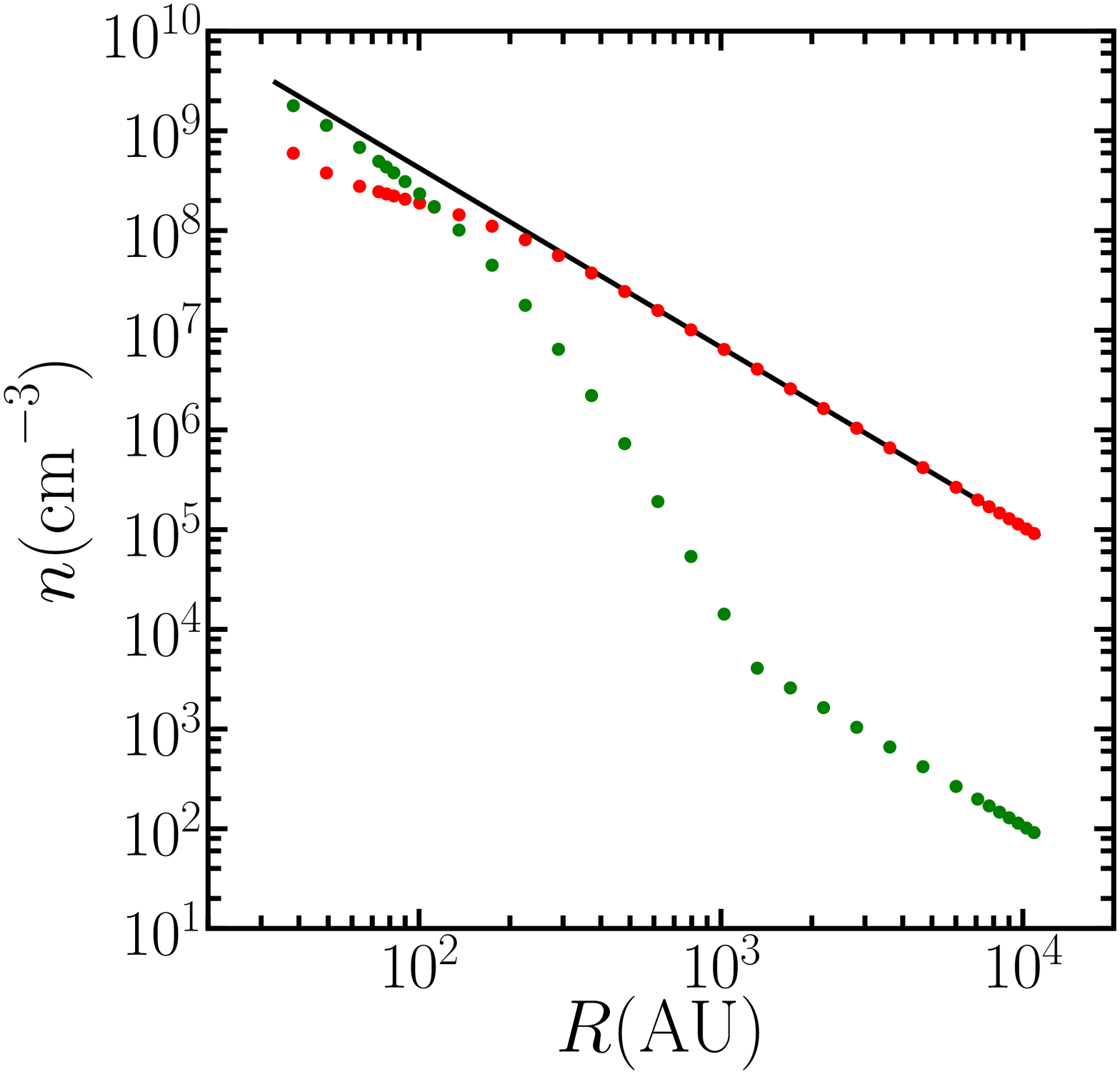}
\includegraphics[width=0.32\textwidth]{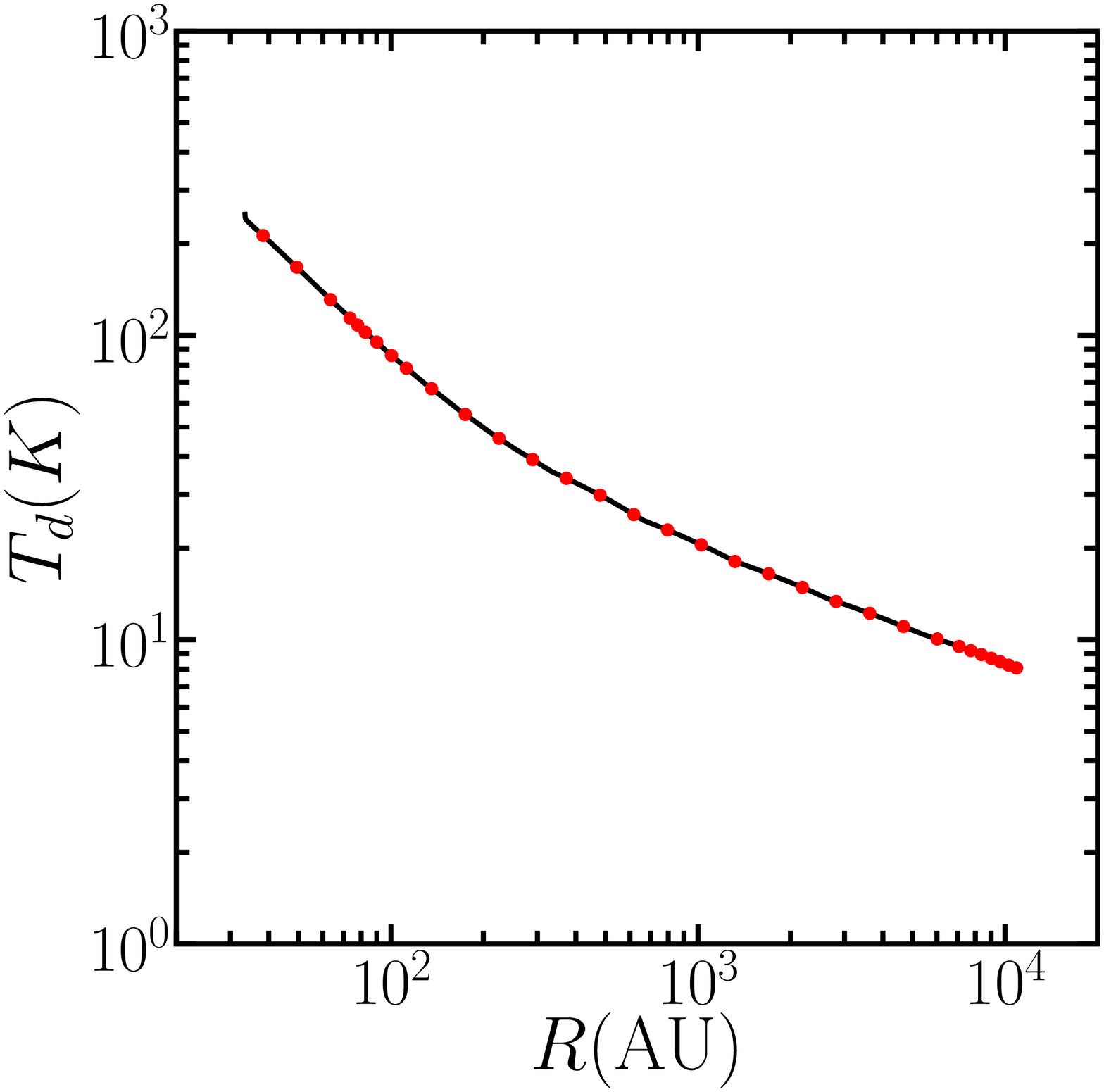}
\includegraphics[width=0.32\textwidth]{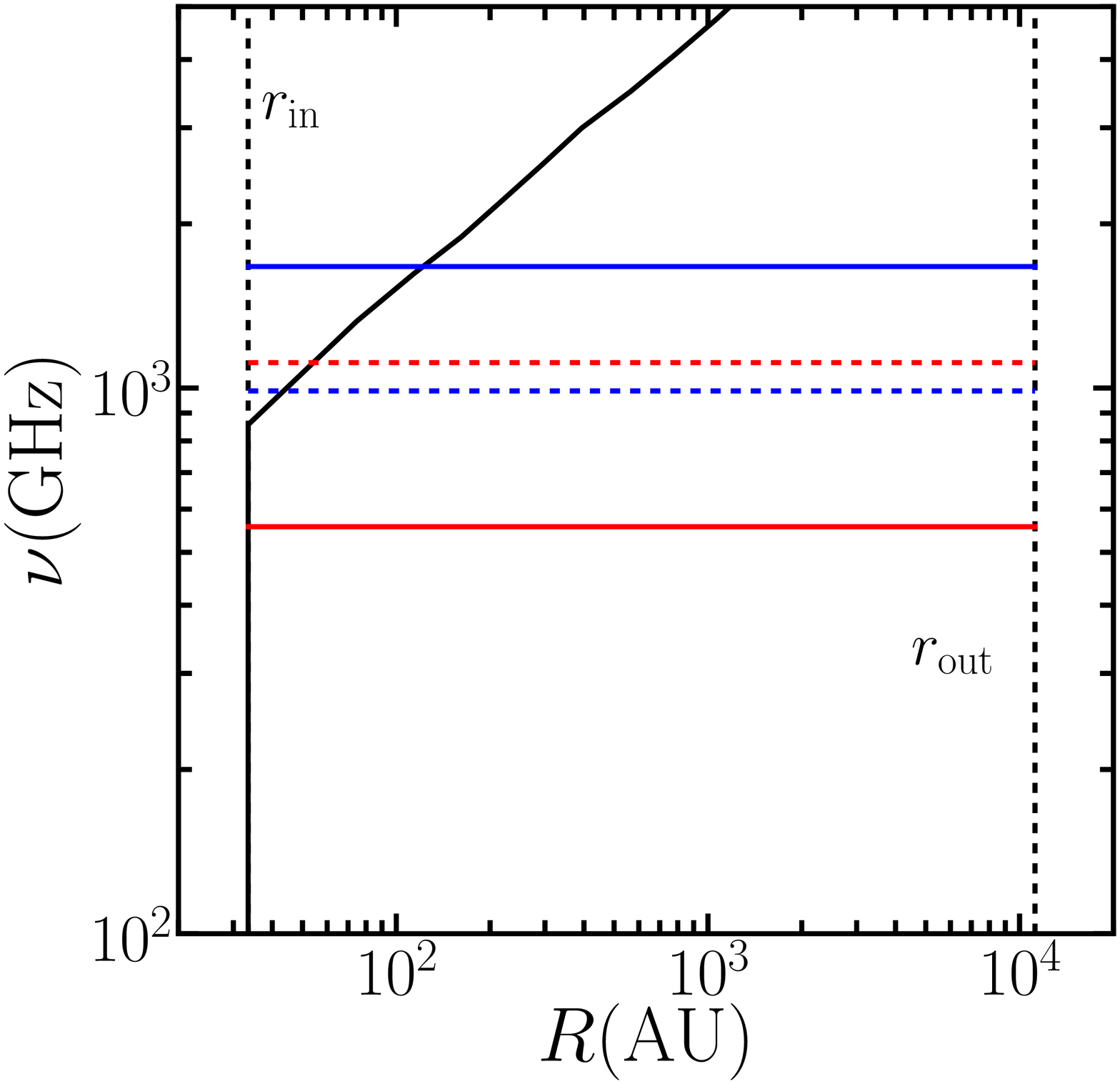}
\caption{Left: Density of H$_{2}$ as a function of radius for IRAS4A. The black line indicates the density power-law from the \textsc{DUSTY} model, while the green and red points indicate the ortho and para-H$_{2}$ densities respectively for each \textsc{RATRAN} cell. Middle: Temperature as a function of radius for the same model as the left plot, where the black line is an interpolation on a fine grid from the \textsc{DUSTY} model and the red points are the temperatures for each \textsc{RATRAN} cell interpolated from the same model. Right: The radius of the continuum $\tau$=1 surface (black) as a function of frequency. The H$_{2}$O 1$_{10}-$1$_{01}$ (solid red), 2$_{12}-$1$_{01}$ (solid blue), 1$_{11}-$0$_{00}$ (dashed red) and 2$_{02}-$1$_{11}$ (dashed blue) lines are all indicated with horizontal lines at the appropriate frequencies.}
\label{F:models_grid}
\end{center}
\end{figure*}

The density is assumed to follow a power-law (i.e. $n\,\propto\,r^{-p}$). The exponent ($p$), dust temperature profile and inner and outer radii were determined by \citet{Kristensen2012} using \textsc{DUSTY} 1-D continuum radiative transfer models \citep{Ivezic1997} to fit the spectral energy distributions (SEDs) from 70\micron{} to 1.3\,mm, scaled to the source bolometric luminosity, and the radial intensity profiles from SCUBA 450\micron{} and 850\micron{} images. For more details, see Appendix C of \citet{Kristensen2012}. We do not fit the wavelength dependance at shorter wavelengths as this primarily probes the small-scale geometry, which would require a 2/3-D continuum model to reproduce correctly. Such a model would have additional free parameters that would be poorly constrained at best, thus adding degeneracy. The DUSTY results used are given in Table~\ref{T:results}. The gas temperature is assumed to be the same as the dust temperature and interpolated onto the cell grid (see Sect.~\ref{S:models_grid}). The ortho and para H$_{2}$ densities are calculated from the density power law assuming a thermal ortho-to-para ratio, but with a lower limit of 10$^{-3}$ \citep{Pagani2009}. The OH5 dust opacities for grains with thin ice mantles from \citet[][]{Ossenkopf1994} are used when calculating the continuum for both the dust and line radiative transfer. An example of the density and temperature profile for IRAS4A is shown in Figure~\ref{F:models_grid}, along with the radius of the continuum $\tau$=1 surface as a function of frequency. Continuum optical depth is not an issue for our observations, as the line optical depth is in all cases much higher than the continuum.
 
\subsection{H$_{2}$O Abundance Structure}
\label{S:models_H2O}

The models also require a profile for how the water abundance with respect to H$_{2}$ varies with radius. As will be shown in Section~\ref{S:chemistry}, water line profiles are particularly sensitive to the shape of this profile. Previous water modelling studies have used parameterised water abundance profiles which either have two or three constant abundances within different zones defined based on temperature and density limits \citep[e.g.][]{vanKempen2008,Coutens2012,Herpin2012}, similar to those often used for other simple molecules such as CO \citep[e.g.][]{Jorgensen2002,Yildiz2012}. However, it was impossible to convincingly reproduce the observed inverse P-Cygni profiles using such a scheme. This will be discussed further in Section.~\ref{S:chemistry}.

\begin{figure}
\begin{center}
\includegraphics[width=0.35\textwidth]{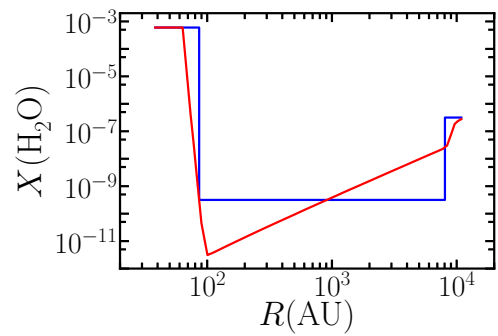}
\caption{Comparison of abundance profiles. The red and blue lines shows the best-fit abundance profiles for IRAS4A using simple water chemistry \citep[SWaN;][]{Schmalzlip} and a drop profile respectively.}
\label{F:models_abundance}
\end{center}
\end{figure}

The SWaN (Simplified WAter Network) chemical model of \citet{Schmalzlip}, based on the chemical description used by \citet{Hollenbach2009}, is therefore used to calculate the water abundance profile as a function of the density, temperature and extinction in the envelope. This takes into account the formation of water on dust grain surfaces through the freeze-out of atomic oxygen and subsequent hydrogenation, the freeze-out of water vapour onto dust grains, the thermal and photodesorption of water ice into the gas phase and the destruction of water vapour through photodissociation. All equations are solved in a time-dependent manner until steady-state is achieved (usually $<$10$^{5}$\,yr). The approach used in SWaN is similar to that used in \citet{Caselli2012}, which followed the earlier work on CO of \citet{Keto2008}.  Use of SWaN naturally leads to a photodesorption layer near the outer edge of the model where the $A_{V}$ is low enough for the interstellar radiation field to penetrate and lower densities lead to less efficient freeze-out, rather than having to add an external (absorbing) layer externally as implemented by \citet{Coutens2012}. The abundance profiles produced by SWaN are consistent with those produced by full chemical networks, unlike the drop abundance profile \citep{Schmalzlip} The best-fit abundance profile for IRAS4A is shown in Figure~\ref{F:models_abundance}. Separate profiles are calculated for each source based on the source physical structure.

The main free parameters which can alter the water abundance profile are the interstellar UV radiation field (ISRF, $G_{\mathrm{isrf}}$) and the cosmic-ray induced UV field ($G_{\mathrm{cr}}$). Both are scaling factors relative to the flux of interstellar photons, which we take to be $F_{\mathrm{isrf}}$=10$^{8}$ photons\,cm$^{-2}$\,s$^{-1}$ \citep[from an energy flux of 1.6$\times$10$^{-3}$ erg\,cm$^{-2}$\,s$^{-1}$ and an energy range of 6$-$13.6\,eV;][]{Hollenbach2009}. All volatile oxygen (i.e. all oxygen not locked up in refractory grains) is assumed to be in the form of water, leading to a total water abundance relative to H$_{2}$ of 6$\times$10$^{-4}$ \citep[from O/H = 3$\times$10$^{-4}$;][]{Meyer1998}. This does not take into account the fraction of oxygen in CO, and is therefore an upper limit. However, the abundance of gas-phase water is relatively insensitive to this quantity when $T\lesssim$100\,K, the primary region probed by our observations. We also assume a photo-desorption yield of 10$^{-3}$ \citep[][]{Oberg2009,Arasa2010} and grains with a radius of 0.1\micron{} \citep[][]{Bergin1995}. A more detailed discussion and comparison with more complex chemical networks will be presented in \citet{Schmalzlip}.

 The water abundance profile and water lines are sensitive to the density at the outer edge, particularly in the 1$_{10}-$1$_{01}$ line which shows significant emission at central velocities from the photodissociation layer if the density at the outer edge is above $\sim$10$^{5}$\pccm{}. As this is not seen in the observations, we set $G_{\mathrm{isrf}}$ to 0 if the outer density is $\geq$10$^{5}$\pccm{} (see Table~\ref{T:results}), as it will almost certainly be attenuated by cloud material at lower densities. In all other cases we set $G_{\mathrm{isrf}}$ to 1. 

 A constant ortho-to-para ratio for water of 3 is also assumed. A lower ortho-to-para ratio (e.g. 1) would lead to an increase (decrease) in the para (ortho) water abundance and thus an increase (decrease) in the intensity of emission and absorption features of all para (ortho) lines. However, most measurements are consistent with a value of 3 \citep[e.g.][]{Emprechtinger2013}. When performing the radiative transfer calculations, we use the latest collisional rate coefficients from \citet{Daniel2011} as downloaded from the Leiden Atomic and Molecular Database \citep[LAMDA;][]{Schoier2005}.

\subsection{Grid}
\label{S:models_grid}

As with all grid-based simulations, it is important that the grid of cells defined in \textsc{RATRAN} resolves the key regions of the envelope. To do this, we begin by defining a grid of 23 cells which are logarithmically spaced in density. Due to the abundance structure of the outer part of the model being so important to the lines we are interested in and because the abundance profile can change significantly over small changes in radii in that region, the outer two cells are then replaced with six cells which have a power-law spacing i.e. the number of cells increases towards the outer edge. Finally, the two cells which cover where the temperature crosses 100\,K are replaced by four cells to better resolve this jump (see Fig.~\ref{F:models_grid}). 

The inner radius for the grid is taken from the \textsc{DUSTY} model. This parameter is not particularly well constrained, but because it is inside the $\tau$=1 surface for all H$_{2}$O lines, it does not affect our results. The outer radius of the model is restricted to either the radius at which the temperature drops below 8\,K or the density drops below 10$^{4}$\pccm{}, whichever is the smaller. This ensures that the model envelopes do not extend to unphysically large radii. Material from the cloud outside this radius along the line of sight has a negligible impact on the line profiles unless a significant foreground cloud layer is present, in which case this layer is not added to the grid but is included later during the creation of synthetic images (see Sect.~\ref{S:models_outflow} for more details).

For Ser-SMM4 and L1157 the full DUSTY model, the SCUBA 850\micron{} images and low-$J$ CO maps \citep{Yildizip} all suggest that the envelope extends out to radii $\gtrsim$30\arcsec{} but the cuts discussed above would lead to a smaller outer radius. In these cases we therefore set the outer radius of the model to be 30\arcsec{} and limit the temperature to be no lower than 8\,K, which corresponds to thermal balance between CO line-cooling and cosmic-ray heating \citep[][]{Evans2001,Galli2002,Bergin2006}, transferred to the dust via gas-grain collisions.

In the case of IRAS4A the model is large enough that the outer layers of the model would encompass IRAS4B and almost touch IRAS4C if these sources are all at a common distance. However, the largest \textit{Herschel} beam is smaller than the model and the separation between the sources, so the properties of the model at radii much larger than the beam size only contribute to the model spectra at the front and back of the model along the line of sight, as is also the case for the observed spectra.

At densities below $\sim$10$^{5}$\pccm{} the dust temperature can rise above the gas temperature \citep{Doty1997}, contrary to our assumption that they are the same. External heating in the outer envelope \citep[e.g.][]{Jorgensen2006} can also have a similar effect which was not considered in the \textsc{DUSTY} models. However, any underestimate in the dust continuum caused by our possible underestimate of the dust temperature at the outer edge of the model will have a negligible effect on the total continuum flux, which is dominated by the inner envelope.

\subsection{Velocity and non-thermal motion profiles}
\label{S:models_v}

While \textsc{RATRAN} takes thermal broadening into account automatically, two other contributions to the line-width must be defined - systematic motions (i.e. infall or expansion) and non-thermal turbulent motions. Due to the observed inverse P-Cygni profiles in the observations, the material within the model is assumed to be in free-fall towards the protostar at the centre of the envelope/disk. The infall velocity ($\varv_{\mathrm{inf}}$) is scaled to the velocity at the inner radius of the \textsc{DUSTY} model ($r_{0}$) such that:

\begin{equation}
\varv_{\mathrm{inf}} (r) = \varv_{1000} \left(\frac{r}{1000\,\mathrm{AU}}\right) ^{-0.5}.
\label{E:v}
\end{equation}

A completely static envelope leads to a symmetric line profile, while increasing infall velocity leads to increasingly red-shifted absorption and blue-shifted emission. Whether this infall is throughout the whole envelope, or just on certain radii will be explored in Section~\ref{S:infall_vs_r}. Rotation is not considered, but would take the form of additional line broadening which increased with decreasing radius. There is little sign of this in the observations and it would impossible to distinguish from turbulent broadening using 1-D models.

In order to take account of turbulence within the envelope, we also define the doppler $b$ within the model, which is the 1/e half-width (0.601\,$\times$\,FWHM). Increasing $b$ broadens both the absorption and emission components of the line profile while decreasing the peak intensity. We set $b$ to be constant as a function of radius, though in some cases variation with radius may play a secondary role (see Section~\ref{S:infall_iras4a}).

\subsection{Additional emission/absorption components}
\label{S:models_outflow}

\begin{figure}
\begin{center}
\includegraphics[width=0.21\textwidth]{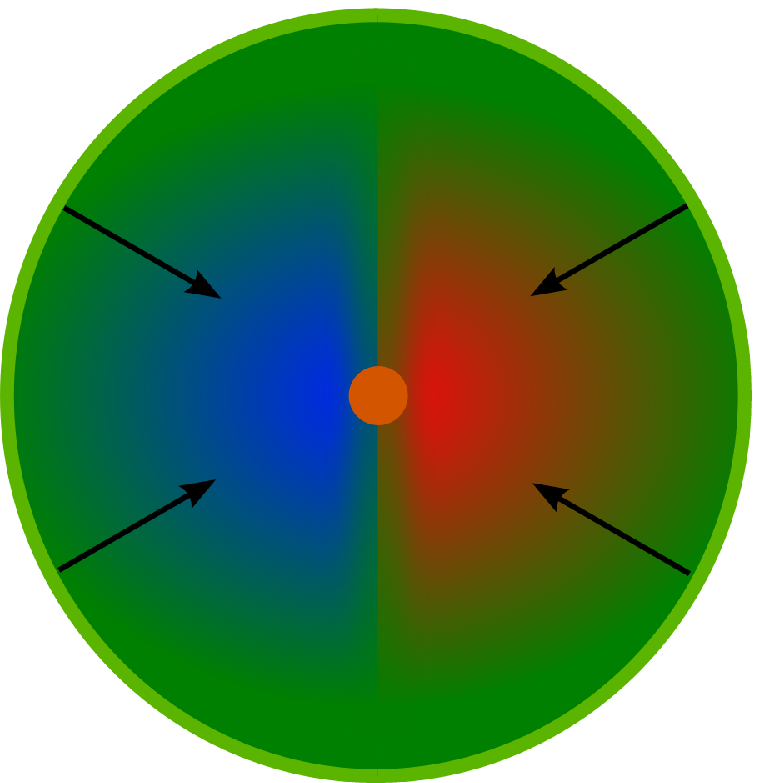}
\includegraphics[width=0.21\textwidth]{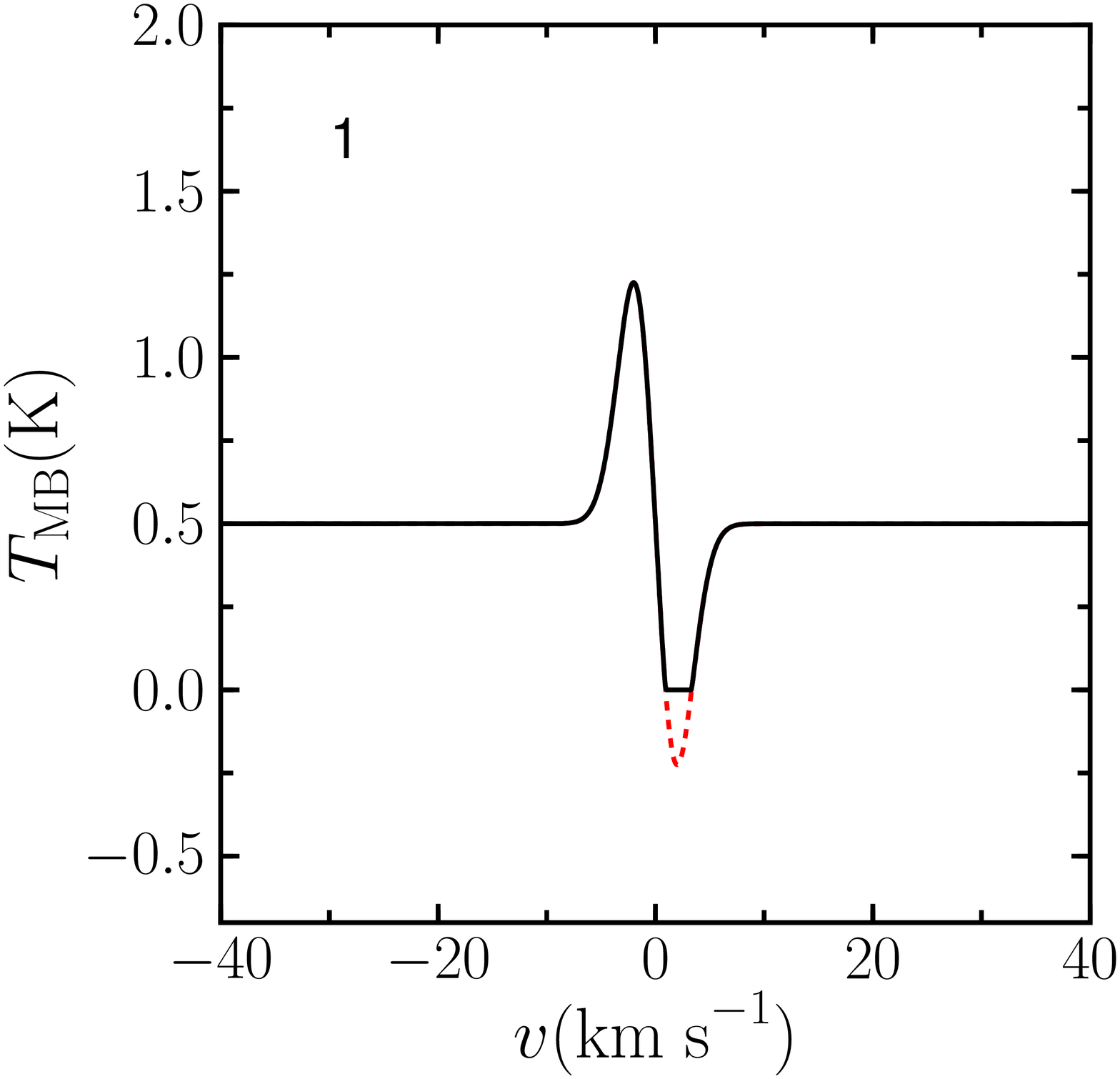}
\includegraphics[width=0.21\textwidth]{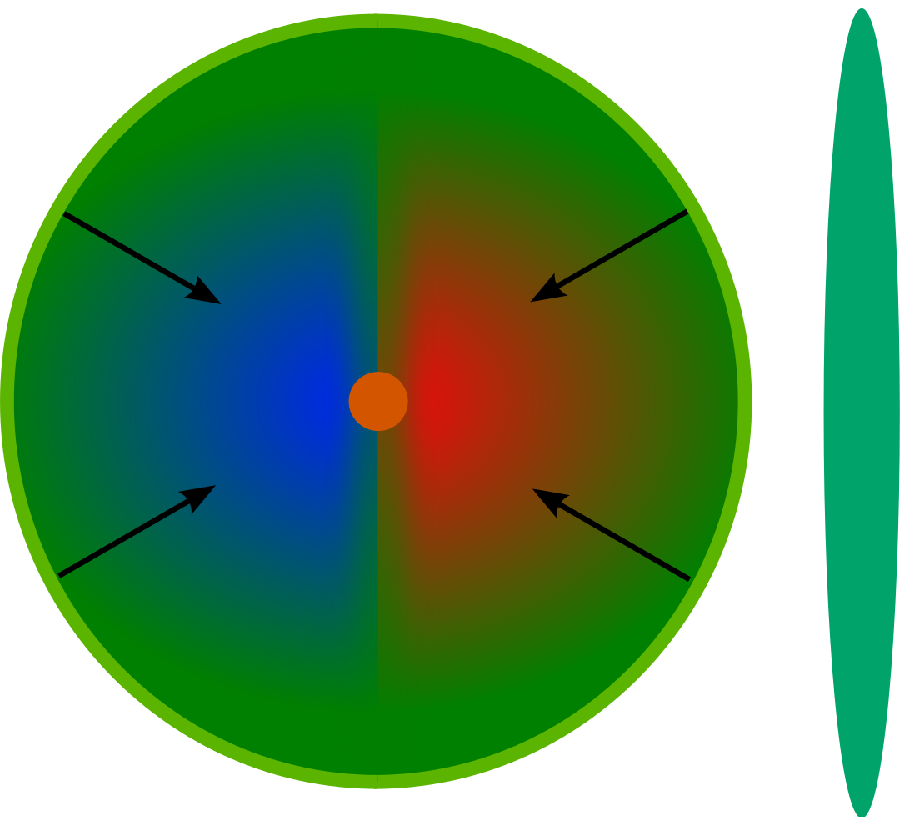}
\includegraphics[width=0.21\textwidth]{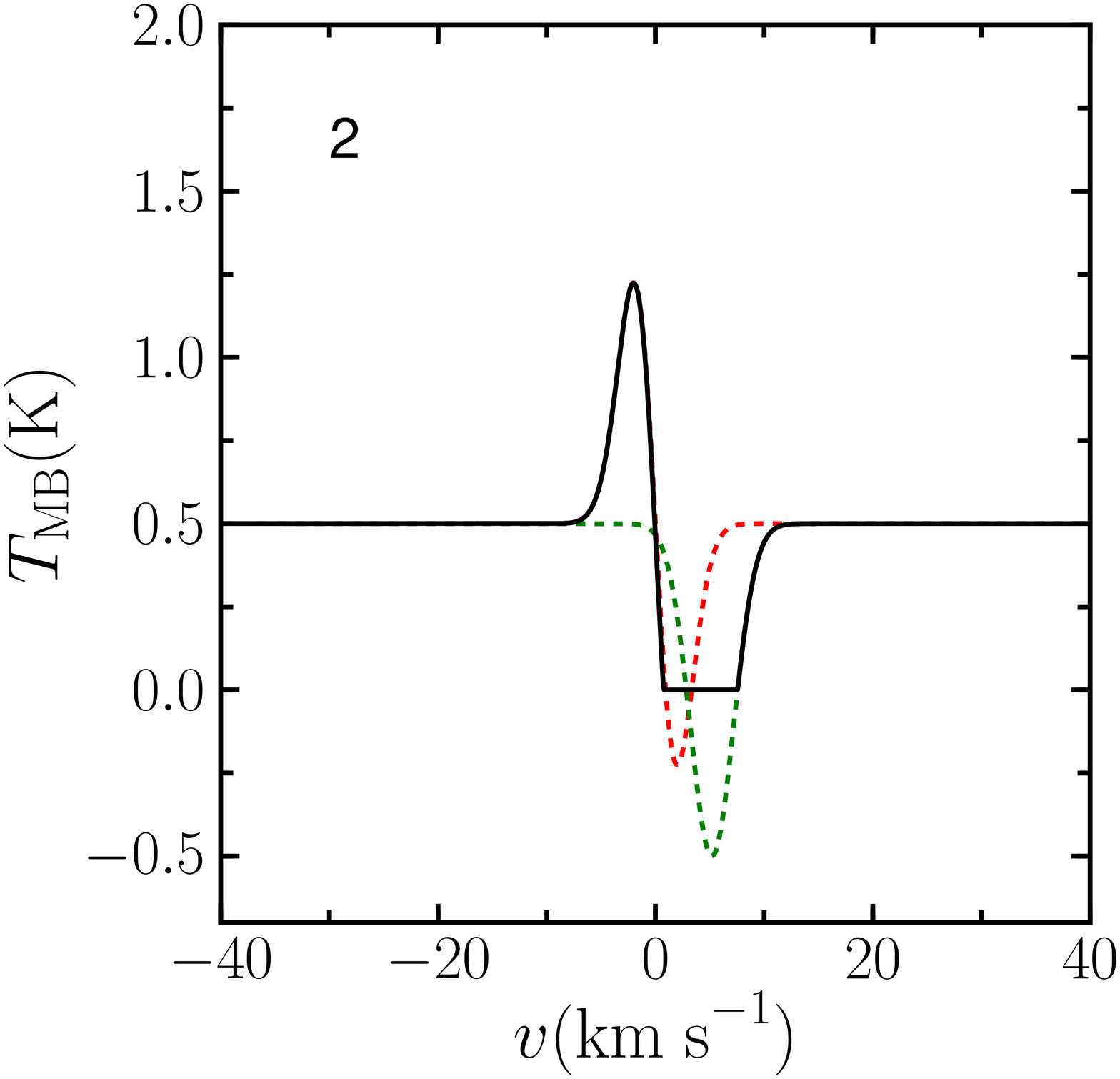}
\includegraphics[width=0.21\textwidth]{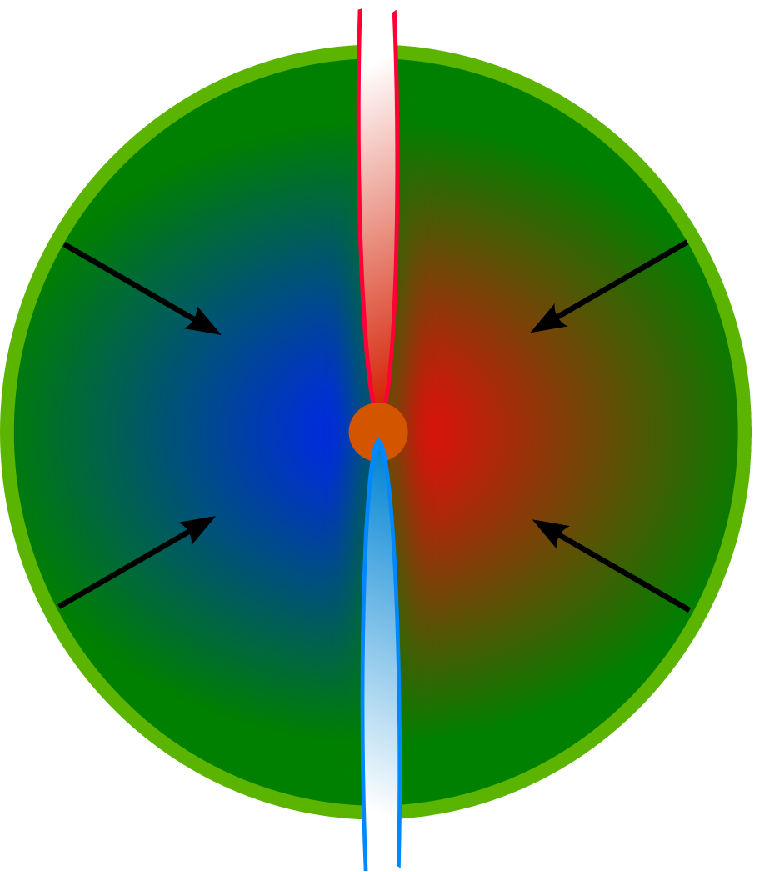}
\includegraphics[width=0.21\textwidth]{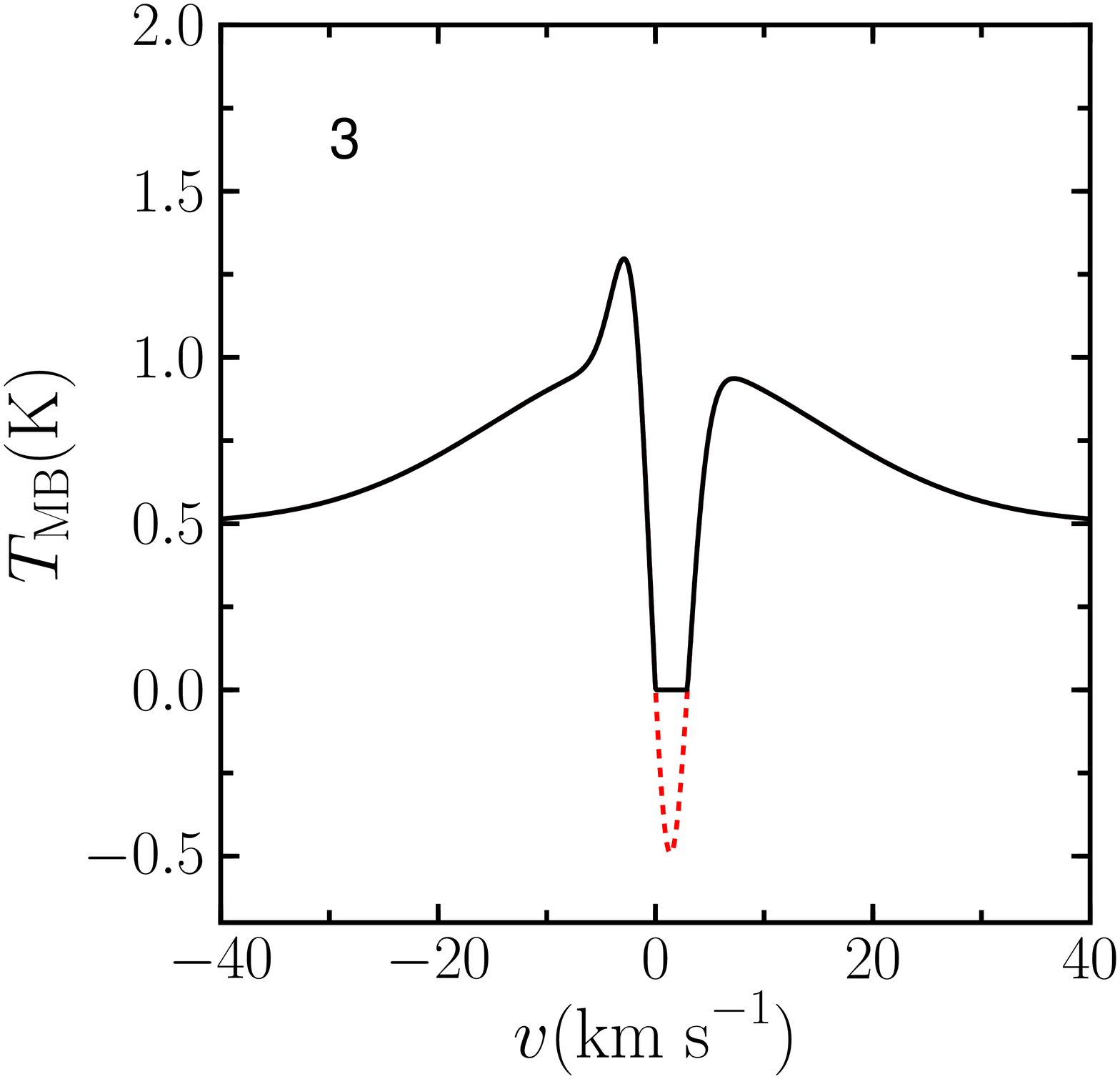}
\caption{Cartoon showing a slice through the envelope structure (left) and the corresponding line profile if viewed from the right-hand side in the plane of the page (right). The line of sight towards the model is from the right in the plane of the page. Top (Case 1): An infalling envelope where the absorption is only against the continuum. The observed line profile is saturated (i.e. flattened at the bottom of the absorption) because the absorption removes all the continuum photons. The red dashed line shows what the absorption part of the profile would look like if the absorption did not saturate. Middle (Case 2): A low-density foreground cloud is present, causing additional absorption which is likely offset from the source velocity, shown by the green dashed line. If this is close to the source velocity it will lead to the absorption being significantly wider than the envelope emission component, as is the case in IRAS4A (c.f. Figure~\ref{F:observations_data}). Bottom (Case 3): Outflow emission is added during ray-tracing in a plane at the centre of the envelope. The outer envelope can then absorb against both the continuum and the outflow, which may or may not lead to saturated absorption. }
\label{F:models_cartoon}
\end{center}
\end{figure}

For absorption to take place, a significant population of molecules must be in the lower level of a transition and a background emission source must provide photons of the correct frequency. In protostellar envelopes the dust continuum emission usually provides these photons, but absorption of line photons from deeper into the envelope or from other emission mechanisms is also possible. In addition, any material along the line of sight such as intervening clouds can also contribute to the absorption, though these clouds must be relatively cold and have low densities as they are not seen in emission and so only contribute to the ground-state absorption. 

Figure~\ref{F:models_cartoon} shows a cartoon of this process. In a simple infalling envelope the emission from the far side of the envelope is blue-shifted while the absorption and emission from the near side of the envelope is red shifted. If the absorption is deep enough, all line and continuum photons are absorbed and so the absorption is saturated. Absorption against the continuum emission is included automatically by \textsc{RATRAN} (c.f. case 1 in Figure~\ref{F:models_cartoon}). 

For foreground clouds, the level populations of the emitting (i.e. envelope) and absorbing (i.e. foreground cloud) material are physically unrelated and so the absorbing layer can be added as an additional source of opacity along the line of sight during the production of synthetic images (c.f. case 2 in Figure~\ref{F:models_cartoon}). In \textsc{RATRAN}, the required variables are the velocity offset and line-width common to all lines along with the brightness temperature and line-centre optical depth of this component for each transition.

As noted in Section~\ref{S:observations}, the absorption in the ground-state lines is saturated at an intensity below the continuum and so must be absorbing both the continuum and the broad outflow component (c.f. case 3 in Figure~\ref{F:models_cartoon}). We do not know the detailed properties of the outflow material on sub-beam scales, so we do not include the outflow in the physical model. Instead, in order to reproduce the observed line profiles in the models we use a similar approach as for foreground layers, by including the observed emission from the outflow after the excitation calculation but during the calculation of model images, with the difference that the emission is added as each ray passes from the back half to the front half of the envelope. This is illustrated in the left-hand cartoon in case 3 of Figure~\ref{F:models_cartoon}. 

During the synthetic image creation, i.e. after the level population calculation, each ray starts at the back (left-hand side) and passes horizontally to the front (right-hand side). The intensity in each velocity channel updated as the ray passes through each cell. To include the outflow, we add a broad gaussian velocity component to the intensity as the ray passes the mid-plane of the model. The emission is then absorbed at some velocities by the front-half of the envelope, in a similar way to the continuum. To create this broad outflow component, we use the line properties (i.e. line-width, brightness temperatures etc.) obtained from the gaussian fit to the data, assuming that the optical depth in this component is $\ll$1. This additional emission is then attenuated by the front half of the envelope. We subsequently remove the broad gaussian emission from after ray-tracing but before the image is convolved to the observing beam, leaving only the absorption.

The main purpose of adding such a simple 'outflow' component is to create a broad emission feature against which absorption by the front half of the envelope can occur. By subsequently subtracting the same broad emission feature so only the absorption remains, we ensure that the details of the broad emission feature are unimportant as long as the overall level of emission was correct so that the opacity of the front half of the envelope results in the right absorption depth in terms of antenna temperature. Including the outflow emission in this way assumes that the emission fills the beam and is unrelated to the excitation of the envelope. While this implementation is artificial and both of these assumptions may not strictly be true, this is probably a good approximation without including a full outflow - something which is a field of study in its own right.

While we assume that the outflow layer has a low optical depth, if it instead has a high optical depth then no emission from the back half of the envelope would be visible. In this case we would see all the red-shifted absorption but very little of the blue-shifted emission. Our assumption should not change the best-fit infall velocities, as these are derived primarily by concentrating on the absorption part of the profile. However, higher optical depth would hide emission and thus we would underestimate the abundance profile which is constrained primarily based on the emission. In reality, the outflow will have high-tau but only in some fraction of the beam while the rest of the emission is unhindered, resulting in something of a compromise situation.

\section{Infall}
\label{S:infall}

\begin{table*}
\begin{center}
\caption[]{Best fit model properties}
\begin{tabular}{lccccccccc}
\hline
\hline\noalign {\smallskip}
&\multicolumn{5}{c}{\textsc{DUSTY}}&&\multicolumn{3}{c}{\textsc{RATRAN}} \\ \cline{2-6} \cline{8-10}
\noalign{\vspace{1mm}}
Source & $p$ & $r_{0}$ & \multicolumn{1}{c}{$n_{0}$} & $r_{\mathrm{out}}$ & $n_{\mathrm{out}}$ && $\varv_{\mathrm{1000}}$ & $b$ & log$_{10}$($G_{\mathrm{cr}}$) \\
& & (AU) & \multicolumn{1}{c}{(10$^{8}$\pccm{})} & (10$^{3}$\,AU) & \multicolumn{1}{c}{(10$^{4}$\pccm{})} && (\kms{}) & (\kms{}) & ($F_{\mathrm{isrf}}$)\\
\hline\noalign {\smallskip}
IRAS4A & 1.8 & 33.5 & 30.5 & 11.2 & \phantom{0}8.7\phantom{\tablefootmark{a}} && 1.1 $\pm$ 0.2\phantom{\tablefootmark{b}} & 0.4 $\pm$ 0.1 & -4.00 $\pm$ 0.25 \\
L1527 & 0.9 & \phantom{0}5.4 & \phantom{0}0.9  & \phantom{0}6.5 & 15.0\tablefootmark{a} && 0.4$\pm$0.2\phantom{\tablefootmark{b}} & 0.9 $\pm$ 0.1 & -4.25 $\pm$ 0.25 \\
BHR71 & 1.7  & 24.8 & \phantom{0}9.4 & 12.4 & \phantom{0}2.4\phantom{\tablefootmark{a}} && 1.3 $\pm$ 0.3\phantom{\tablefootmark{b}} & 0.2 $\pm$ 0.1 & -4.50 $\pm$ 0.25 \\
IRAS15398 & 1.4 & \phantom{0}6.1 & 19.4 & \phantom{0}4.9 & 17.1\tablefootmark{a} && 0.9 $\pm$ 0.2\phantom{\tablefootmark{b}} & 0.7 $\pm$ 0.1 & -4.25 $\pm$ 0.25 \\
GSS30-IRS1 & 1.6 & 16.2 & \phantom{0}1.2 & \phantom{0}5.9 & \phantom{0}1.0\phantom{\tablefootmark{a}} && 1.0 $\pm$ 0.3\phantom{\tablefootmark{b}} & 0.8 $\pm$ 0.1 & -3.50 $\pm$ 0.50\\
Ser-SMM4  & 1.0 & 12.1 & \phantom{0}7.9 & \phantom{0}12.5 & 43.1\tablefootmark{a} && 6.0 $\pm$ 0.6\tablefootmark{b} & 0.7 $\pm$ 0.1 & -4.00 $\pm$ 0.25 \\
L1157 & 1.6 & 14.4 & 17.4 & \phantom{0}9.8 & \phantom{0}5.1\phantom{\tablefootmark{a}} && 1.4 $\pm$ 0.5\phantom{\tablefootmark{b}} & 0.3 $\pm$ 0.1 & -4.75 $\pm$ 0.25 \\
\hline\noalign {\smallskip}
\label{T:results}
\end{tabular}
\tablefoot{\tablefoottext{a}{$G_{\mathrm{isrf}}$=0 due to high outer density}
  \tablefoottext{b}{$\varv_{1000}$ is the infall velocity extrapolated to 1000\,AU. In the case of Ser-SMM4 this seems too high to be reasonable. The implications of these results are discussed further in Section~\ref{S:infall_sample}.}}
\end{center}
\end{table*}

\textsc{RATRAN} models as described above were run simultaneously for ortho and para H$_{2}$O for a range of values of $\varv_{1000}$, $b$ and $G_{\mathrm{cr}}$. Using a $\chi^{2}$ statistic to find the best-fit model is not straightforward because of the variation in $S/N$ between the lines. We therefore find the best fit by varying the different parameters and using by-eye comparison to decide which model is the most consistent with the data, within the uncertainties of the spectra. The properties of the best-fit model for all sources are presented in Table~\ref{T:results}, along with approximate uncertainties obtained by changing each parameter individually until a noticeably worse fit was achieved. The columns give the density power law ($p$), inner and outer model radii and densities ($r_{0}$,$r_{\mathrm{out}}$,$n_{0}$ and $n_{\mathrm{out}}$), the infall velocity at 1000\,AU ($\varv_{1000}$), the doppler $b$ and $G_{\mathrm{cr}}$. Note that though we extrapolate the velocity field to the inner radius of the model, in the case of Ser-SMM4 this results in unfeasibly large velocities. This will be discussed more in Section~\ref{S:infall_sample}.

\subsection{IRAS4A: a representative model}
\label{S:infall_iras4a}

\begin{figure}
\begin{center}
\includegraphics[width=0.43\textwidth]{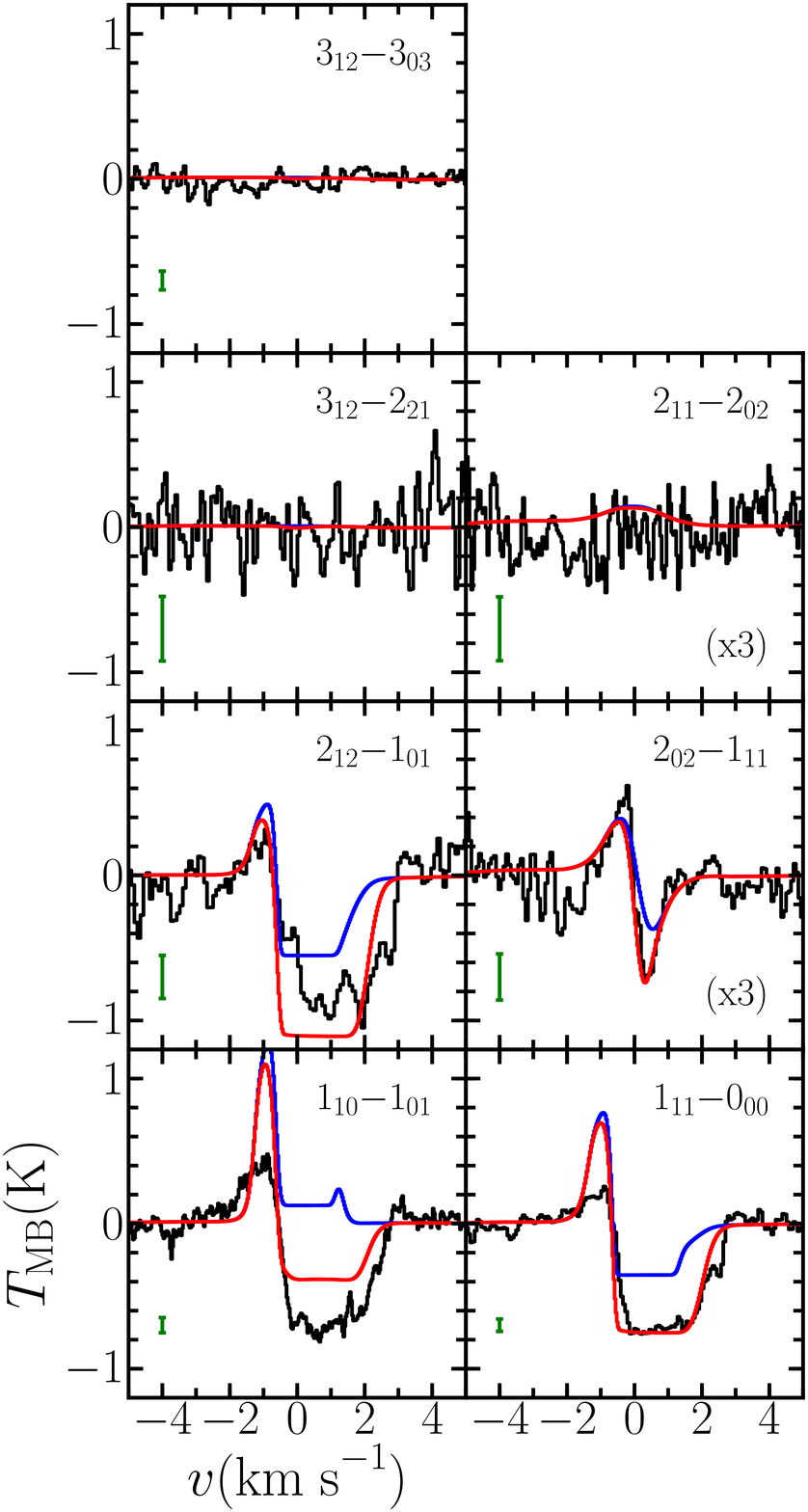}
\caption{Comparison between the observed (black) and best-fit model H$_{2}$O line profiles for IRAS4A both with (red) and without (blue) including absorption against the outflow and by foreground clouds as described in Section~\ref{S:models_outflow}. The green error-bar in the lower left corner of each plot indicates the 1$\sigma_{\mathrm{rms}}$ uncertainty in the observations. For some lines, both the model and the data are scaled up by a factor indicated in the panel to aid comparison.}
\label{F:infall_iras4a_fit}
\end{center}
\end{figure}

Let us now examine in detail the best-fit model for IRAS4A, and variations in parameter space around it, as this source shows an inverse P-Cygni profile in the excited 2$_{02}-$1$_{11}$ line. This rules out foreground absorption as the explanation for the inverse P-Cygni profiles because absorption in this line in typical cloud conditions ($T\sim10$\,K, $n\sim10^{4}$\pccm{}, $X_{\mathrm{H}_{2}\mathrm{O}}\sim$10$^{-7}$) would require water column densities of order 8$\times$10$^{15}$\,cm$^{-2}$, and therefore path-lengths greater than 2.5\,pc, so there must be infall in the envelope. In addition to absorption against both the outflow and continuum (see Fig.~\ref{F:observations_data}), IRAS4A also shows absorption in the ground-state lines due to a low density foreground cloud which is redshifted by 0.8\kms{} with respect to the systemic velocity of IRAS4A (i.e. at $\varv_{\mathrm{LSR}}$=8.0\kms{}). This cloud component was identified in CO observations by \citet[][]{Liseau1988} and has recently also been tentatively detected in O$_{2}$ emission by \citet{Yildiz2013b} with \textit{Herschel}. They find a line FWHM of 1.3\kms{} and total column density of $N({H_{2}})$=10$^{22}$\,cm$^{-2}$ with temperature in the range 20$-$70\,K and density in the range 7$-$2$\times$10$^{3}$\pccm{}. Using the average density and temperature, we used the non-LTE slab-modelling code \textsc{RADEX} \citep{vanderTak2007} to calculate optical depths and peak line intensities to be input into \textsc{RATRAN} as a foreground layer. We find that the total water abundance in this layer must be of order 10$^{-8}$ in order to reproduce the width of the observed absorption. Such an abundance is typical of gas-phase chemistry \citep[e.g.][]{McElroy2013}. Figure~\ref{F:infall_iras4a_fit} shows a comparison between the best-fit model (parameters given in Table~\ref{T:results}) and the observations both with and without including the foreground and outflow absorption to illustrate the impact this has on the model line-profiles. 

The best model fits the observed absorption and emission in the 2$_{02}-$1$_{11}$ and 2$_{12}-$1$_{01}$ lines very well, and reproduces the observed non-detections in the 2$_{11}-$2$_{02}$, 3$_{12}-$2$_{21}$ and 3$_{12}-$3$_{03}$ lines. The absorption in the 1$_{11}-$0$_{00}$ line is well reproduced though there is more emission in the model than in the observations. The emission in the model 1$_{10}-$1$_{01}$ line is more intense and narrower than the observations by approximately a factor of two, while the absorption is still underproduced. The latter is most likely due to small differences between the strength of the continuum and outflow component in the models with respect to the observations. The absorption observed in the 2$_{02}-$1$_{11}$ is offset from the source velocity by 0.3\kms{}, which is inconsistent with it being due to the foreground cloud at 0.8\kms{}, and so the line profile can only be reproduced with an infalling envelope.

\begin{figure*}
\begin{center}
\includegraphics[width=0.48\textwidth]{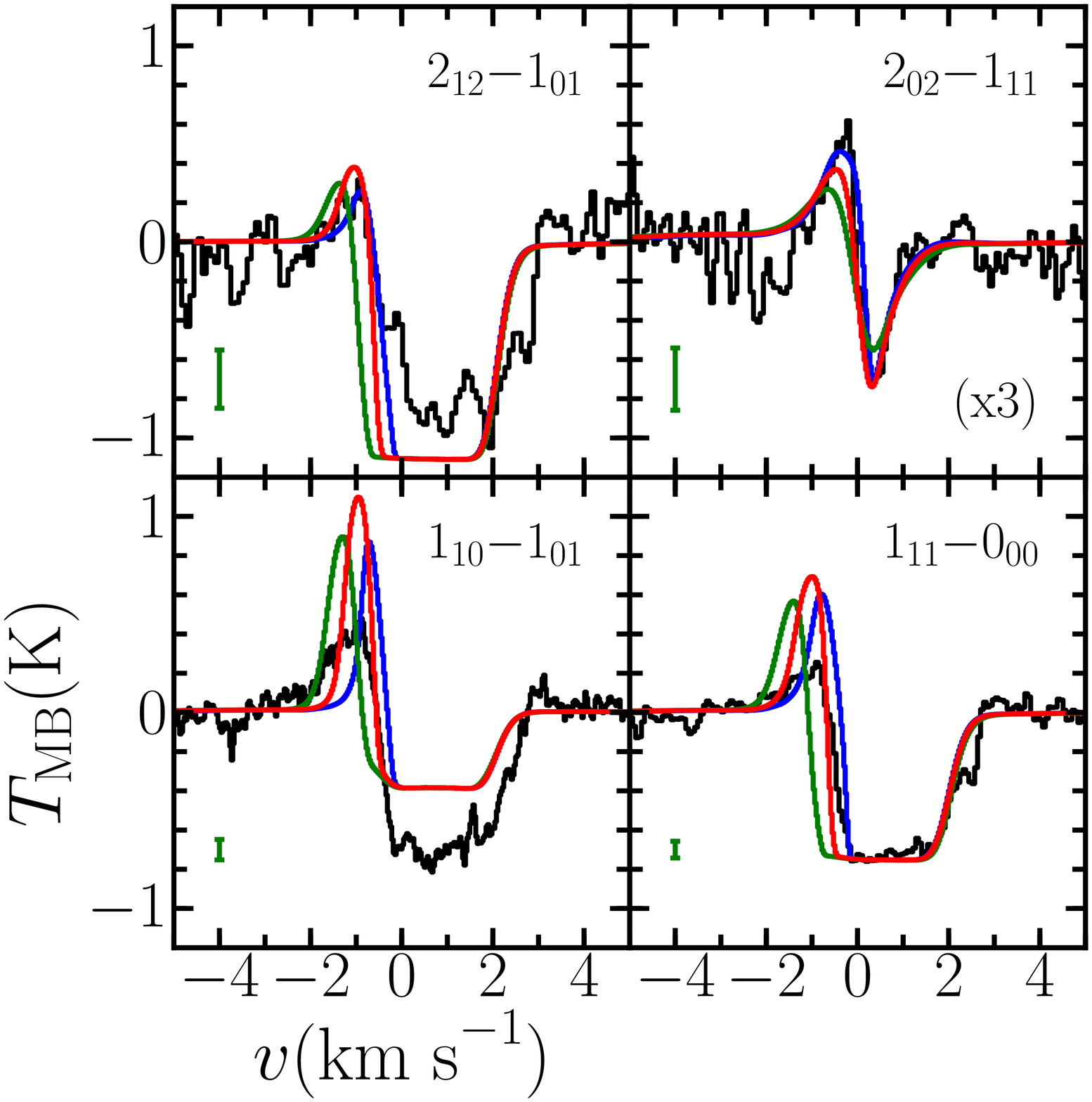}
\includegraphics[width=0.48\textwidth]{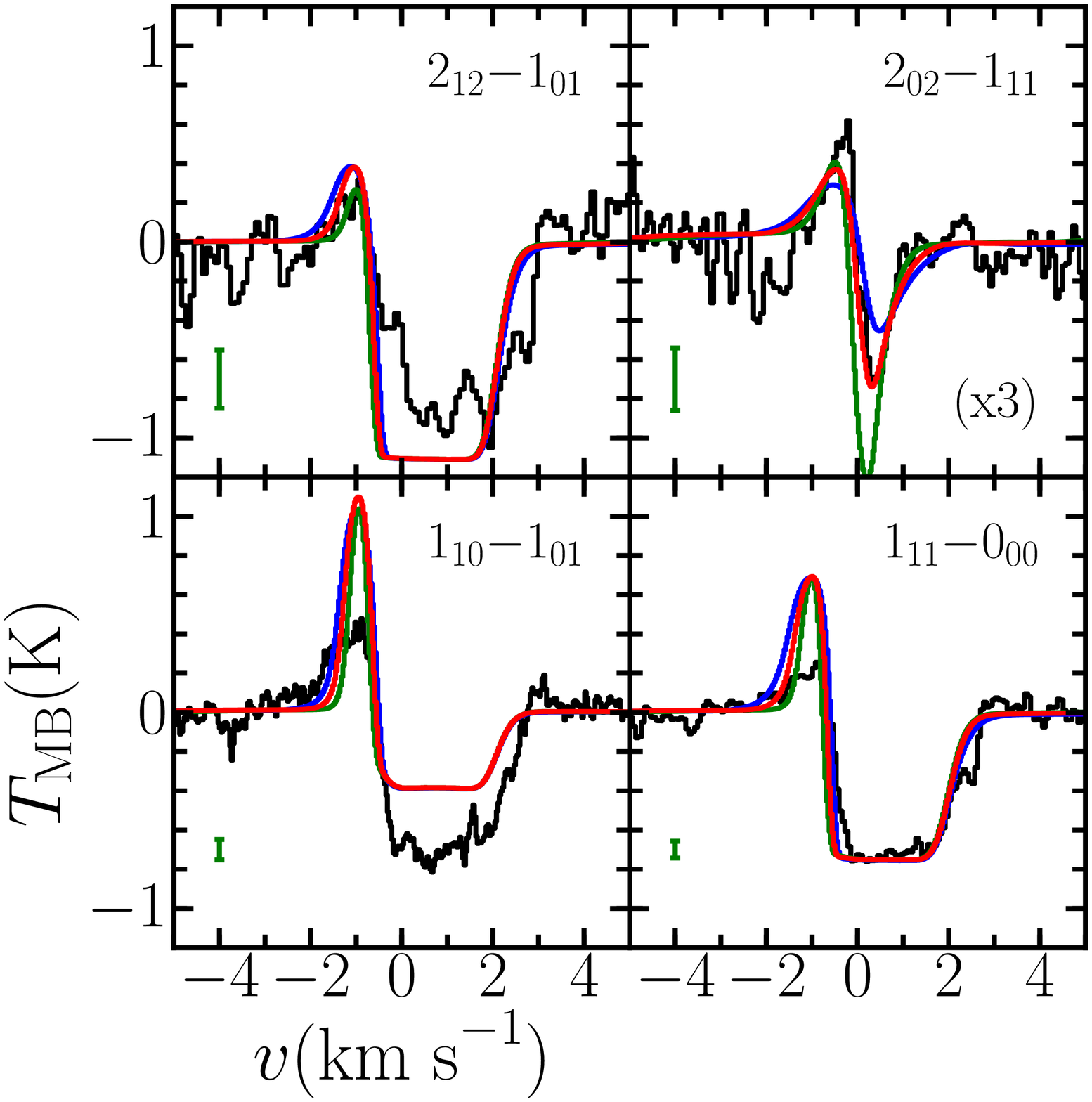}
\caption{Left: Comparison of model H$_{2}$O line profiles for IRAS4A with $b$ varying between 0.2\kms{} (blue), 0.4\kms{} (red, best-fit) and 0.6\kms{} (green). Right: Comparison of model H$_{2}$O line profiles for IRAS4A with $\varv_{1000}$ varying between 0.7\kms{} (blue), 1.1\kms{} (red, best-fit) and 1.5\kms{} (green). All other variables take the best-fit values and are held constant. The green error bar and scaling factors are the same as in Figure~\ref{F:infall_iras4a_fit}.}
\label{F:infall_iras4a_vfield}
\end{center}
\end{figure*}

To explore what parameters the different lines are sensitive to, we show in Figure~\ref{F:infall_iras4a_vfield} how the model line-profiles vary with different values for $b$ and $\varv_{1000}$. For the remainder of the paper we only show lines which are sensitive to the parameter under consideration. The 1$_{11}-$0$_{00}$ and 1$_{10}-$1$_{01}$ lines are sensitive to $b$ but not strongly sensitive to $\varv_{1000}$ while the 2$_{02}-$1$_{11}$ and 2$_{12}-$1$_{01}$ lines are sensitive to both parameters. While increasing $b$ would improve the fit to the emission component of the 1$_{10}-$1$_{01}$ line, this broadens the absorption in the other lines too much.

\begin{figure}
\begin{center}
\includegraphics[width=0.35\textwidth]{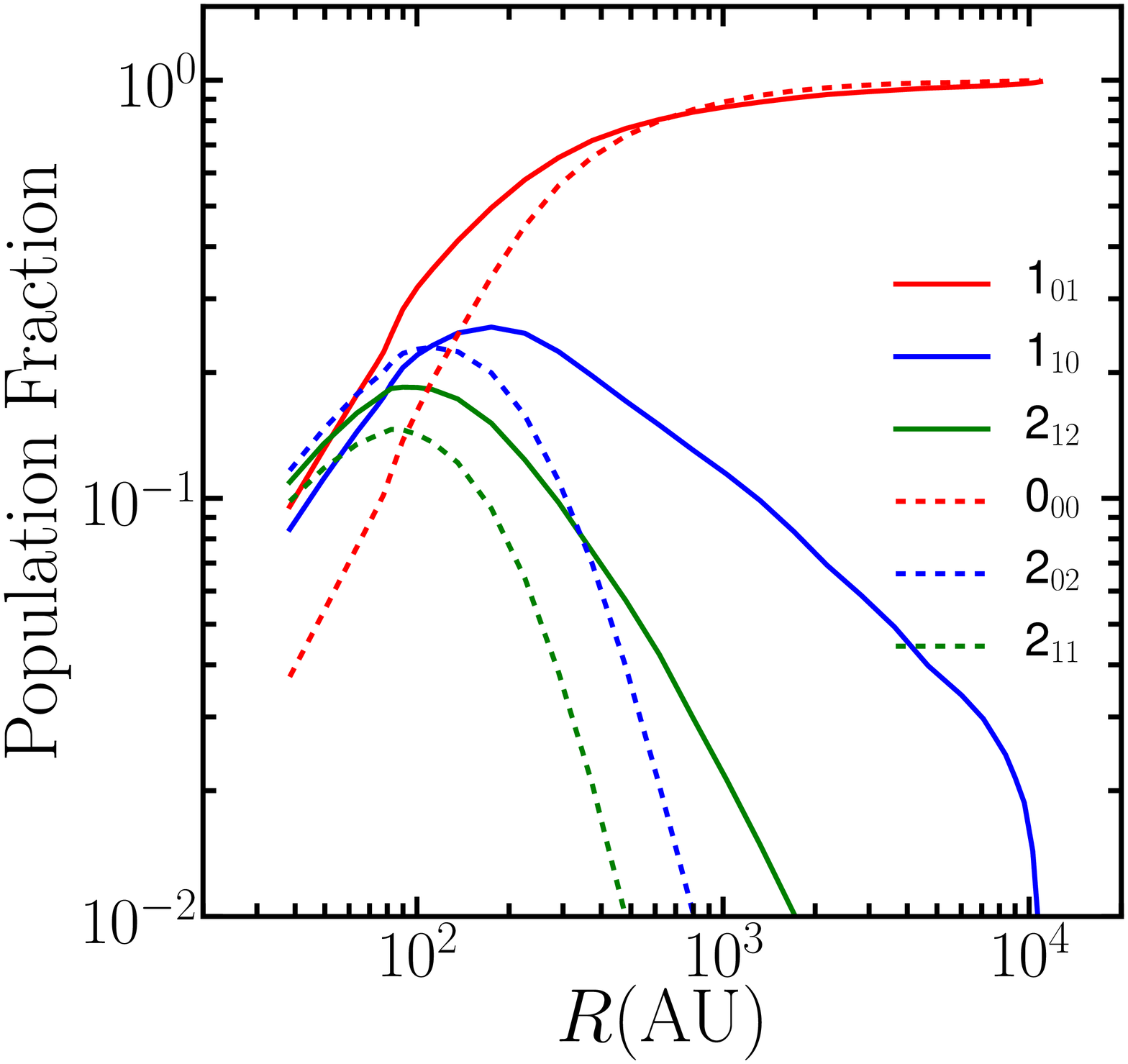}
\includegraphics[width=0.35\textwidth]{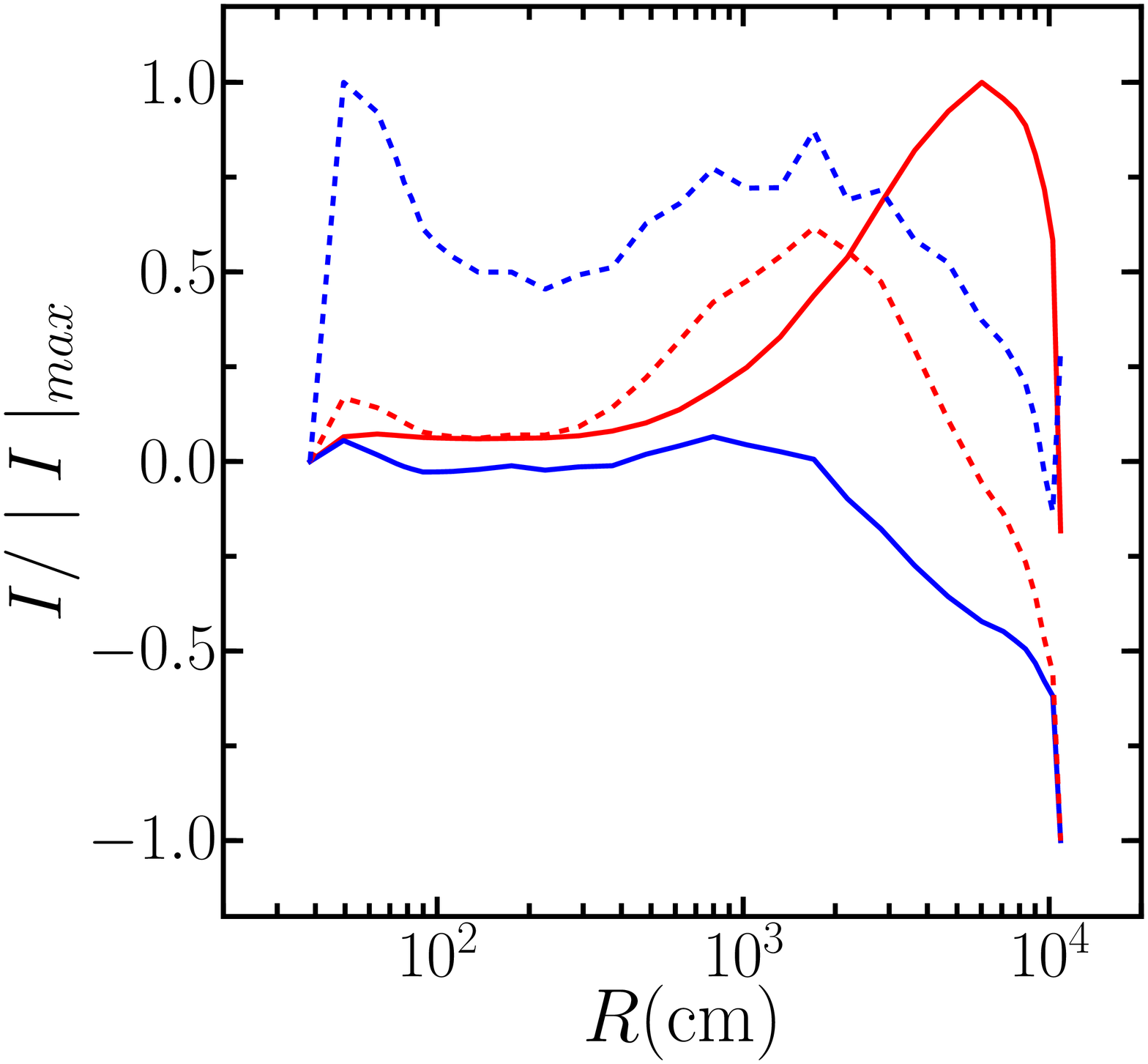}
\caption{Top: Population fraction as a function of radius for the ortho (solid) and para (dashed) levels contributing to the four detected lines for the best-fit model for IRAS4A. Bottom: Cumulative contribution ($I(r)$) for the H$_{2}$O 1$_{10}-$1$_{01}$ (solid red), 2$_{12}-$1$_{01}$ (solid blue), 1$_{11}-$0$_{00}$ (dashed red) and 2$_{02}-$1$_{11}$ (dashed blue) lines for the best-fit model relative to the maximum absolute integrated intensity in each line ($\mid I\mid_{\mathrm{max}}$).}
\label{F:infall_iras4a_contribution}
\end{center}
\end{figure}

The sensitivity of each line to infall and/or turbulence can be understood if we consider which regions within the envelope these lines probe. The top panel of Figure~\ref{F:infall_iras4a_contribution} shows the fraction of water molecules at a given radius in the first three energy levels of ortho and para water. The majority of water molecules in the outer envelope are in the ground-state, meaning that the excited lines are emitted in a much smaller region and so are subject to stronger beam dilution, explaining why no emission or absorption is detected in the 2$_{11}-$2$_{02}$, 3$_{12}-$2$_{21}$ and 3$_{12}-$3$_{03}$ lines. The lower panel of Figure~\ref{F:infall_iras4a_contribution} shows the cumulative integrated intensity of the four detected lines within a given radius in the model ($I(r)$) relative to the maximum absolute integrated intensity in each line ($\mid I\mid_{\mathrm{max}}$). This is calculated for each radius by taking the level populations from the full model and re-running the ray-tracing and beam convolution with all cells outside that radius removed. The figure therefore gives a measure of how different radii within the model contribute to the emission and absorption for each line. The foreground layer was only included for the full model, hence the drop at the outer edge in the three affected lines. The emission from the 1$_{10}-$1$_{01}$ line peaks around $\sim$7000\,AU, outside of which absorption dominates, while the emission for the other lines peaks between 700 and 2000\,AU.

\begin{figure}
\begin{center}
\includegraphics[width=0.4\textwidth]{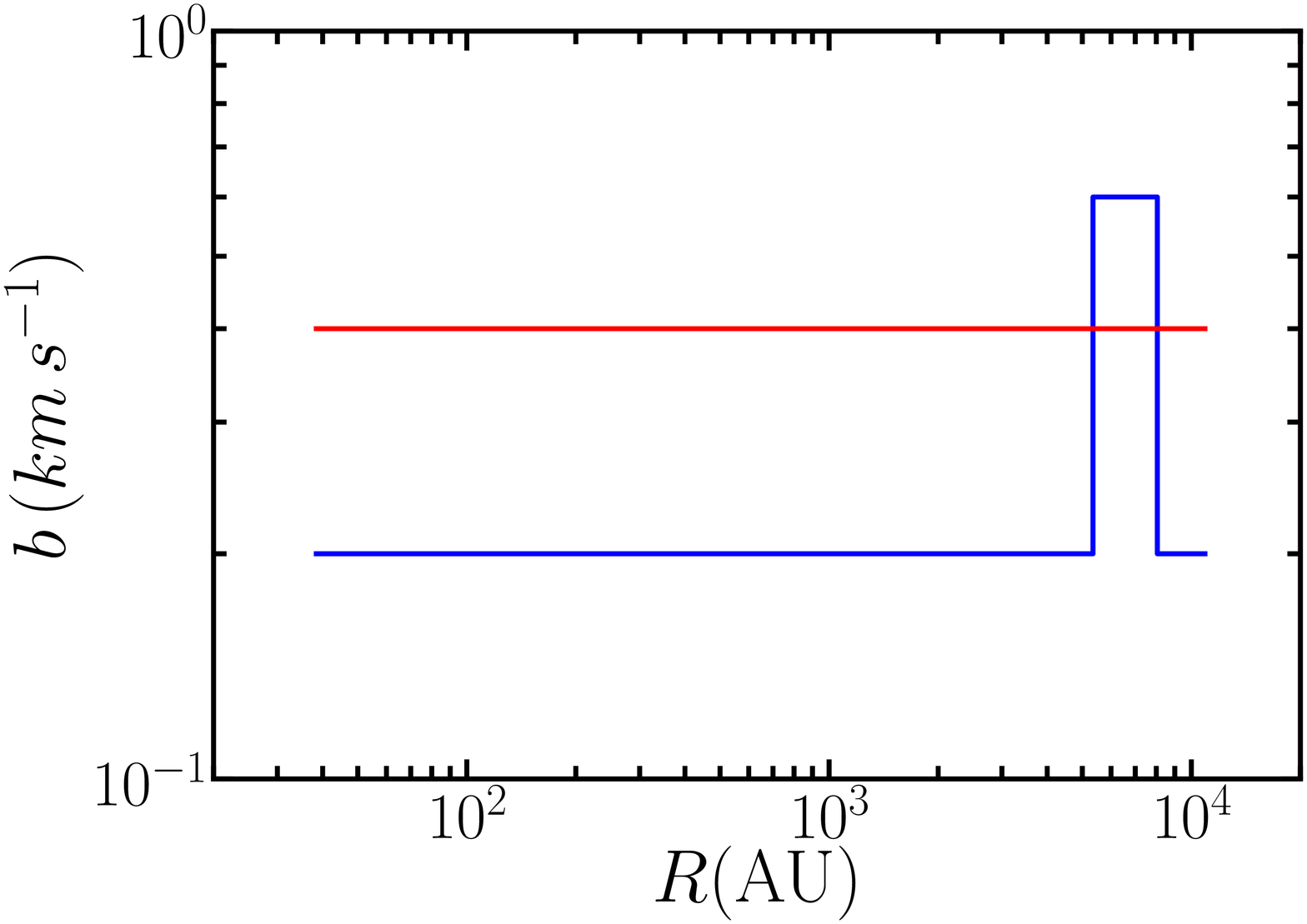}
\includegraphics[width=0.45\textwidth]{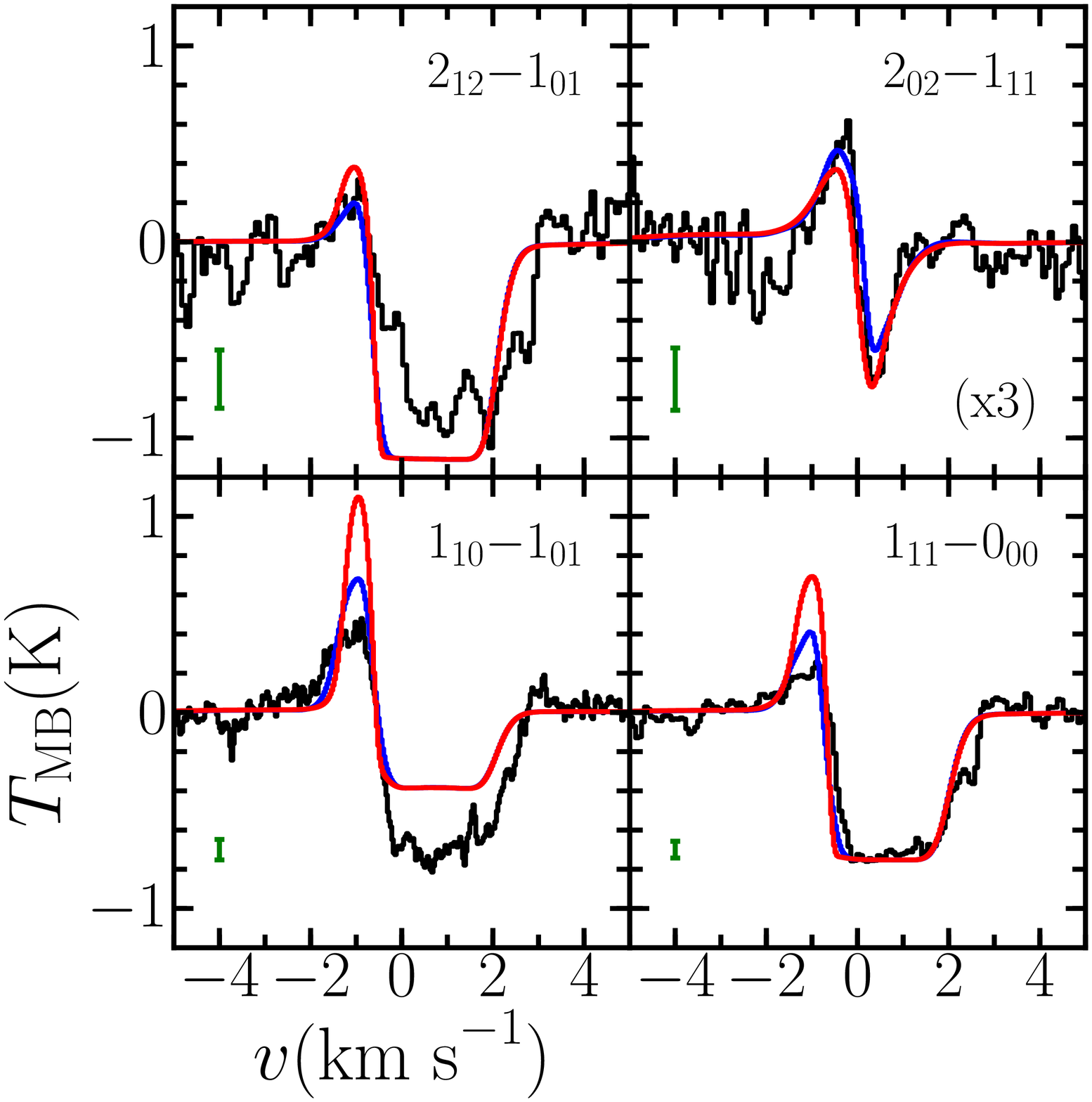}
\caption{Top: The best-fit model for IRAS4A with a constant $b$ profile (red) and one with a jump between 5000 and 8000\,AU (blue) but with all other variables held at the best-fit values. Bottom: A comparison of the resulting line profiles with the data. The green error bar and scaling factors are the same as in Figure~\ref{F:infall_iras4a_fit}.}
\label{F:infall_iras4a_variabledb}
\end{center}
\end{figure}

The 1$_{10}-$1$_{01}$ line is therefore much less sensitive to $\varv_{1000}$ because at the radii it probes, $\varv_{0}$ $\lesssim b$. The best-fit model does not reproduce the line-width of the emission from the 1$_{10}-$1$_{01}$ line well but the $b$=0.5\kms{} model has a broader emission component. A higher turbulent line-width is therefore required in the region where the line emission comes from. This zone of higher turbulence cannot extend all the way to the outer edge as this would still lead to much broader absorption than is observed, which comes from the very outer layers. It also cannot extend to the inner edge of the model because the other lines would then be broader (c.f. Fig.~\ref{F:infall_iras4a_vfield}). 

Figure~\ref{F:infall_iras4a_variabledb} shows a model with a jump in $b$ from 0.2 to 0.6\kms{} between 5000 and 8000\,AU, confirming this reasoning. This also solves the overproduction of emission in the 1$_{10}-$1$_{01}$ and 1$_{11}-$0$_{00}$ lines. A decrease in turbulence to smaller envelope radii might be expected due to the increasing density allowing turbulence to be damped more efficiently, and this has been observed before towards the high-mass YSO W43-MM1 \citep[][]{Herpin2012} and in the B5 region of the Perseus molecular cloud \citep{Pineda2010}. However, it is unclear what physical mechanism would cause such a region of higher turbulence inside the envelope without causing a shock, which would cause variations in the infall velocity profile which we do not see in this region (see Section~\ref{S:infall_vs_r}). The most likely candidate is that this is caused by the outflow driving turbulence in the wall of the outflow cavity, which is at $\sim$45-60\deg{} to the line of sight \citep{Yildiz2012}. However, a detailed investigation would require 2/3-D modelling including a physical model for the outflow which is beyond the scope of this paper.

For consistency of analysis and because this step model results in a worse fit to the other lines we do not consider this model in the following sections, reverting to the best-fit model with a constant $b$. Since we are primarily interested in the infall velocity throughout the envelope, and the 1$_{10}-$1$_{01}$ is primarily sensitive to motion on large scales due to a combination of beam dilution and optical depth effects, this will not impact that analysis.

\begin{figure}
\begin{center}
\includegraphics[width=0.43\textwidth]{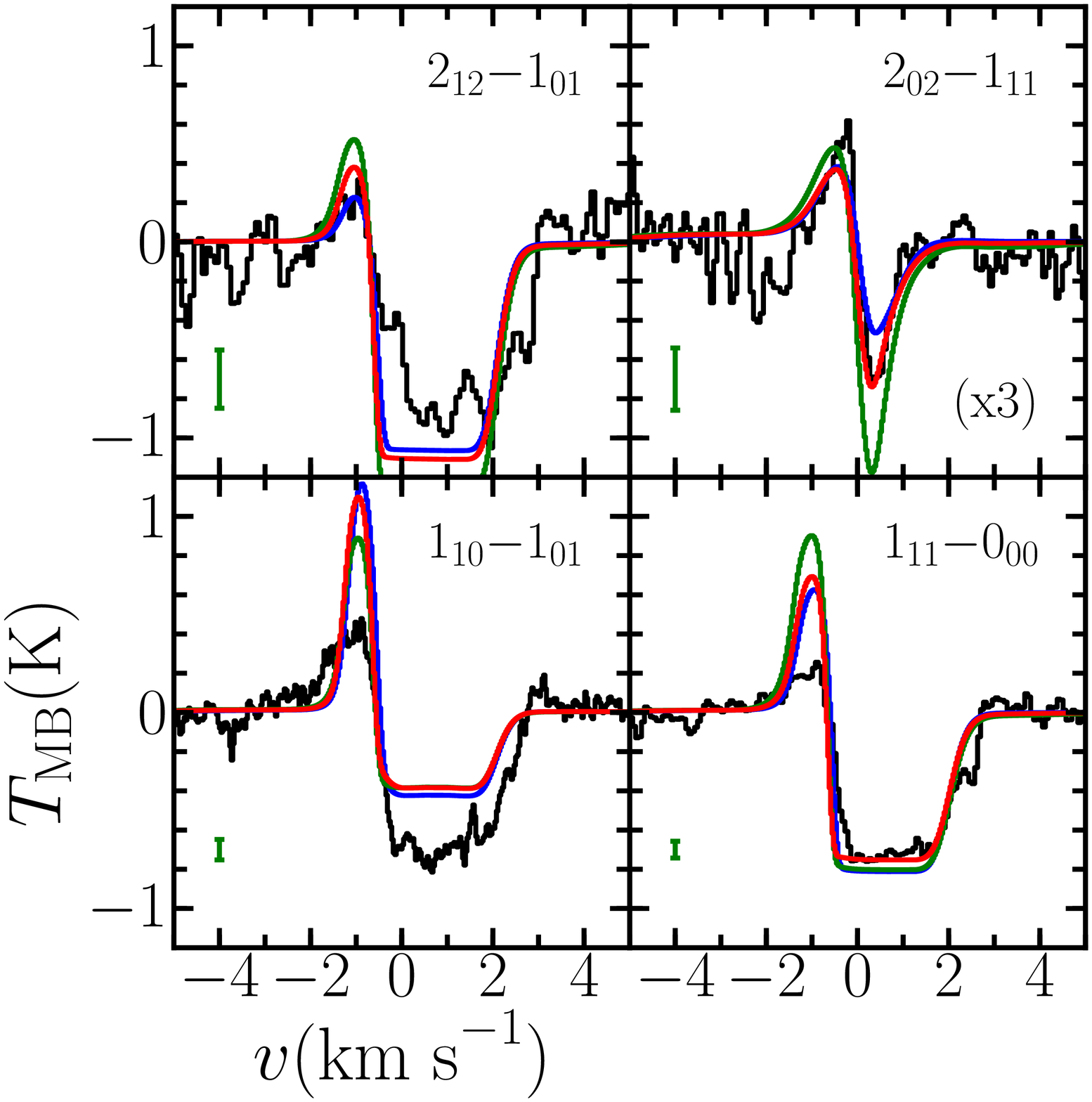}
\caption{Comparison of model H$_{2}$O line profiles for IRAS4A between the best-fit source model (red) and ones with a density power-law index ($p$) of 1.5 (blue) and 2.0 (green). All other variables take the best-fit values and are held constant. The green error bar and scaling factors are the same as in Figure~\ref{F:infall_iras4a_fit}.}
\label{F:infall_iras4a_alpha}
\end{center}
\end{figure}

As discussed in Sect.~\ref{S:models}, our physical structure assumes spherical symmetry but some of these sources have elongated or flattened envelopes which could lead to a lower or higher density power-law index ($p$) than is in fact the case. We show the difference changing $p$ to 1.5 and 2.0, as opposed to the value of 1.8 determined by \citet{Kristensen2012} in Figure~\ref{F:infall_iras4a_alpha}. The inner radius and density were recalculated for these models but the optical depth at 100\micron{} and ratio of inner to outer radius were kept the same as in the best-fit case. As can be seen, while the change in density profile does indeed lead to changes in the model water line profiles, these are similar to or smaller than the changes caused by varying other free parameters. It is therefore unlikely that non-spherical envelope structures would cause the best-fit values to be outside the estimated uncertainties, or that infall is observed in these sources.

\subsection{Infall vs. radius}
\label{S:infall_vs_r}

\begin{figure}
\begin{center}
\includegraphics[width=0.4\textwidth]{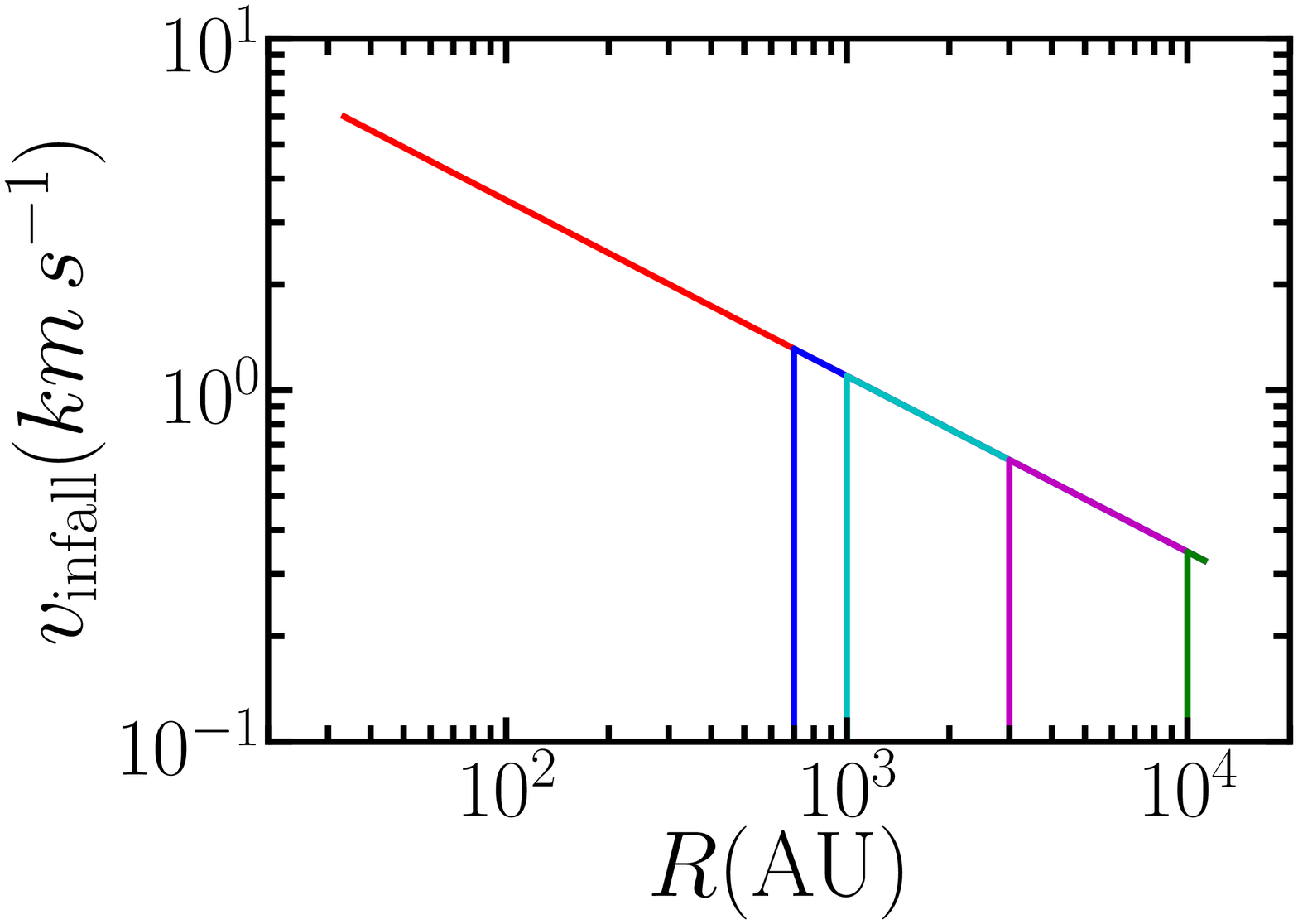}
\includegraphics[width=0.45\textwidth]{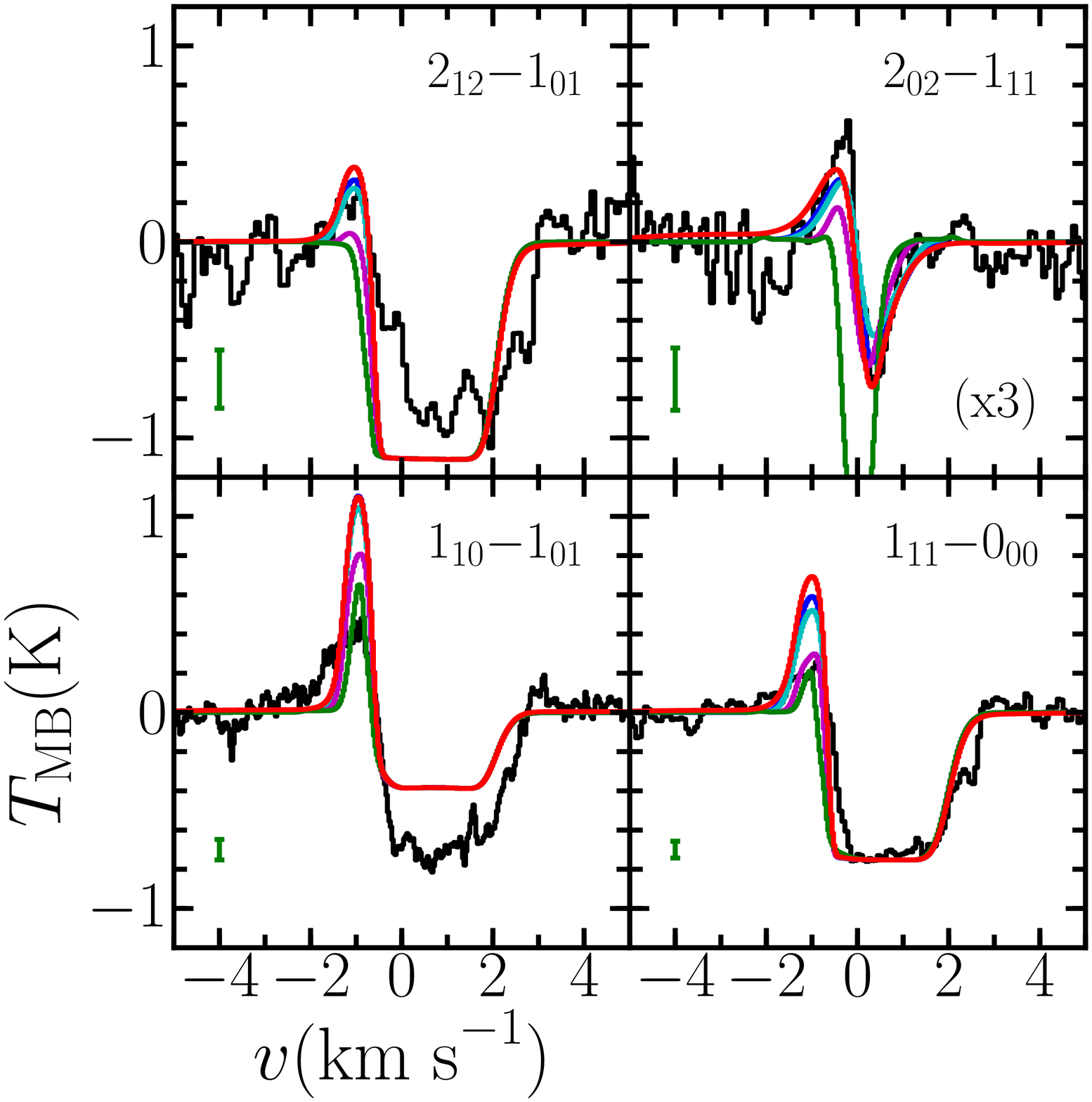}
\caption{Top: Velocity profile for the best-fit model for IRAS4A (red), and those with the infall stopped inside 700 (blue), 1000 (cyan), 3000 (magenta) and 10000\,AU (green). Outside the stop radius the infall profile is the same as in the best-fit model. Bottom: Comparison of resulting line profiles for these different velocity profiles. The green error bar and scaling factors are the same as in Figure~\ref{F:infall_iras4a_fit}.}
\label{F:infall_vs_r_rin}
\end{center}
\end{figure}

Let us now consider the envelope velocity profile more closely, using IRAS4A as a working example. The optical depth in a given line decreases with offset from the source velocity. Thus even those lines which become optically thick quickly at the line centre probe a range of radii within the model in the line wings. However, there are limits to how far into the envelope our observations can probe the infalling material (c.f. Figs~\ref{F:infall_iras4a_vfield} and \ref{F:infall_iras4a_contribution}). Figure~\ref{F:infall_vs_r_rin} shows the velocity profiles and the resulting line profiles for models where the infall velocity inside 700, 1000, 3000 and 10000\,AU was set to zero. For the model where infall stops at 10000\,AU, the absorption in the 2$_{02}-$1$_{11}$ line is too deep and shifted too far towards the systemic velocity to be consistent with the observations. The emission in both the 2$_{02}-$1$_{11}$ and 2$_{12}-$1$_{01}$ lines is also too weak for the models where infall is stopped at 10000 or 3000\,AU. There is little difference within the uncertainties in the data between the best-fit model and those where infall stops at $r\leq$1000\,AU, though the model where infall stops at 700\,AU is a slightly better fit to the absorption for the 2$_{02}-$1$_{11}$ line. Thus, infall in IRAS4A must continue to radii of at least 1000\,AU, consistent with the interferometric observations of \citet{DiFrancesco2001}. The observations are not able to constrain whether infall really stops at or inside this radius, but they can constrain the velocity profile outside this radius.

 \begin{figure}
\begin{center}
\includegraphics[width=0.4\textwidth]{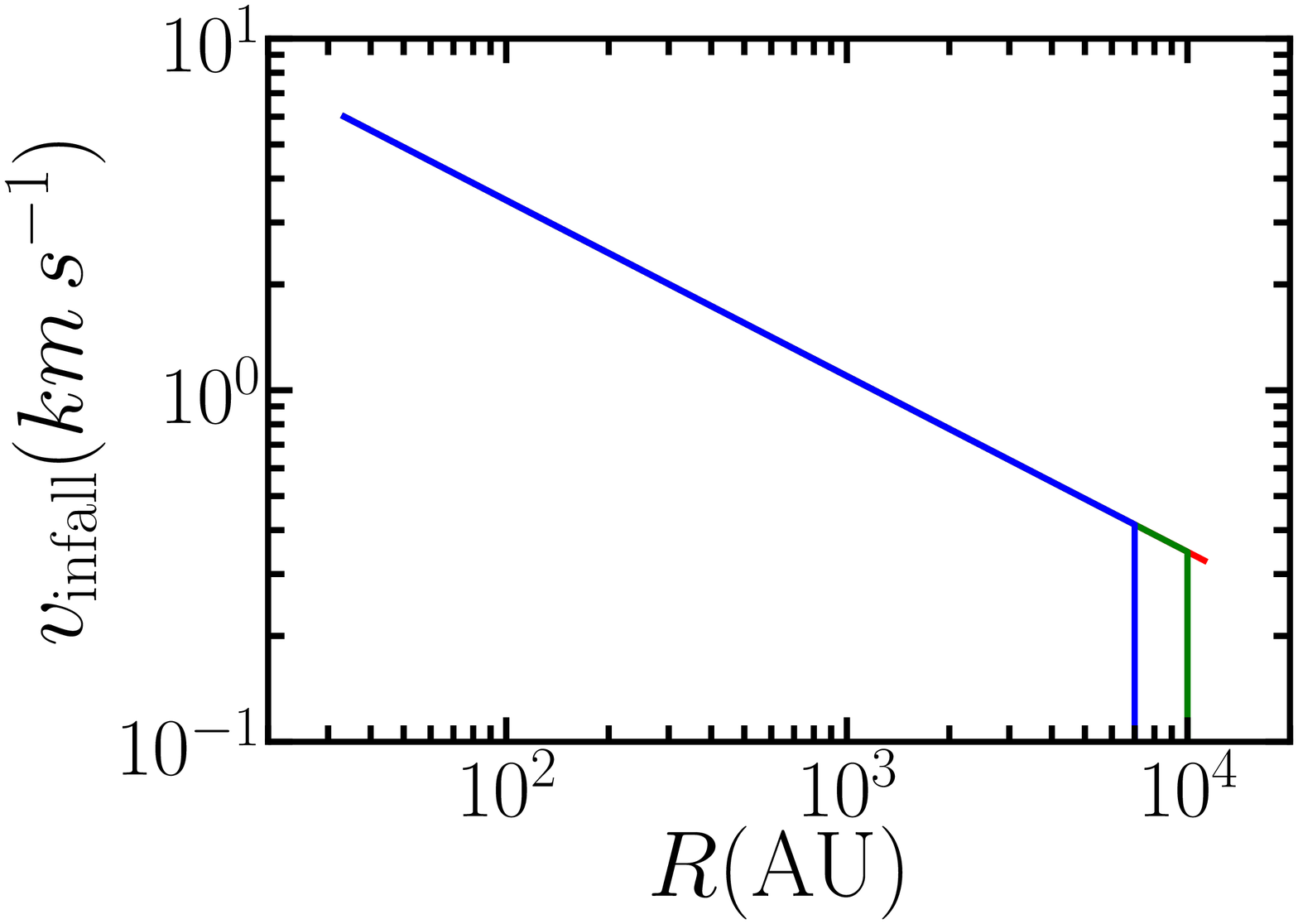}
\includegraphics[width=0.45\textwidth]{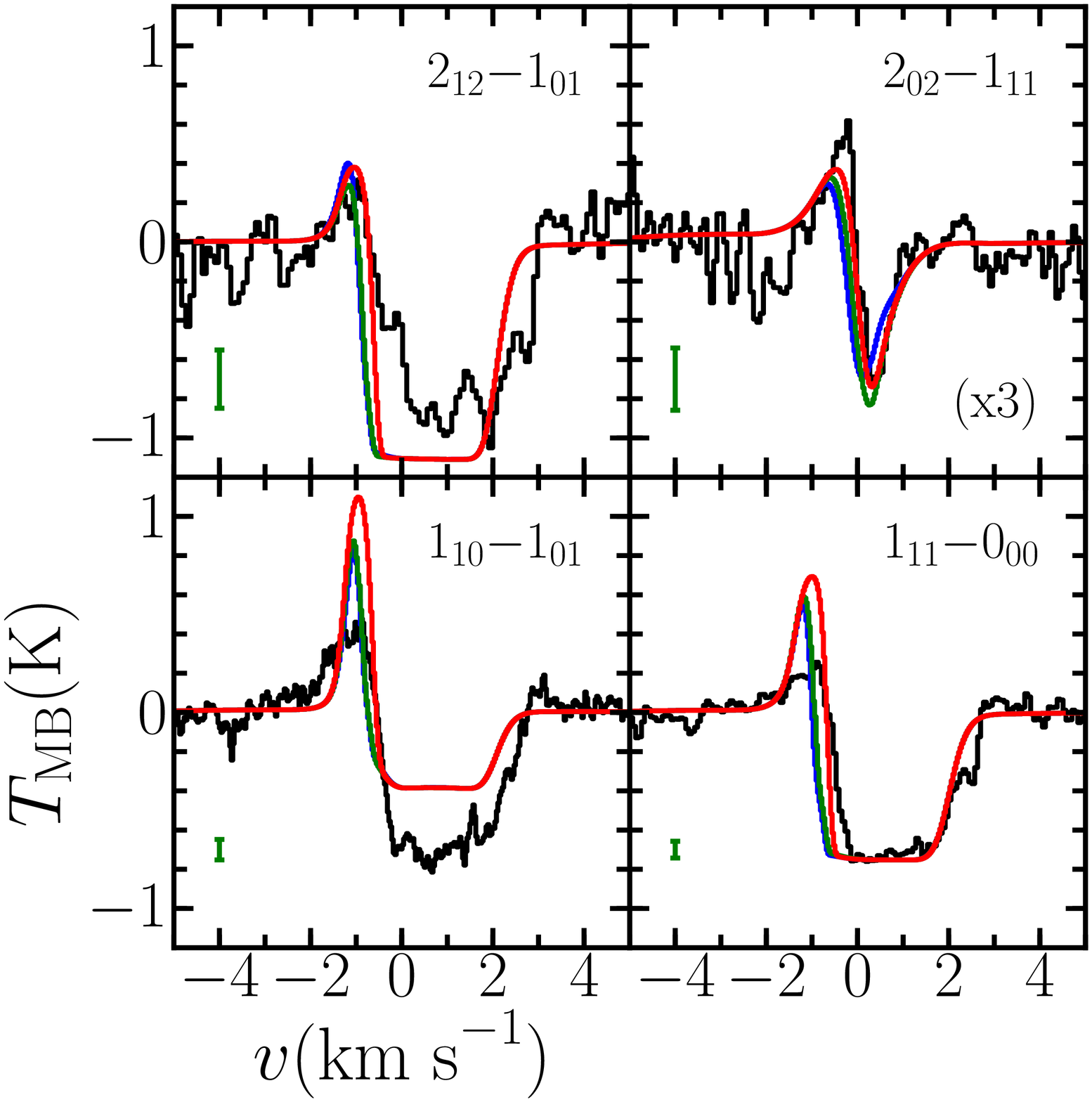}
\caption{As in Figure~\ref{F:infall_vs_r_rin} but with the infall stopped outside 7000 (blue) and 10000\,AU (green).}
\label{F:infall_vs_r_rout}
\end{center}
\end{figure}

If infall proceeds in an inside-out fashion \citep{Shu1977}, the outer-most radius at which infall still takes place, the free-fall radius ($r_{ff}$), is also of interest. Figure~\ref{F:infall_vs_r_rout} shows a comparison of the velocity and line profiles for the best-fit model with models where the infall velocity outside 7000 and 10000\,AU was set to zero. Both models have too much absorption near the source velocity compared to the observations, so in IRAS4A infall must extend at least to the outer radius considered by the model. Assuming that infall proceeds in gravitational free-fall at the sound speed from the inside outwards, the free-fall timescale since infall began is given by:

\begin{equation}
t_{\mathrm{ff}} = 2435\left(\frac{p}{2.0}\right)\left(\frac{r_{0}}{10\,\mathrm{AU}}\right)^{-1}\left(\frac{n_{0}}{10^{9}\,\mathrm{cm}^{-3}}\right)^{-1}\int_{r_0}^{r_{\mathrm{ff}}}\left(\frac{r_{\mathrm{ff}}}{r_{0}}\right) ^{\frac{p-2}{2}} dr ~\mathrm{yr},
\label{E:tff}
\end{equation}

\noindent taking into account the variation of density through the envelope. $r_{\mathrm{ff}}$ is given and AU and $p$ is again the slope of the density power-law which is given for our sources in Table~\ref{T:results}. For IRAS4A this gives a free-fall timescale of $\gtrsim$3$\times$10$^{5}$~yr.

Therefore, for IRAS4A infall must be taking place at least from 1000\,AU to the outer edge of the model 11200\,AU. We define the minimum radius outside of which infall must take place as the minimum detectible infall radius ($r_{\mathrm{mdi}}$), for which we can then calculate the corresponding velocity ($\varv_{\mathrm{mdi}}$) using Eqn.~\ref{E:v} and $\varv_{1000}$. Having obtained the infall velocity, this can be used to calculate the central gravitational mass ($M_{\mathrm{g}}$) from:

\begin{equation}
M_{\mathrm{g}} = \frac{\varv^{2}_{\mathrm{mdi}}\,r_{\mathrm{mdi}}}{2G} = 0.56\,\left(\frac{\varv_{1000}}{1\,\mathrm{km\,s}^{-1}}\right)^{2}~\rm{M}_{\odot},
\label{E:Mc}
\end{equation}

\noindent and the instantaneous mass infall rate at $r_{\mathrm{mdi}}$:

\begin{equation}
\dot{M}_{\mathrm{inf}} = 4\pi\,\mu\,m_{H}\,r_{\mathrm{mdi}}^{2}\,n_{\mathrm{mdi}}\,\varv_{\mathrm{mdi}},
\end{equation}

\noindent where $\mu$ is the mean molecular weight, for which we assume 2.8 from \citet{Kauffmann2008}. When Eqn.~\ref{E:v} and the power-law density dependence are taken into account, this becomes:

\begin{multline}
\dot{M}_{\mathrm{inf}} = 2.08\times10^{-8}\left(\frac{n_{0}}{10^{9}\,\rm{cm}^{-3}}\right)\\\left(\frac{\varv_{1000}}{1 \rm{km}\,\rm{s}^{-1}}\right)\left(\frac{r_{\mathrm{mdi}}}{1000 \rm{AU}}\right)^{1.5}\left(\frac{r_{\mathrm{mdi}}}{r_{0}}\right) ^{-p} \rm{M}_{\odot}\,\mathrm{yr}^{-1}.
\label{E:Mdotin}
\end{multline}

\noindent Due to the assumption of spherical symmetry, $\dot{M}_{\mathrm{inf}}$ is primarily a measure of the accretion of material from the envelope onto any disk-like structure rather than necessarily the accretion rate of material directly onto the central object ($\dot{M}_{\mathrm{acc}}$). Therefore, $\dot{M}_{\mathrm{inf}}$ is not expected to be directly related to the stellar luminosity, though higher $\dot{M}_{\mathrm{inf}}$ probably leads to higher $\dot{M}_{\mathrm{acc}}$. Dividing $M_{\mathrm{g}}$ by $\dot{M}_{\mathrm{inf}}$ gives the infall timescale ($t_{\mathrm{inf}}$), assuming that the accretion is steady over time at the rate measured at $r_{\mathrm{mdi}}$. If $p$~$\neq$1.5 the accretion will, by definition, not be constant over time, but this assumption enables a simple analytic estimate of the infall timescale, such that:

\begin{multline}
t_{\mathrm{inf}} = 2.7\times10^{7}\left(\frac{n_{0}}{10^{9}\,\rm{cm}^{-3}}\right)^{-1}\\\left(\frac{\varv_{1000}}{1 \rm{km}\,\rm{s}^{-1}}\right)\left(\frac{r_{\mathrm{mdi}}}{1000 \rm{AU}}\right)^{-1.5}\left(\frac{r_{\mathrm{mdi}}}{r_{0}}\right) ^{p} \,\mathrm{yr}.
\label{E:tinf}
\end{multline}

\noindent The values of $M_{\mathrm{g}}$, $r_{\mathrm{mdi}}$, $\dot{M}_{\mathrm{inf}}$, $t_{\mathrm{ff}}$ and $t_{\mathrm{inf}}$ are given in Table~\ref{T:infall}. The free-fall and infall timescales give characteristic timescales for collapse of the envelope under different sets of assumptions, which probably lead to them being accurate to a factor of a few at best. That $r_{\mathrm{ff}}$ is always the outer edge of the model might be an indication that the collapse is global, as in e.g. the Larson-Penston solution \citep{Larson1969,Penston1969}, or outside-in rather than inside-out. Whether constant mass infall and accretion is a good assumption will be discussed in Section~\ref{S:discussion}, as well as the wider implications of our results and comparison to other studies.

\begin{table*}
\begin{center}
\caption[]{Infall properties}
\begin{tabular}{lccccc}
\hline
\hline \noalign {\smallskip}
Source & $r_{\mathrm{mdi}}$ & $M_{\mathrm{g}}$ & $\dot{M}_{\mathrm{inf}}$ & $t_{\mathrm{inf}}$ & $t_{\mathrm{ff}}$ \\
 &  (10$^{3}$\,AU) & (\msol{}) & (10$^{-5}$ \msol{}\,yr$^{-1}$) & (10$^{4}$ yr) & (10$^{4}$ yr)  \\
\hline\noalign {\smallskip}
IRAS4A  & 1 & 0.68 & 15.4 & 0.44 & 10.4 \\
L1527   & 5  & 0.08 & \phantom{0}1.6 & 0.47 & \phantom{0}7.6 \\
BHR71   & 3 & 0.90 & \phantom{0}3.7 & 2.42 & 19.7 \\
IRAS15398   & 3 & 0.50 & \phantom{0}3.4 & 1.46 & \phantom{0}7.4 \\
L1157 & 3 & 1.17  & \phantom{0}5.3 & 2.22 & 13.5\\
\hline\noalign {\smallskip}
\label{T:infall}
\end{tabular}
\end{center}
\end{table*}

\subsection{Whole sample}
\label{S:infall_sample}

The results for the other sources which show inverse P-Cygni profiles can be divided into two groups, with comparison between the best-fit models and the data shown in Figures~\ref{F:infall_sample_l1527} to \ref{F:infall_sample_l1157}. IRAS15398, L1527, L1157 and BHR71 can all be fit with models similar to IRAS4A with similarly reasonable infall velocities and $b$ values. Though the model for BHR71 produces too much emission in the 1$_{11}-$0$_{00}$ line, the absorption component of this line is well fit, as is the 1$_{10}-$1$_{01}$ line. Non-detections are confirmed in the higher-excited lines. The reasons for this will be discussed in Section~\ref{S:chemistry}, however this results in the infall velocities being less well constrained because the ground-state lines are less sensitive to the infall. It also means that our observations only probe infall down to $\sim$3000\,AU (i.e. core to envelope) scales in these sources. 

Inspection of archival single-dish HCO$^{+}$ ($J$=4$-$3) observations for all these sources, as well as IRAS4A, show asymmetric blue-shifted profiles \citep{Gregersen1997,Hogerheijde2000a,vanKempen2009c}, suggesting that infall likely continues beyond this radius in the dense inner envelope. For L1157 the HCO$^{+}$ observations present a more complex picture, as the $J$=3$-$2 spectrum presented by \citet{Gregersen1997} shows an asymmetric blue-shifted profile while the $J$=4$-$3 is also asymmetric but with slightly stronger emission in the red than the blue peak. \citet{Tobin2012a} concluded that the L1157 envelope is elongated with the velocity field, consistent with material flowing along the filament onto the envelope on similar scales to our model. Material at the outer edge of the envelope must be infalling in all sources.

On the other hand, Ser-SMM4 requires a much higher infall velocity which seem inconsistent with infall on protostellar scales, particularly given the fact that it would result in a gravitational mass more than an order of magnitude larger than the envelope mass for this heavily embedded source. Indeed, the velocity for this model at the outer radius is large enough that even the ground-state lines are dominated by bulk motions rather than turbulence. The single-dish HCO$^{+}$ ($J$=4$-$3) observations of Ser-SMM4 are inconclusive, with a very faint second peak redshifted by $\sim$ 1.5\kms{} from the source velocity \citep{Ramchandani2012}.

\citet{DuarteCabral2010} found two emission components separated by $\sim$1.5\kms{} in C$^{18}$O ($J$=1$-$0) observations towards the Serpens Main cloud, which \citet{DuarteCabral2011} successfully modelled as a collision of two clouds. SMM4 is situated on the boundary between these two velocity components which correspond to the absorption and emission components observed in the water ground-state lines. In addition, the velocity at the outer edge of the best-fit model for SMM4 is $\sim$1.3\kms{}. Our observations of Ser-SMM4 would therefore seem more consistent with tracing this cloud-scale flow or a foreground cloud rather than infall on envelope scales.

\begin{figure}
\begin{center}
\includegraphics[width=0.4\textwidth]{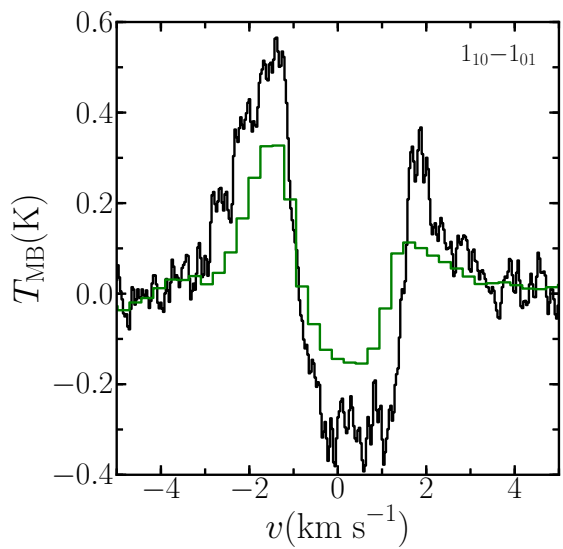}
\caption{Comparison of continuum and outflow subtracted observations of the H$_{2}$O 1$_{10}-$1$_{01}$ line towards GSS30 (black) with similar observations towards VLA1623 (green) from position 1 in Figure~1 of \citet[][]{Bjerkeli2012}.}
\label{F:whole_sample_gss30_vla1623}
\end{center}
\end{figure}

Judging by the fit values alone, GSS30 would seem to lie in the former group. However, this source actually lies near the prominent outflow VLA1623 and a diffuse layer has been detected in front of the Ophiuchus cloud where both sources are situated \citep{vanKempen2009a}. \citet{Bjerkeli2012} observe inverse P-Cygni profiles across the whole of their H$_{2}$O 1$_{10}-$1$_{01}$ map towards VLA1623 which are very similar to those observed towards GSS30 (see Figure~\ref{F:whole_sample_gss30_vla1623}). Large-scale flows or foreground clouds would therefore seem to be the cause of the observed line profiles, which is also more consistent with GSS30's older evolutionary stage.

\section{Chemistry}
\label{S:chemistry}

\begin{figure}
\begin{center}
\includegraphics[width=0.45\textwidth]{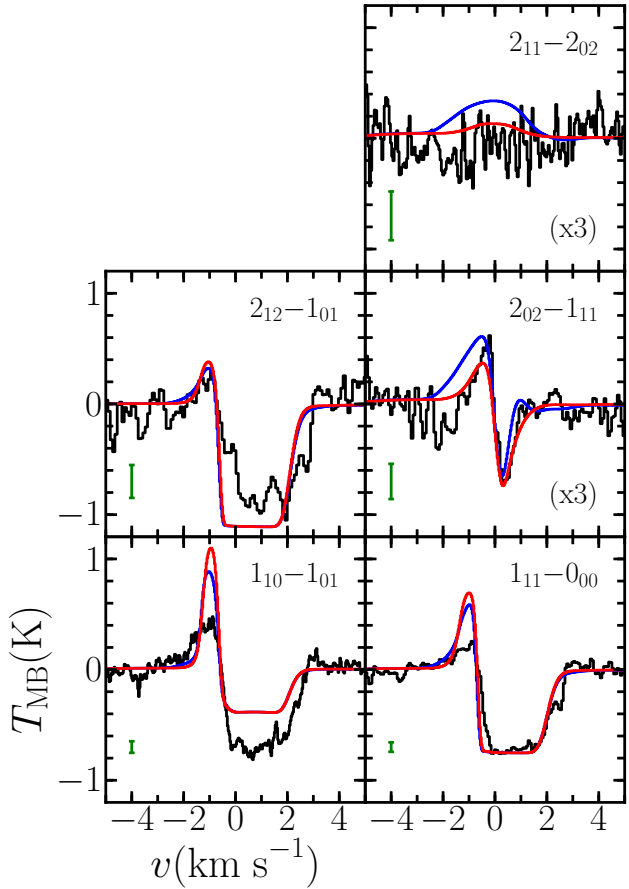}
\caption{Comparison of resulting line profiles for IRAS4A between the best-fit simple chemical abundance model (red) and a drop abundance profile (blue) as shown in Figure~\ref{F:models_abundance}. All other variables use the best-fit values. The green error bar and scaling factors are the same as in Figure~\ref{F:infall_iras4a_fit}.}
\label{F:results_iras4a_constant}
\end{center}
\end{figure}

As discussed in Section~\ref{S:models_H2O}, most previous attempts to model water lines have used drop or jump abundance profiles, rather than the simplified chemical model employed here. Figure~\ref{F:results_iras4a_constant} shows a comparison between line profiles for the best-fit model with the best possible fit using a drop abundance profile. The adopted abundance profiles are shown in Figure~\ref{F:models_abundance}. For the drop profile, the inner abundance is held constant at 6$\times$10$^{-4}$ where $T\geq$100\,K, consistent with the simple water profile. The photodesorption layer extends up to $A_{\mathrm{V}}$=5 but is still within the grid rather than as a separate layer as in \citet{Coutens2012}. The abundance in the outer envelope and photodesorption layer are free parameters. The best fit was obtained for $X_{\mathrm{out}}$=3$\times$10$^{-10}$ and $X_{pdl}$=3$\times$10$^{-7}$. The difference between the two models is relatively small for the lower excited line, but the step profile results in unobserved emission in the 2$_{11}-$2$_{02}$ line and has a much too wide emission component and too narrow absorption component in the 2$_{02}-$1$_{11}$ line. Indeed, only a very small set of profiles produce absorption below the continuum in the 2$_{02}-$1$_{11}$ line and none reproduce the equal intensity and width in the absorption and emission components observed.

This comparison also serves to show that if the gas-phase water abundance were higher in the inner, warmer part of the envelope then the 2$_{02}-$1$_{11}$ and 2$_{11}-$2$_{02}$ lines would be detected. This is true for the other sources as well, so that our non-detection of these lines is confirmed by the models also confirms that freeze-out of water is strong in the inner envelope below the $T$=100\,K radius. That we detect the 2$_{02}-$1$_{11}$ line in IRAS4A is probably due to a unique combination of the density and temperature profiles leading to just the right abundance and excitation structure. The non-detection of the 3$_{12}-$2$_{21}$ and 3$_{12}-$3$_{03}$ lines in all sources is because these lines are primarily sensitive to the $T\geq$100\,K region, which for IRAS4A has a diameter of $\sim$1\arcsec{} and so is too heavily beam-diluted.

This sensitivity of the water observations to the water abundance profile means that at least simple water chemistry including density-dependent freezeout must be taken into account when performing modelling of these lines. These observations do not probe the hot core (i.e. $T\geq$100\,K) region, for which we refer the reader to \citet[][]{Persson2012} and \citet[][]{Visser2013}.

\begin{figure}
\begin{center}
\includegraphics[width=0.48\textwidth]{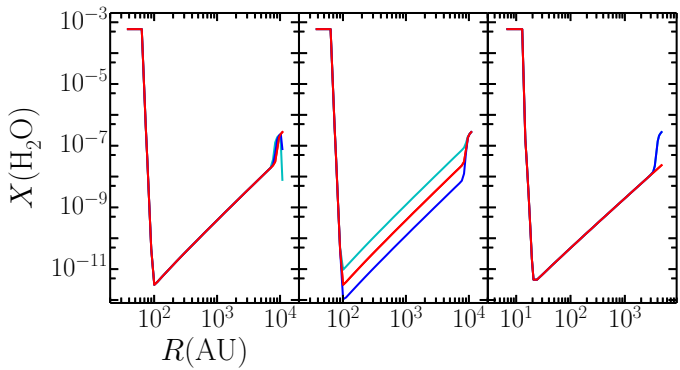}
\caption{Left: The abundance profile of water with respect to H$_{2}$ for IRAS4A for varying the interstellar radiation field ($G_{\mathrm{isrf}}$) between 1 (red), 10 (black) and 100 (blue).  Middle: As for the left plot, but for cosmic-ray UV rates ($G_{\mathrm{cr}}$) of 3$\times$10$^{-5}$ (blue), 10$^{-4}$ (red) and 3$\times$10$^{-4}$ (cyan). Right: The abundance profile of water for IRAS15398 with $G_{\mathrm{irsf}}$=0 (red) and $G_{\mathrm{irsf}}$=1 (blue).}
\label{F:chemistry_abundance}
\end{center}
\end{figure}

Let us now consider the impact of changing the key variables involved in the simple water chemistry model. Figure~\ref{F:chemistry_abundance} shows how the abundance profile varies with varying interstellar radiation field and cosmic-ray UV rate. As might be expected, increasing $G_{\mathrm{isrf}}$ from the standard value increases the destruction of water in the very outer layer of the model and thus decreases the abundance. Increasing or decreasing $G_{\mathrm{cr}}$ leads to more or less water in the gas phase respectively due to increasing or decreasing the rate of cosmic-ray induced photodesorption.

\begin{figure}
\begin{center}
\includegraphics[width=0.45\textwidth]{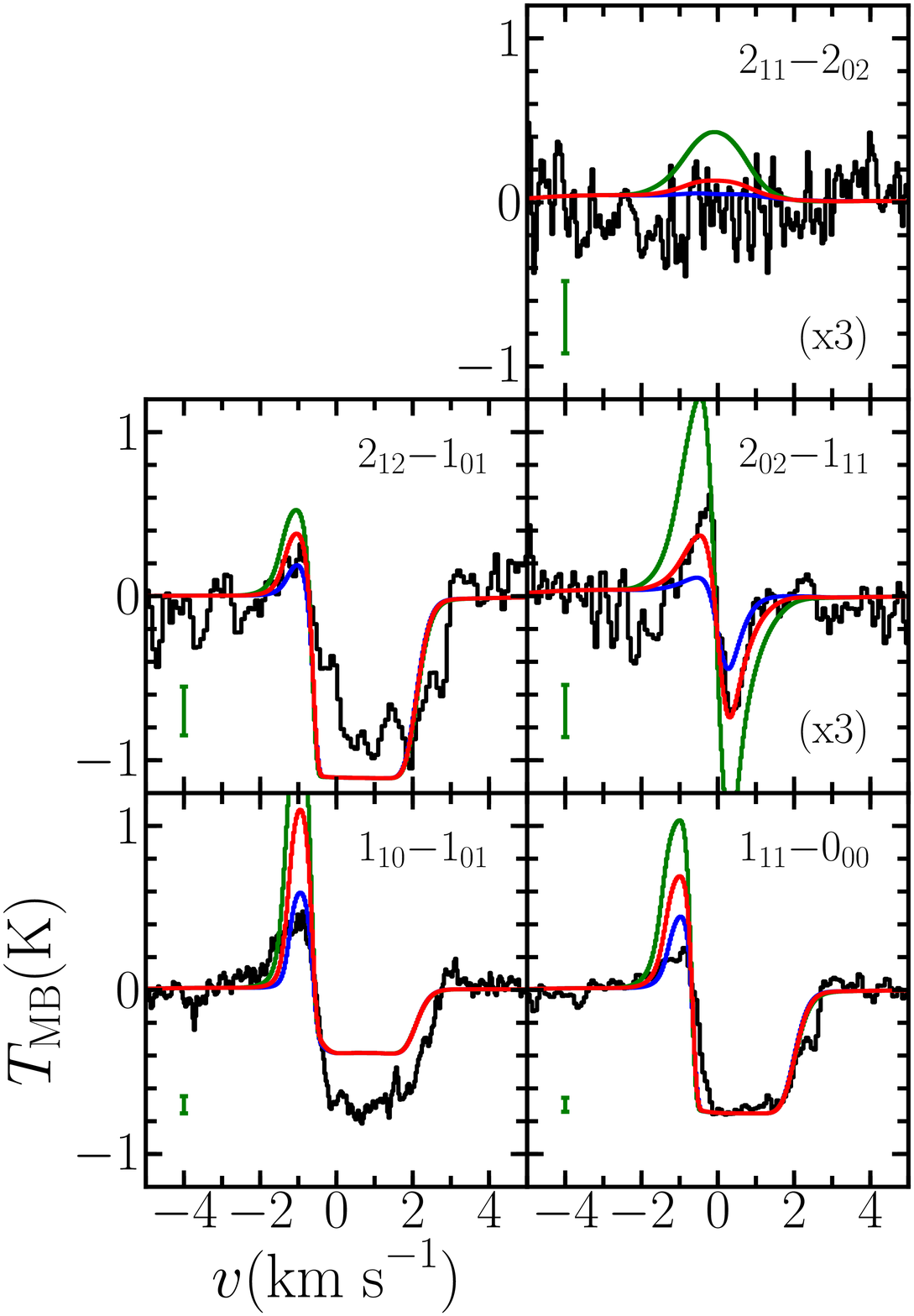}
\includegraphics[width=0.45\textwidth]{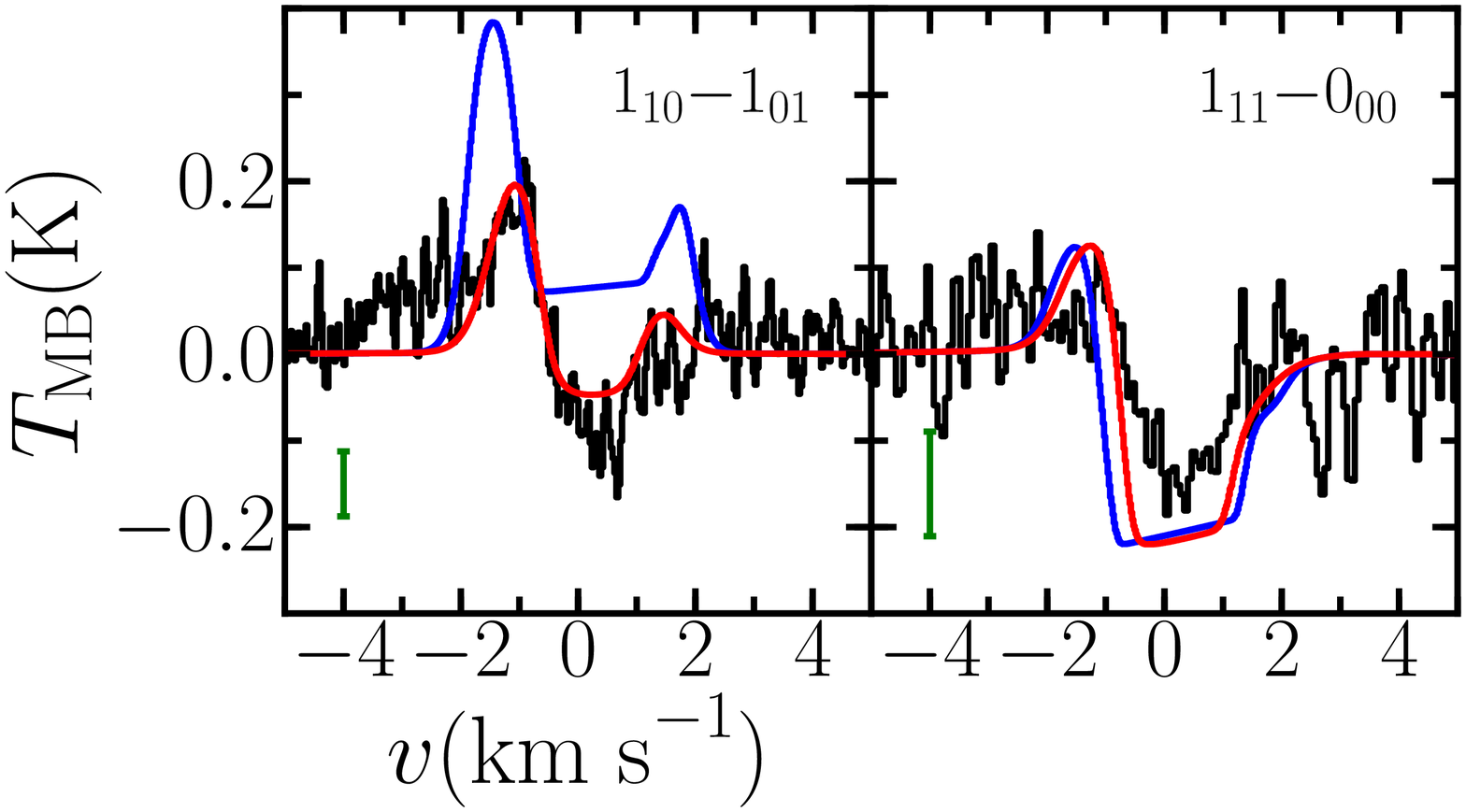}
\caption{Top: Comparison of model H$_{2}$O line profiles for IRAS4A using the abundance profiles shown in the right-hand panel of Figure~\ref{F:chemistry_abundance} with $G_{\mathrm{cr}}$ varying between 3$\times$10$^{-5}$ (blue), 10$^{-4}$ (red) and 3$\times$10$^{-4}$ (green). All other variables use the best-fit values. Bottom: Comparison of model H$_{2}$O line profiles for IRAS15398 with $G_{\mathrm{isrf}}$=0 (red) and $G_{\mathrm{isrf}}$=1 (blue). The green error bar and scaling factors are the same as in Figure~\ref{F:infall_iras4a_fit}.}
\label{F:chemistry_iras4a_isrfcr}
\end{center}
\end{figure}

The resulting model line profiles are shown in Figure~\ref{F:chemistry_iras4a_isrfcr}. Increasing $G_{\mathrm{cr}}$ leads to a detectable emission feature in the 2$_{11}-$2$_{02}$ line, which is not observed, and too broad emission and absorption in the 2$_{02}-$1$_{11}$ line. Decreasing $G_{\mathrm{cr}}$ leads to too little emission in the 2$_{02}-$1$_{11}$ and 2$_{12}-$1$_{01}$ lines. Increasing $G_{\mathrm{isrf}}$ has a minimal effect on the line profiles. This is because the water affected by this change is closest to the source velocity and predominantly in the ground state where the absorption is already saturated. For those sources where the outer envelope density is above 10$^{5}$\pccm{} there is a significant difference between $G_{\mathrm{isrf}}$=0 and $G_{\mathrm{isrf}}$=1 (see lower plot of Figure~\ref{F:chemistry_iras4a_isrfcr}). This is because the photodesorption layer caused by $G_{\mathrm{isrf}}$=1 results in a significant jump in water abundance (see right-hand panel of figure~\ref{F:chemistry_abundance}) and the density is high enough that photodissociation only occurs in a very thin shell at the outer edge.

The sensitivity of the water line profiles not only to the radial abundance profile but also to its absolute scaling makes it possible to constrain the cosmic-ray induced UV field. As discussed in Section~\ref{S:models_H2O}, the cosmic-ray induced UV flux is scaled by $G_{\mathrm{cr}}$ to the interstellar photon flux, which leads to photodesorption of H$_{2}$O with an efficiency ($\eta$) which we assume to be 10$^{-3}$. We find that the canonical value for the cosmic-ray induced UV flux \citep[i.e. $G_{\mathrm{cr}}$=10$^{-4}$,][]{Prasad1983,Gredel1989,Shen2004} gives the best fit, in contrast to the 1 dex enhancement (i.e. $G_{\mathrm{cr}}$=10$^{-3}$) suggested by \citet{Hollenbach2009} and used for the prestellar core L1544 by \citet{Caselli2012}. Indeed, even an enhancement of 0.25 dex is already a noticeably poorer fit to our observations. Formally, we are only able to constrain the product of $G_{\mathrm{cr}}$ and $\eta$. However, given that $\eta$ is well constrained from laboratory experiments and theoretical calculations \citep{Oberg2009,Arasa2010}, this parameter is already well constrained.

Most sources require a cosmic-ray induced UV field in the range (0.2-1)$\times$10$^{4}$ photons\,cm$^{-2}$\,s$^{-1}$, which is consistent with estimates for standard molecular cloud conditions \citep{Shen2004}. The exception is GSS30 which requires $G_{\mathrm{cr}}$=3$\times$10$^{-4}$, but as discussed above the observed line profiles likely trace a foreground cloud or cloud-scale flow. In addition, the region is disrupted by the outflow from VLA1623 and is embedded in a weak PDR powered by S1 and/or HD147889. Both of these factors could lead to abundance enhancements which are not accounted for in our model. 

For those sources which require lower values of $G_{\mathrm{cr}}$, it should be pointed out that this does not necessarily mean that they actually have lower cosmic-ray rates, but rather this is the main variable within our model which varies the abundance and thus the emission intensity. It could be instead that the outflow blocks some fraction of the emission from the back half of the envelope (see Section~\ref{S:models_outflow}). Indeed, for BHR71 there is still a significant excess of emission in the 1$_{11}-$0$_{00}$ line which cannot be improved by varying any of the model parameters. The outflow of BHR71 has an hourglass shape at the base in CO observations \citep{Parise2006,vanKempen2009b} which would fill more of the 19\arcsec{} beam of the 1$_{11}-$0$_{00}$ line than the 38\arcsec{} beam of the 1$_{10}-$1$_{01}$ line. Thus the blocking of emission by the outflow may be the cause of the lower intensity rather than a physically lower value of $G_{\mathrm{cr}}$. As the absorption is both less sensitive to the abundance and takes place in the part of the envelope between the outflow and the observer, the good fit by the models to this part of the line profiles still constrains the velocity profile down to $r_{\mathrm{mdi}}$.

\section{Discussion}
\label{S:discussion}

Let us now discuss the consistency of our results with those obtained using other tracers. For IRAS4A, \citet{DiFrancesco2001} calculated an infall velocity of 0.68\kms{} at 2700\,AU from interferometric observations of H$_{2}$CO using a 2-slab model. Our model gives 0.67\kms{} at this radius, fully consistent with the results of \citet{DiFrancesco2001} given their assumptions. The value of $b$ required to fit the water observations of IRAS4A is lower than the 0.8\kms{} used by \citet{Yildiz2012} for various C$^{18}$O lines. While those authors use the same source structure as used here, they only considered a static model which leaves turbulence as the dominant component to the line width. On the other hand, \citet{Jorgensen2004} show that a pure infall (i.e. $b$=0) can also be ruled out, so the observed C$^{18}$O linewidths must be due to a combination of infall and turbulence. For the sample as a whole we find best-fit values of $b$ in the range 0.2 to 0.9 \kms{}, corresponding to $\sigma_{\mathrm{NT}}$ values in the range 0.15 to 0.64 \kms{}.

At the minimum detectible infall radius of 1000\,AU, the temperature in the model is 20\,K which corresponds to a sound speed of 0.26\kms{} while the infall velocity is 1.1\kms{}, so the infall has a Mach number of 4.3. This flow will not result in a shock as long as the mass infall rate does not decrease with decreasing radius. However, the outflow timescale derived by \citet{Yildiz2012} ($\sim$6$-$10$\times$10$^{4}$\,yr) and the free-fall time of the envelope (2$\times$10$^{5}$\,yr) are approximately an order of magnitude larger than the infall timescale ($\gtrsim$30000\,yr), suggesting that something must slow the infalling material in the inner $\sim$1000\,AU. Whether there is a shock formed as material falls onto a flattened disk-like structure will depend on how gradually the radial infall is mediated. Conservation of angular momentum could help to enhance circular motions in the infalling material. In addition, IRAS4A is known to have an hour-glass shaped magnetic field on $\sim$400\,AU scales \citep{Girart2006} which could cause a decrease in the component of motion perpendicular to the axis of the magnetic field. While evidence for a disk shock due to infall onto a disk was claimed due to the existence of hot water in the neighbouring source NGC1333-IRAS4B by \citet{Watson2007}, this was subsequently refuted due to the spatial offset of the emission from the central source by \citet{Herczeg2012}. CH$_{3}$OH masers have been claimed to be excited in disk shocks in high-mass star forming regions \citep{Torstensson2011}, but have not been detected towards low-mass sources. Evidence for such a shock is therefore limited or otherwise difficult to disentangle from shocks related to the outflow.

The stability of any central condensation depends on how well it can mediate the competing processes which serve to increase or decrease its mass over time. The instantaneous mass infall rate we determine for IRAS4A (2.0$\times$10$^{-4}$\msol{}\,yr$^{-1}$) is higher by approximately one order of magnitude than the maximum mass accretion rate in simulations of forming protostars \citep[e.g.][]{Dunham2012}. From conservation of energy, the maximum radius that a star can have is:

\begin{equation}
R_{*} = 61 R_{\odot}\,\left(\frac{M_{*}}{M_{\odot}}\right),
\end{equation}

\noindent  \citep{Palla1999}, where $M_{*}$ is the stellar mass of the protostar. However, using this equation means that the mass drops out of the mass accretion rate calculation. Thus if we assume that all the luminosity of the star is due to accretion, the maximum possible mass accretion rate for IRAS4A is 2$\times$10$^{-5}$\msol{}\,yr$^{-1}$, which is an order of magnitude lower than the measured mass infall rate.
 
The mass infall rate is also 3.5 orders of magnitude higher than the mass-loss rate of 5$\times$10$^{-8}$\msol{}\,yr$^{-1}$ in the atomic jet obtained from the [OI] emission by \citet{Karska2013} and at least one order of magnitude higher than the mass outflow rate from low-$J$ CO observations of swept-up gas in the molecular outflow \citep[][]{Yildiz2012,Yildizip}. This suggests that whatever flattened structure exists around IRAS4A, it is gaining mass significantly faster than it is likely to be accreted onto the central protostar or ejected via winds and/or outflows and so is unlikely to be gravitationally stable. How long this has been the case is unclear.

Indeed, IRAS4A is also a known binary with a separation of $\sim$400\,AU \citep{Choi2004}, likely formed through turbulent fragmentation of the initial prestellar core \citep[e.g.][]{Offner2010}. The presence of a binary provides additional forces on material in the inner regions, possibly enhancing instability of any circumbinary or circumstellar disk(s). Such instability could lead to episodic accretion, which has been suggested for some time \citep[e.g.][]{Kenyon1990} and has been predicted to occur due to gravitational instability of a disk caused by mass-loading \citep[e.g.][]{Vorobyov2009,Vorobyov2010} in a similar scenario to that taking place in IRAS4A. It could even lead to the formation of additional companions through rotational fragmentation of the disk \citep[e.g.][]{Burkert1993}.

L1527 has similar mass infall and outflow rates, though the maximum mass accretion rate is an order of magnitude smaller. This is primarily due to the small infall velocity resulting in a small gravitational mass though this source shows one of the strongest asymmetries in HCO$^{+}$ ($J$=4$-$3) and ($J$=3$-$2) observations \citep{Hogerheijde2000a}. The insensitivity of the water observations to the infall velocity in this source, primarily due to the relatively flat density profile, therefore makes it likely that the infall velocity obtained for L1527 is a lower limit. IRAS15398 is similar to IRAS4A in that the mass infall rate is an order of magnitude higher than the mass outflow and maximum mass accretion rate. BHR71 and L1157 have similar mass infall, accretion and outflow rates and so may be in more stable configurations. Infall in BHR71, L1157 and IRAS15398 is supersonic at $r_{\mathrm{mdi}}$. A larger sample and detailed mapping of the velocity field down to smaller radii is required to tell whether this is related to initial conditions or an evolutionary effect.

The gravitational masses ($M_{\mathrm{g}}$) given in Table~\ref{T:infall} were calculated assuming that the infall speed is solely dependant on the mass of the protostar, and range from 1.17 to 0.08\msol{}. The lowest value is for L1527, which is half an order of magnitude lower than obtained by \citet{Tobin2012b} based on interferometric observations. As discussed above, the infall velocity for L1527 is particularly uncertain, and the velocity corresponding to the result obtained by \citet{Tobin2012b} does lie within the uncertainties. However, there is also the possibility that infall has only recently started in the outer parts of the envelope as probed by our observations or that the outer layers are not gravitationally unstable.

Finally, it is at this point worth considering why we do not observe infall signatures towards all 29 WISH sources. Four sources show regular P-Cygni profiles, indicative of envelope expansion but what of the remaining 18? Low $S/N$ may play a role in a few observations, but in most cases outflow emission is observed and so red-shifted absorption should be detectible. NGC1333-4B and DK-Cha both have pole-on outflows which would make detection of the envelope almost impossible. Similarly, the broad absorption towards Serpens SMM1 and SMM3 is undoubtedly caused by the same cloud-scale effects as in SMM4 but because neither is in the region where the two cloud components overlap there is no inverse P-Cygni profile. The more evolved sources may simply not have enough envelope remaining to show inverse or regular P-Cygni profiles. For the remaining younger sources, it could be that the outer layers of the envelope have yet to start infalling, or that the outflow blocks or disrupt enough of the emission and absorption to make identification of infall with water impossible. Alternatively, infall on large-scales may have already stopped or been reversed, in which case this phase is short-lived.

\section{Conclusions}
\label{S:conclusions}

We have presented 1-D non-LTE radiative transfer models of water in seven protostars which show infall signatures in observations. Though these models are spherically-symmetric and include a simplified implementation of the outflow, they nevertheless reproduce the observations well while being constrained by them. Water proves to be an excellent tracer of both the dynamics and chemistry of protostellar envelopes, particularly when both ground-state high-$\tau$ lines and more moderately optically thick excited lines are observed together. We find that a step abundance model cannot reproduce the observations while a simple chemical network with reasonable parameters is consistent with the observations. When this is done we find best-fit values for the cosmic-ray induced UV flux in the range (1-5)$\times$10$^{4}$ times lower than the interstellar UV radiation field, which is in line with estimates for standard conditions in molecular clouds. 

Infall of material from the core into the envelope is observed in four sources from $\sim$10000 to 3000\,AU scales and in one source, NGC1333-IRAS4A, down to at least $\sim$1000\,AU. This is only possible for IRAS4A due to the detection in the moderately optically thick 2$_{02}-$1$_{11}$ line, so future studies should include mid-$\tau$ lines where possible in preference to higher-$\tau$ lines. In four of the five sources this infall is supersonic and infall in all sources must take place at the outer edge of the envelope, which may be evidence that collapse is global or outside-in rather than inside-out. The mass infall rates derived for IRAS4A and IRAS15398 are so high that any disk-like structure cannot be gravitationally stable, which may result in episodic accretion or fragmentation. However, not all observations showing infall signatures in water can be reproduced with 1-D infalling envelope models. In the case of Serpens-SMM4 and GSS30, the observations are more likely tracing large-scale cloud flows. Using slab-model analysis alone cannot distinguish between these two scenarios. While care must be taken when interpreting inverse P-Cygni or infall profiles, the combination of slab and 1-D envelope models is able to constrain the radial velocity distribution, provided that multiple transitions of a molecule which probes a range of temperatures and optical depths are used.


\begin{acknowledgements}

The authors would like to thank the anonymous referee, whose comments helped improved the clarity and content of the paper. We also thank Floris van der Tak, Mario Tafalla, Paola Caselli, Neal Evans and Audrey Coutens for their valuable comments on the manuscript, and Daniel Harsono, Catherine Walsh, Pamela Klaassen and Tobias Albertsson for stimulating discussions. JCM is funded by grant 614.001.008 from the Netherlands Organisation for Scientific Research (NWO). Astrochemistry in Leiden is supported by the Netherlands Research School for Astronomy (NOVA), by a Spinoza grant and by the European Community's Seventh Framework Programme FP7/2007-2013 under grant agreement 238258 (LASSIE). HIFI has been designed and built by a consortium of institutes and university departments from across Europe, Canada and the United States under the leadership of SRON Netherlands Institute for Space Research, Groningen, The Netherlands and with major contributions from Germany, France and the US. Consortium members are: Canada: CSA, U.Waterloo; France: CESR, LAB, LERMA, IRAM; Germany: KOSMA, MPIfR, MPS; Ireland, NUI Maynooth; Italy: ASI, IFSI-INAF, Osservatorio Astrofisico di Arcetri- INAF; Netherlands: SRON, TUD; Poland: CAMK, CBK; Spain: Observatorio Astron{\'o}mico Nacional (IGN), Centro de Astrobiolog{\'i}a (CSIC-INTA). Sweden: Chalmers University of Technology - MC2, RSS $\&$ GARD; Onsala Space Observatory; Swedish National Space Board, Stockholm University - Stockholm Observatory; Switzerland: ETH Zurich, FHNW; USA: Caltech, JPL, NHSC.

\end{acknowledgements}

\bibliography{wish_ipcbib}

\newcommand{\noop}[1]{}
\begin{thebibliography}{95}
\expandafter\ifx\csname natexlab\endcsname\relax\def\natexlab#1{#1}\fi

\bibitem[{{Arasa} {et~al.}(2010){Arasa}, {Andersson}, {Cuppen}, {van Dishoeck},
  \& {Kroes}}]{Arasa2010}
{Arasa}, C., {Andersson}, S., {Cuppen}, H.~M., {van Dishoeck}, E.~F., \&
  {Kroes}, G.-J. 2010, \jcp, 132, 184510

\bibitem[{{Attard} {et~al.}(2009){Attard}, {Houde}, {Novak}, {Li},
  {Vaillancourt}, {Dowell}, {Davidson}, \& {Shinnaga}}]{Attard2009}
{Attard}, M., {Houde}, M., {Novak}, G., {et~al.} 2009, \apj, 702, 1584

\bibitem[{{Bergin} {et~al.}(2003){Bergin}, {Kaufman}, {Melnick}, {Snell}, \&
  {Howe}}]{Bergin2003}
{Bergin}, E.~A., {Kaufman}, M.~J., {Melnick}, G.~J., {Snell}, R.~L., \& {Howe},
  J.~E. 2003, \apj, 582, 830

\bibitem[{{Bergin} {et~al.}(1995){Bergin}, {Langer}, \&
  {Goldsmith}}]{Bergin1995}
{Bergin}, E.~A., {Langer}, W.~D., \& {Goldsmith}, P.~F. 1995, \apj, 441, 222

\bibitem[{{Bergin} {et~al.}(2006){Bergin}, {Maret}, {van der Tak}, {Alves},
  {Carmody}, \& {Lada}}]{Bergin2006}
{Bergin}, E.~A., {Maret}, S., {van der Tak}, F.~F.~S., {et~al.} 2006, \apj,
  645, 369

\bibitem[{{Bjerkeli} {et~al.}(2012){Bjerkeli}, {Liseau}, {Larsson}, {Rydbeck},
  {Nisini}, {Tafalla}, {Antoniucci}, {Benedettini}, {Bergman}, {Cabrit},
  {Giannini}, {Melnick}, {Neufeld}, {Santangelo}, \& {van
  Dishoeck}}]{Bjerkeli2012}
{Bjerkeli}, P., {Liseau}, R., {Larsson}, B., {et~al.} 2012, \aap, 546, A29

\bibitem[{{Brinch} {et~al.}(2009){Brinch}, {J{\o}rgensen}, \&
  {Hogerheijde}}]{Brinch2009}
{Brinch}, C., {J{\o}rgensen}, J.~K., \& {Hogerheijde}, M.~R. 2009, \aap, 502,
  199

\bibitem[{{Burkert} \& {Bodenheimer}(1993)}]{Burkert1993}
{Burkert}, A. \& {Bodenheimer}, P. 1993, \mnras, 264, 798

\bibitem[{{Caselli} {et~al.}(2012){Caselli}, {Keto}, {Bergin}, {Tafalla},
  {Aikawa}, {Douglas}, {Pagani}, {Y{\'{\i}}ld{\'{\i}}z}, {van der Tak},
  {Walmsley}, {Codella}, {Nisini}, {Kristensen}, \& {van
  Dishoeck}}]{Caselli2012}
{Caselli}, P., {Keto}, E., {Bergin}, E.~A., {et~al.} 2012, \apjl, 759, L37

\bibitem[{{Cassen} \& {Moosman}(1981)}]{Cassen1981}
{Cassen}, P. \& {Moosman}, A. 1981, \icarus, 48, 353

\bibitem[{{Choi}(2001)}]{Choi2001}
{Choi}, M. 2001, \apj, 553, 219

\bibitem[{{Choi} {et~al.}(2004){Choi}, {Kamazaki}, {Tatematsu}, \&
  {Panis}}]{Choi2004}
{Choi}, M., {Kamazaki}, T., {Tatematsu}, K., \& {Panis}, J.-F. 2004, \apj, 617,
  1157

\bibitem[{{Coutens} {et~al.}(2012){Coutens}, {Vastel}, {Caux}, {Ceccarelli},
  {Bottinelli}, {Wiesenfeld}, {Faure}, {Scribano}, \& {Kahane}}]{Coutens2012}
{Coutens}, A., {Vastel}, C., {Caux}, E., {et~al.} 2012, \aap, 539, A132

\bibitem[{{Daniel} {et~al.}(2011){Daniel}, {Dubernet}, \&
  {Grosjean}}]{Daniel2011}
{Daniel}, F., {Dubernet}, M.-L., \& {Grosjean}, A. 2011, \aap, 536, A76

\bibitem[{{de Graauw} {et~al.}(2010){de Graauw}, {Helmich}, {Phillips},
  {Stutzki}, {Caux}, {Whyborn}, {Dieleman}, {Roelfsema}, {Aarts}, {Assendorp},
  {Bachiller}, {Baechtold}, {Barcia}, {Beintema}, {Belitsky}, {Benz}, {Bieber},
  {Boogert}, {Borys}, {Bumble}, {Ca{\"i}s}, {Caris}, {Cerulli-Irelli},
  {Chattopadhyay}, {Cherednichenko}, {Ciechanowicz}, {Coeur-Joly}, {Comito},
  {Cros}, {de Jonge}, {de Lange}, {Delforges}, {Delorme}, {den Boggende},
  {Desbat}, {Diez-Gonz{\'a}lez}, {di Giorgio}, {Dubbeldam}, {Edwards},
  {Eggens}, {Erickson}, {Evers}, {Fich}, {Finn}, {Franke}, {Gaier}, {Gal},
  {Gao}, {Gallego}, {Gauffre}, {Gill}, {Glenz}, {Golstein}, {Goulooze},
  {Gunsing}, {G{\"u}sten}, {Hartogh}, {Hatch}, {Higgins}, {Honingh}, {Huisman},
  {Jackson}, {Jacobs}, {Jacobs}, {Jarchow}, {Javadi}, {Jellema}, {Justen},
  {Karpov}, {Kasemann}, {Kawamura}, {Keizer}, {Kester}, {Klapwijk}, {Klein},
  {Kollberg}, {Kooi}, {Kooiman}, {Kopf}, {Krause}, {Krieg}, {Kramer},
  {Kruizenga}, {Kuhn}, {Laauwen}, {Lai}, {Larsson}, {Leduc}, {Leinz}, {Lin},
  {Liseau}, {Liu}, {Loose}, {L{\'o}pez-Fernandez}, {Lord}, {Luinge}, {Marston},
  {Mart{\'{\i}}n-Pintado}, {Maestrini}, {Maiwald}, {McCoey}, {Mehdi}, {Megej},
  {Melchior}, {Meinsma}, {Merkel}, {Michalska}, {Monstein}, {Moratschke},
  {Morris}, {Muller}, {Murphy}, {Naber}, {Natale}, {Nowosielski}, {Nuzzolo},
  {Olberg}, {Olbrich}, {Orfei}, {Orleanski}, {Ossenkopf}, {Peacock}, {Pearson},
  {Peron}, {Phillip-May}, {Piazzo}, {Planesas}, {Rataj}, {Ravera}, {Risacher},
  {Salez}, {Samoska}, {Saraceno}, {Schieder}, {Schlecht}, {Schl{\"o}der},
  {Schm{\"u}lling}, {Schultz}, {Schuster}, {Siebertz}, {Smit}, {Szczerba},
  {Shipman}, {Steinmetz}, {Stern}, {Stokroos}, {Teipen}, {Teyssier}, {Tils},
  {Trappe}, {van Baaren}, {van Leeuwen}, {van de Stadt}, {Visser}, {Wildeman},
  {Wafelbakker}, {Ward}, {Wesselius}, {Wild}, {Wulff}, {Wunsch}, {Tielens},
  {Zaal}, {Zirath}, {Zmuidzinas}, \& {Zwart}}]{deGraauw2010}
{de Graauw}, T., {Helmich}, F.~P., {Phillips}, T.~G., {et~al.} 2010, \aap, 518,
  L6

\bibitem[{{Di Francesco} {et~al.}(2001){Di Francesco}, {Myers}, {Wilner},
  {Ohashi}, \& {Mardones}}]{DiFrancesco2001}
{Di Francesco}, J., {Myers}, P.~C., {Wilner}, D.~J., {Ohashi}, N., \&
  {Mardones}, D. 2001, \apj, 562, 770

\bibitem[{{Doty} \& {Neufeld}(1997)}]{Doty1997}
{Doty}, S.~D. \& {Neufeld}, D.~A. 1997, \apj, 489, 122

\bibitem[{{Duarte-Cabral} {et~al.}(2011){Duarte-Cabral}, {Dobbs}, {Peretto}, \&
  {Fuller}}]{DuarteCabral2011}
{Duarte-Cabral}, A., {Dobbs}, C.~L., {Peretto}, N., \& {Fuller}, G.~A. 2011,
  \aap, 528, A50

\bibitem[{{Duarte-Cabral} {et~al.}(2010){Duarte-Cabral}, {Fuller}, {Peretto},
  {Hatchell}, {Ladd}, {Buckle}, {Richer}, \& {Graves}}]{DuarteCabral2010}
{Duarte-Cabral}, A., {Fuller}, G.~A., {Peretto}, N., {et~al.} 2010, \aap, 519,
  A27

\bibitem[{{Dunham} \& {Vorobyov}(2012)}]{Dunham2012}
{Dunham}, M.~M. \& {Vorobyov}, E.~I. 2012, \apj, 747, 52

\bibitem[{{Dzib} {et~al.}(2010){Dzib}, {Loinard}, {Mioduszewski}, {Boden},
  {Rodr{\'{\i}}guez}, \& {Torres}}]{Dzib2010}
{Dzib}, S., {Loinard}, L., {Mioduszewski}, A.~J., {et~al.} 2010, \apj, 718, 610

\bibitem[{{Emprechtinger} {et~al.}(2013){Emprechtinger}, {Lis}, {Rolffs},
  {Schilke}, {Monje}, {Comito}, {Ceccarelli}, {Neufeld}, \& {van der
  Tak}}]{Emprechtinger2013}
{Emprechtinger}, M., {Lis}, D.~C., {Rolffs}, R., {et~al.} 2013, \apj, 765, 61

\bibitem[{{Evans}(1999)}]{Evans1999}
{Evans}, II, N.~J. 1999, \araa, 37, 311

\bibitem[{{Evans} {et~al.}(2001){Evans}, {Rawlings}, {Shirley}, \&
  {Mundy}}]{Evans2001}
{Evans}, II, N.~J., {Rawlings}, J.~M.~C., {Shirley}, Y.~L., \& {Mundy}, L.~G.
  2001, \apj, 557, 193

\bibitem[{{Foster} \& {Chevalier}(1993)}]{Foster1993}
{Foster}, P.~N. \& {Chevalier}, R.~A. 1993, \apj, 416, 303

\bibitem[{{Fuller} {et~al.}(2005){Fuller}, {Williams}, \&
  {Sridharan}}]{Fuller2005}
{Fuller}, G.~A., {Williams}, S.~J., \& {Sridharan}, T.~K. 2005, \aap, 442, 949

\bibitem[{{Galli} {et~al.}(2002){Galli}, {Walmsley}, \& {Gon{\c
  c}alves}}]{Galli2002}
{Galli}, D., {Walmsley}, M., \& {Gon{\c c}alves}, J. 2002, \aap, 394, 275

\bibitem[{{Girart} {et~al.}(2006){Girart}, {Rao}, \& {Marrone}}]{Girart2006}
{Girart}, J.~M., {Rao}, R., \& {Marrone}, D.~P. 2006, Science, 313, 812

\bibitem[{{Gredel} {et~al.}(1989){Gredel}, {Lepp}, {Dalgarno}, \&
  {Herbst}}]{Gredel1989}
{Gredel}, R., {Lepp}, S., {Dalgarno}, A., \& {Herbst}, E. 1989, \apj, 347, 289

\bibitem[{{Gregersen} {et~al.}(1997){Gregersen}, {Evans}, {Zhou}, \&
  {Choi}}]{Gregersen1997}
{Gregersen}, E.~M., {Evans}, II, N.~J., {Zhou}, S., \& {Choi}, M. 1997, \apj,
  484, 256

\bibitem[{{Herczeg} {et~al.}(2012){Herczeg}, {Karska}, {Bruderer},
  {Kristensen}, {van Dishoeck}, {J{\o}rgensen}, {Visser}, {Wampfler}, {Bergin},
  {Y{\i}ld{\i}z}, {Pontoppidan}, \& {Gracia-Carpio}}]{Herczeg2012}
{Herczeg}, G.~J., {Karska}, A., {Bruderer}, S., {et~al.} 2012, \aap, 540, A84

\bibitem[{{Herpin} {et~al.}(2012){Herpin}, {Chavarr{\'{\i}}a}, {van der Tak},
  {Wyrowski}, {van Dishoeck}, {Jacq}, {Braine}, {Baudry}, {Bontemps}, \&
  {Kristensen}}]{Herpin2012}
{Herpin}, F., {Chavarr{\'{\i}}a}, L., {van der Tak}, F., {et~al.} 2012, \aap,
  542, A76

\bibitem[{{Hogerheijde} \& {Sandell}(2000)}]{Hogerheijde2000a}
{Hogerheijde}, M.~R. \& {Sandell}, G. 2000, \apj, 534, 880

\bibitem[{{Hogerheijde} \& {van der Tak}(2000)}]{Hogerheijde2000b}
{Hogerheijde}, M.~R. \& {van der Tak}, F.~F.~S. 2000, \aap, 362, 697

\bibitem[{{Hollenbach} {et~al.}(2009){Hollenbach}, {Kaufman}, {Bergin}, \&
  {Melnick}}]{Hollenbach2009}
{Hollenbach}, D., {Kaufman}, M.~J., {Bergin}, E.~A., \& {Melnick}, G.~J. 2009,
  \apj, 690, 1497

\bibitem[{{Ivezi\'{c}} \& {Elitzur}(1997)}]{Ivezic1997}
{Ivezi\'{c}}, Z. \& {Elitzur}, M. 1997, \mnras, 287, 799

\bibitem[{{J{\o}rgensen} {et~al.}(2005){J{\o}rgensen}, {Bourke}, {Myers},
  {Sch{\"o}ier}, {van Dishoeck}, \& {Wilner}}]{Jorgensen2005}
{J{\o}rgensen}, J.~K., {Bourke}, T.~L., {Myers}, P.~C., {et~al.} 2005, \apj,
  632, 973

\bibitem[{{J{\o}rgensen} {et~al.}(2006){J{\o}rgensen}, {Johnstone}, {van
  Dishoeck}, \& {Doty}}]{Jorgensen2006}
{J{\o}rgensen}, J.~K., {Johnstone}, D., {van Dishoeck}, E.~F., \& {Doty}, S.~D.
  2006, \aap, 449, 609

\bibitem[{{J{\o}rgensen} {et~al.}(2002){J{\o}rgensen}, {Sch{\"o}ier}, \& {van
  Dishoeck}}]{Jorgensen2002}
{J{\o}rgensen}, J.~K., {Sch{\"o}ier}, F.~L., \& {van Dishoeck}, E.~F. 2002,
  \aap, 389, 908

\bibitem[{{J{\o}rgensen} {et~al.}(2004){J{\o}rgensen}, {Sch{\"o}ier}, \& {van
  Dishoeck}}]{Jorgensen2004}
{J{\o}rgensen}, J.~K., {Sch{\"o}ier}, F.~L., \& {van Dishoeck}, E.~F. 2004,
  \aap, 416, 603

\bibitem[{{Karska} {et~al.}(2013){Karska}, {Herczeg}, {van Dishoeck},
  {Wampfler}, {Kristensen}, {Goicoechea}, {Visser}, {Nisini}, {San
  Jos{\'e}-Garc{\'{\i}}a}, {Bruderer}, {{\'S}niady}, {Doty}, {Fedele},
  {Y{\i}ld{\i}z}, {Benz}, {Bergin}, {Caselli}, {Herpin}, {Hogerheijde},
  {Johnstone}, {J{\o}rgensen}, {Liseau}, {Tafalla}, {van der Tak}, \&
  {Wyrowski}}]{Karska2013}
{Karska}, A., {Herczeg}, G.~J., {van Dishoeck}, E.~F., {et~al.} 2013, \aap,
  552, A141

\bibitem[{{Kauffmann} {et~al.}(2008){Kauffmann}, {Bertoldi}, {Bourke}, {Evans},
  \& {Lee}}]{Kauffmann2008}
{Kauffmann}, J., {Bertoldi}, F., {Bourke}, T.~L., {Evans}, II, N.~J., \& {Lee},
  C.~W. 2008, \aap, 487, 993

\bibitem[{{Kenyon} {et~al.}(1990){Kenyon}, {Hartmann}, {Strom}, \&
  {Strom}}]{Kenyon1990}
{Kenyon}, S.~J., {Hartmann}, L.~W., {Strom}, K.~M., \& {Strom}, S.~E. 1990,
  \aj, 99, 869

\bibitem[{{Keto} \& {Caselli}(2008)}]{Keto2008}
{Keto}, E. \& {Caselli}, P. 2008, \apj, 683, 238

\bibitem[{{Kristensen} {et~al.}(2013){Kristensen}, {van Dishoeck}, {Benz},
  {Bruderer}, {Visser}, \& {Wampfler}}]{Kristensen2013}
{Kristensen}, L.~E., {van Dishoeck}, E.~F., {Benz}, A.~O., {et~al.} 2013,
  ArXiv:1307.1710

\bibitem[{{Kristensen} {et~al.}(2012){Kristensen}, {van Dishoeck}, {Bergin},
  {Visser}, {Y{\i}ld{\i}z}, {San Jose-Garcia}, {J{\o}rgensen}, {Herczeg},
  {Johnstone}, {Wampfler}, {Benz}, {Bruderer}, {Cabrit}, {Caselli}, {Doty},
  {Harsono}, {Herpin}, {Hogerheijde}, {Karska}, {van Kempen}, {Liseau},
  {Nisini}, {Tafalla}, {van der Tak}, \& {Wyrowski}}]{Kristensen2012}
{Kristensen}, L.~E., {van Dishoeck}, E.~F., {Bergin}, E.~A., {et~al.} 2012,
  \aap, 542, A8

\bibitem[{{Kristensen} {et~al.}(2010){Kristensen}, {Visser}, {van Dishoeck},
  {Y{\i}ld{\i}z}, {Doty}, {Herczeg}, {Liu}, {Parise}, {J{\o}rgensen}, {van
  Kempen}, {Brinch}, {Wampfler}, {Bruderer}, {Benz}, {Hogerheijde}, {Deul},
  {Bachiller}, {Baudry}, {Benedettini}, {Bergin}, {Bjerkeli}, {Blake},
  {Bontemps}, {Braine}, {Caselli}, {Cernicharo}, {Codella}, {Daniel}, {de
  Graauw}, {di Giorgio}, {Dominik}, {Encrenaz}, {Fich}, {Fuente}, {Giannini},
  {Goicoechea}, {Helmich}, {Herpin}, {Jacq}, {Johnstone}, {Kaufman}, {Larsson},
  {Lis}, {Liseau}, {Marseille}, {McCoey}, {Melnick}, {Neufeld}, {Nisini},
  {Olberg}, {Pearson}, {Plume}, {Risacher}, {Santiago-Garc{\'{\i}}a},
  {Saraceno}, {Shipman}, {Tafalla}, {Tielens}, {van der Tak}, {Wyrowski},
  {Beintema}, {de Jonge}, {Dieleman}, {Ossenkopf}, {Roelfsema}, {Stutzki}, \&
  {Whyborn}}]{Kristensen2010}
{Kristensen}, L.~E., {Visser}, R., {van Dishoeck}, E.~F., {et~al.} 2010, \aap,
  521, L30

\bibitem[{{Larson}(1969)}]{Larson1969}
{Larson}, R.~B. 1969, \mnras, 145, 271

\bibitem[{{Liseau} {et~al.}(1988){Liseau}, {Sandell}, \& {Knee}}]{Liseau1988}
{Liseau}, R., {Sandell}, G., \& {Knee}, L.~B.~G. 1988, \aap, 192, 153

\bibitem[{{Mardones} {et~al.}(1997){Mardones}, {Myers}, {Tafalla}, {Wilner},
  {Bachiller}, \& {Garay}}]{Mardones1997}
{Mardones}, D., {Myers}, P.~C., {Tafalla}, M., {et~al.} 1997, \apj, 489, 719

\bibitem[{{McElroy} {et~al.}(2013){McElroy}, {Walsh}, {Markwick}, {Cordiner},
  {Smith}, \& {Millar}}]{McElroy2013}
{McElroy}, D., {Walsh}, C., {Markwick}, A.~J., {et~al.} 2013, \aap, 550, A36

\bibitem[{{Menten} {et~al.}(1987){Menten}, {Serabyn}, {Guesten}, \&
  {Wilson}}]{Menten1987}
{Menten}, K.~M., {Serabyn}, E., {Guesten}, R., \& {Wilson}, T.~L. 1987, \aap,
  177, L57

\bibitem[{{Meyer} {et~al.}(1998){Meyer}, {Jura}, \& {Cardelli}}]{Meyer1998}
{Meyer}, D.~M., {Jura}, M., \& {Cardelli}, J.~A. 1998, \apj, 493, 222

\bibitem[{{Mottram} {et~al.}(\noop{3002}in prep.){Mottram}, {Kristensen}, {van
  Dishoeck}, {Another}, {Another}, {Another}, {Another}, {Another}, {Another},
  {Another}, {Another}, {Another}, \& {Another}}]{Mottramip}
{Mottram}, J.~C., {Kristensen}, L.~E., {van Dishoeck}, E.~F., {et~al.}
  \noop{3002}in prep.

\bibitem[{{Myers} {et~al.}(2000){Myers}, {Evans}, \& {Ohashi}}]{Myers2000}
{Myers}, P.~C., {Evans}, II, N.~J., \& {Ohashi}, N. 2000, Protostars and
  Planets IV, 217

\bibitem[{{{\"O}berg} {et~al.}(2009){{\"O}berg}, {Linnartz}, {Visser}, \& {van
  Dishoeck}}]{Oberg2009}
{{\"O}berg}, K.~I., {Linnartz}, H., {Visser}, R., \& {van Dishoeck}, E.~F.
  2009, \apj, 693, 1209

\bibitem[{{Offner} {et~al.}(2010){Offner}, {Kratter}, {Matzner}, {Krumholz}, \&
  {Klein}}]{Offner2010}
{Offner}, S.~S.~R., {Kratter}, K.~M., {Matzner}, C.~D., {Krumholz}, M.~R., \&
  {Klein}, R.~I. 2010, \apj, 725, 1485

\bibitem[{{Ossenkopf} \& {Henning}(1994)}]{Ossenkopf1994}
{Ossenkopf}, V. \& {Henning}, T. 1994, \aap, 291, 943

\bibitem[{{Ott}(2010)}]{Ott2010}
{Ott}, S. 2010, in Astronomical Society of the Pacific Conference Series, Vol.
  434, Astronomical Data Analysis Software and Systems XIX, ed. Y.~{Mizumoto},
  K.-I. {Morita}, \& M.~{Ohishi}, 139

\bibitem[{{Pagani} {et~al.}(2009){Pagani}, {Vastel}, {Hugo}, {Kokoouline},
  {Greene}, {Bacmann}, {Bayet}, {Ceccarelli}, {Peng}, \&
  {Schlemmer}}]{Pagani2009}
{Pagani}, L., {Vastel}, C., {Hugo}, E., {et~al.} 2009, \aap, 494, 623

\bibitem[{{Palla}(1999)}]{Palla1999}
{Palla}, F. 1999, in NATO ASIC Proc. 540: The Origin of Stars and Planetary
  Systems, ed. C.~J. {Lada} \& N.~D. {Kylafis}, 375

\bibitem[{{Parise} {et~al.}(2006){Parise}, {Belloche}, {Leurini}, {Schilke},
  {Wyrowski}, \& {G{\"u}sten}}]{Parise2006}
{Parise}, B., {Belloche}, A., {Leurini}, S., {et~al.} 2006, \aap, 454, L79

\bibitem[{{Penston}(1969)}]{Penston1969}
{Penston}, M.~V. 1969, \mnras, 144, 425

\bibitem[{{Persson} {et~al.}(2012){Persson}, {J{\o}rgensen}, \& {van
  Dishoeck}}]{Persson2012}
{Persson}, M.~V., {J{\o}rgensen}, J.~K., \& {van Dishoeck}, E.~F. 2012, \aap,
  541, A39

\bibitem[{{Pilbratt} {et~al.}(2010){Pilbratt}, {Riedinger}, {Passvogel},
  {Crone}, {Doyle}, {Gageur}, {Heras}, {Jewell}, {Metcalfe}, {Ott}, \&
  {Schmidt}}]{Pilbratt2010}
{Pilbratt}, G.~L., {Riedinger}, J.~R., {Passvogel}, T., {et~al.} 2010, \aap,
  518, L1

\bibitem[{{Pineda} {et~al.}(2010){Pineda}, {Goodman}, {Arce}, {Caselli},
  {Foster}, {Myers}, \& {Rosolowsky}}]{Pineda2010}
{Pineda}, J.~E., {Goodman}, A.~A., {Arce}, H.~G., {et~al.} 2010, \apjl, 712,
  L116

\bibitem[{{Prasad} \& {Tarafdar}(1983)}]{Prasad1983}
{Prasad}, S.~S. \& {Tarafdar}, S.~P. 1983, \apj, 267, 603

\bibitem[{{Ramchandani}(2012)}]{Ramchandani2012}
{Ramchandani}, J. 2012, MSc. Thesis, Leiden University

\bibitem[{{Roelfsema} {et~al.}(2012){Roelfsema}, {Helmich}, {Teyssier},
  {Ossenkopf}, {Morris}, {Olberg}, {Shipman}, {Risacher}, {Akyilmaz},
  {Assendorp}, {Avruch}, {Beintema}, {Biver}, {Boogert}, {Borys}, {Braine},
  {Caris}, {Caux}, {Cernicharo}, {Coeur-Joly}, {Comito}, {de Lange},
  {Delforge}, {Dieleman}, {Dubbeldam}, {de Graauw}, {Edwards}, {Fich},
  {Flederus}, {Gal}, {di Giorgio}, {Herpin}, {Higgins}, {Hoac}, {Huisman},
  {Jarchow}, {Jellema}, {de Jonge}, {Kester}, {Klein}, {Kooi}, {Kramer},
  {Laauwen}, {Larsson}, {Leinz}, {Lord}, {Lorenzani}, {Luinge}, {Marston},
  {Mart{\'{\i}}n-Pintado}, {McCoey}, {Melchior}, {Michalska}, {Moreno},
  {M{\"u}ller}, {Nowosielski}, {Okada}, {Orlea{\'n}ski}, {Phillips}, {Pearson},
  {Rabois}, {Ravera}, {Rector}, {Rengel}, {Sagawa}, {Salomons},
  {S{\'a}nchez-Su{\'a}rez}, {Schieder}, {Schl{\"o}der}, {Schm{\"u}lling},
  {Soldati}, {Stutzki}, {Thomas}, {Tielens}, {Vastel}, {Wildeman}, {Xie},
  {Xilouris}, {Wafelbakker}, {Whyborn}, {Zaal}, {Bell}, {Bjerkeli}, {De Beck},
  {Cavali{\'e}}, {Crockett}, {Hily-Blant}, {Kama}, {Kaminski}, {Lefl{\'o}ch},
  {Lombaert}, {de Luca}, {Makai}, {Marseille}, {Nagy}, {Pacheco}, {van der
  Wiel}, {Wang}, \& {Y{\i}ld{\i}z}}]{Roelfsema2012}
{Roelfsema}, P.~R., {Helmich}, F.~P., {Teyssier}, D., {et~al.} 2012, \aap, 537,
  A17

\bibitem[{{Schmalzl} {et~al.}(\noop{3003}in prep.){Schmalzl}, {Visser},
  {Walsh}, {Another}, {Another}, {Another}, {Another}, {Another}, {Another},
  {Another}, {Another}, {Another}, \& {Another}}]{Schmalzlip}
{Schmalzl}, M., {Visser}, R., {Walsh}, C., {et~al.} \noop{3003}in prep.

\bibitem[{{Sch{\"o}ier} {et~al.}(2005){Sch{\"o}ier}, {van der Tak}, {van
  Dishoeck}, \& {Black}}]{Schoier2005}
{Sch{\"o}ier}, F.~L., {van der Tak}, F.~F.~S., {van Dishoeck}, E.~F., \&
  {Black}, J.~H. 2005, \aap, 432, 369

\bibitem[{{Shen} {et~al.}(2004){Shen}, {Greenberg}, {Schutte}, \& {van
  Dishoeck}}]{Shen2004}
{Shen}, C.~J., {Greenberg}, J.~M., {Schutte}, W.~A., \& {van Dishoeck}, E.~F.
  2004, \aap, 415, 203

\bibitem[{{Shu}(1977)}]{Shu1977}
{Shu}, F.~H. 1977, \apj, 214, 488

\bibitem[{{Terebey} {et~al.}(1984){Terebey}, {Shu}, \& {Cassen}}]{Terebey1984}
{Terebey}, S., {Shu}, F.~H., \& {Cassen}, P. 1984, \apj, 286, 529

\bibitem[{{Tobin} {et~al.}(2012{\natexlab{a}}){Tobin}, {Hartmann}, {Bergin},
  {Chiang}, {Looney}, {Chandler}, {Maret}, \& {Heitsch}}]{Tobin2012a}
{Tobin}, J.~J., {Hartmann}, L., {Bergin}, E., {et~al.} 2012{\natexlab{a}},
  \apj, 748, 16

\bibitem[{{Tobin} {et~al.}(2012{\natexlab{b}}){Tobin}, {Hartmann}, {Chiang},
  {Wilner}, {Looney}, {Loinard}, {Calvet}, \& {D'Alessio}}]{Tobin2012b}
{Tobin}, J.~J., {Hartmann}, L., {Chiang}, H.-F., {et~al.} 2012{\natexlab{b}},
  \nat, 492, 83

\bibitem[{{Torstensson} {et~al.}(2011){Torstensson}, {van Langevelde},
  {Vlemmings}, \& {Bourke}}]{Torstensson2011}
{Torstensson}, K.~J.~E., {van Langevelde}, H.~J., {Vlemmings}, W.~H.~T., \&
  {Bourke}, S. 2011, \aap, 526, A38

\bibitem[{{Ulrich}(1976)}]{Ulrich1976}
{Ulrich}, R.~K. 1976, \apj, 210, 377

\bibitem[{{van der Tak} {et~al.}(2007){van der Tak}, {Black}, {Sch{\"o}ier},
  {Jansen}, \& {van Dishoeck}}]{vanderTak2007}
{van der Tak}, F.~F.~S., {Black}, J.~H., {Sch{\"o}ier}, F.~L., {Jansen}, D.~J.,
  \& {van Dishoeck}, E.~F. 2007, \aap, 468, 627

\bibitem[{{van der Tak} {et~al.}(2013){van der Tak}, {Chavarr{\'{\i}}a},
  {Herpin}, {Wyrowski}, {Walmsley}, {van Dishoeck}, {Benz}, {Bergin},
  {Caselli}, {Hogerheijde}, {Johnstone}, {Kristensen}, {Liseau}, {Nisini}, \&
  {Tafalla}}]{vanderTak2013}
{van der Tak}, F.~F.~S., {Chavarr{\'{\i}}a}, L., {Herpin}, F., {et~al.} 2013,
  \aap, 554, A83

\bibitem[{{van Dishoeck} {et~al.}(2011){van Dishoeck}, {Kristensen}, {Benz},
  {Bergin}, {Caselli}, {Cernicharo}, {Herpin}, {Hogerheijde}, {Johnstone},
  {Liseau}, {Nisini}, {Shipman}, {Tafalla}, {van der Tak}, {Wyrowski},
  {Aikawa}, {Bachiller}, {Baudry}, {Benedettini}, {Bjerkeli}, {Blake},
  {Bontemps}, {Braine}, {Brinch}, {Bruderer}, {Chavarr{\'{\i}}a}, {Codella},
  {Daniel}, {de Graauw}, {Deul}, {di Giorgio}, {Dominik}, {Doty}, {Dubernet},
  {Encrenaz}, {Feuchtgruber}, {Fich}, {Frieswijk}, {Fuente}, {Giannini},
  {Goicoechea}, {Helmich}, {Herczeg}, {Jacq}, {J{\o}rgensen}, {Karska},
  {Kaufman}, {Keto}, {Larsson}, {Lefloch}, {Lis}, {Marseille}, {McCoey},
  {Melnick}, {Neufeld}, {Olberg}, {Pagani}, {Pani{\'c}}, {Parise}, {Pearson},
  {Plume}, {Risacher}, {Salter}, {Santiago-Garc{\'{\i}}a}, {Saraceno},
  {St{\"a}uber}, {van Kempen}, {Visser}, {Viti}, {Walmsley}, {Wampfler}, \&
  {Y{\i}ld{\i}z}}]{vanDishoeck2011}
{van Dishoeck}, E.~F., {Kristensen}, L.~E., {Benz}, A.~O., {et~al.} 2011,
  \pasp, 123, 138

\bibitem[{{van Kempen} {et~al.}(2008){van Kempen}, {Doty}, {van Dishoeck},
  {Hogerheijde}, \& {J{\o}rgensen}}]{vanKempen2008}
{van Kempen}, T.~A., {Doty}, S.~D., {van Dishoeck}, E.~F., {Hogerheijde},
  M.~R., \& {J{\o}rgensen}, J.~K. 2008, \aap, 487, 975

\bibitem[{{van Kempen} {et~al.}(2009{\natexlab{a}}){van Kempen}, {van
  Dishoeck}, {G{\"u}sten}, {Kristensen}, {Schilke}, {Hogerheijde}, {Boland},
  {Menten}, \& {Wyrowski}}]{vanKempen2009b}
{van Kempen}, T.~A., {van Dishoeck}, E.~F., {G{\"u}sten}, R., {et~al.}
  2009{\natexlab{a}}, \aap, 507, 1425

\bibitem[{{van Kempen} {et~al.}(2009{\natexlab{b}}){van Kempen}, {van
  Dishoeck}, {Hogerheijde}, \& {G{\"u}sten}}]{vanKempen2009c}
{van Kempen}, T.~A., {van Dishoeck}, E.~F., {Hogerheijde}, M.~R., \&
  {G{\"u}sten}, R. 2009{\natexlab{b}}, \aap, 508, 259

\bibitem[{{van Kempen} {et~al.}(2009{\natexlab{c}}){van Kempen}, {van
  Dishoeck}, {Salter}, {Hogerheijde}, {J{\o}rgensen}, \&
  {Boogert}}]{vanKempen2009a}
{van Kempen}, T.~A., {van Dishoeck}, E.~F., {Salter}, D.~M., {et~al.}
  2009{\natexlab{c}}, \aap, 498, 167

\bibitem[{{Visser} {et~al.}(2013){Visser}, {J{\o}rgensen}, {Kristensen}, {van
  Dishoeck}, \& {Bergin}}]{Visser2013}
{Visser}, R., {J{\o}rgensen}, J.~K., {Kristensen}, L.~E., {van Dishoeck},
  E.~F., \& {Bergin}, E.~A. 2013, \apj, 769, 19

\bibitem[{{Vorobyov}(2009)}]{Vorobyov2009}
{Vorobyov}, E.~I. 2009, \apj, 704, 715

\bibitem[{{Vorobyov} \& {Basu}(2010)}]{Vorobyov2010}
{Vorobyov}, E.~I. \& {Basu}, S. 2010, \apj, 719, 1896

\bibitem[{{Watson} {et~al.}(2007){Watson}, {Bohac}, {Hull}, {Forrest},
  {Furlan}, {Najita}, {Calvet}, {D'Alessio}, {Hartmann}, {Sargent}, {Green},
  {Kim}, \& {Houck}}]{Watson2007}
{Watson}, D.~M., {Bohac}, C.~J., {Hull}, C., {et~al.} 2007, \nat, 448, 1026

\bibitem[{{Y{\i}ld{\i}z} {et~al.}(2013{\natexlab{a}}){Y{\i}ld{\i}z},
  {Acharyya}, {Goldsmith}, {van Dishoeck}, {Melnick}, {Snell}, {Liseau},
  {Chen}, {Pagani}, {Bergin}, {Caselli}, {Herbst}, {Kristensen}, {Visser},
  {Lis}, \& {Gerin}}]{Yildiz2013b}
{Y{\i}ld{\i}z}, U.~A., {Acharyya}, K., {Goldsmith}, P.~F., {et~al.}
  2013{\natexlab{a}}, ArXiv:1307.8031

\bibitem[{{Y{\i}ld{\i}z} {et~al.}(\noop{3001}in prep.){Y{\i}ld{\i}z},
  {Kristensen}, {van Dishoeck}, {Another}, {Another}, {Another}, {Another},
  {Another}, {Another}, {Another}, {Another}, {Another}, \&
  {Another}}]{Yildizip}
{Y{\i}ld{\i}z}, U.~A., {Kristensen}, L.~E., {van Dishoeck}, E.~F., {et~al.}
  \noop{3001}in prep.

\bibitem[{{Y{\i}ld{\i}z} {et~al.}(2012){Y{\i}ld{\i}z}, {Kristensen}, {van
  Dishoeck}, {Belloche}, {van Kempen}, {Hogerheijde}, {G{\"u}sten}, \& {van der
  Marel}}]{Yildiz2012}
{Y{\i}ld{\i}z}, U.~A., {Kristensen}, L.~E., {van Dishoeck}, E.~F., {et~al.}
  2012, \aap, 542, A86

\bibitem[{{Y{\i}ld{\i}z} {et~al.}(2013{\natexlab{b}}){Y{\i}ld{\i}z},
  {Kristensen}, {van Dishoeck}, {San Jos{\'e}-Garc{\'{\i}}a}, {Karska},
  {Harsono}, {Tafalla}, {Fuente}, {Visser}, {J{\o}rgensen}, \&
  {Hogerheijde}}]{Yildiz2013a}
{Y{\i}ld{\i}z}, U.~A., {Kristensen}, L.~E., {van Dishoeck}, E.~F., {et~al.}
  2013{\natexlab{b}}, \aap, 556, A89

\bibitem[{{Zhou}(1992)}]{Zhou1992}
{Zhou}, S. 1992, \apj, 394, 204

\bibitem[{{Zhou} {et~al.}(1993){Zhou}, {Evans}, {Koempe}, \&
  {Walmsley}}]{Zhou1993}
{Zhou}, S., {Evans}, II, N.~J., {Koempe}, C., \& {Walmsley}, C.~M. 1993, \apj,
  404, 232

\end{thebibliography}

\bibliographystyle{aa}

\clearpage
\appendix

\section{Supplementary material}

\begin{table*}
\caption[]{Observation identification numbers}
\begin{center}
\begin{tabular}{lccccccc}
\hline
\hline \noalign {\smallskip}
\centering
Source &  H$_{2}$O 1$_{10}$-1$_{01}$ & H$_{2}$O 2$_{12}$-1$_{01}$ & H$_{2}$O 3$_{12}$-2$_{21}$ & H$_{2}$O 3$_{12}$-3$_{03}$ & H$_{2}$O 1$_{11}$-0$_{00}$ & H$_{2}$O 2$_{02}$-1$_{11}$ & H$_{2}$O 2$_{11}$-2$_{02}$ \\
\hline\noalign {\smallskip}
NGC1333-IRAS4A  & 1342202065 & 1342203951 & 1342191721 & 1342201796 & 1342191656 & 1342191605 & 1342191749 \\
L1527  & 1342192524 & $-$ & 1342203256 & $-$ & 1342216335 & 1342203156 & 1342203214 \\
BHR71  & 1342201677 & $-$ & 1342201732 & $-$ & 1342200764 & 1342215915 & 1342201542 \\
IRAS15398  & 1342213732 & $-$ & 1342214446 & $-$ & 1342214414 & 1342203165 & 1342204795 \\
GSS30-IRS1  & 1342205302 & $-$ & 1342214442 & $-$ & 1342214413 & 1342203163 & 1342204796 \\
Ser SMM4  & 1342208577 & $-$ & 1342207700 & 1342207381 & 1342207380 & 1342194993 & 1342194562 \\
L1157  & 1342196407 & $-$ & 1342198346 & $-$ & 1342200763 & 1342197970 & 1342201551 \\
\hline\noalign {\smallskip}
\label{T:obsids}
\end{tabular}
\end{center}
\end{table*}

\begin{figure}
\begin{center}
\includegraphics[width=0.4\textwidth]{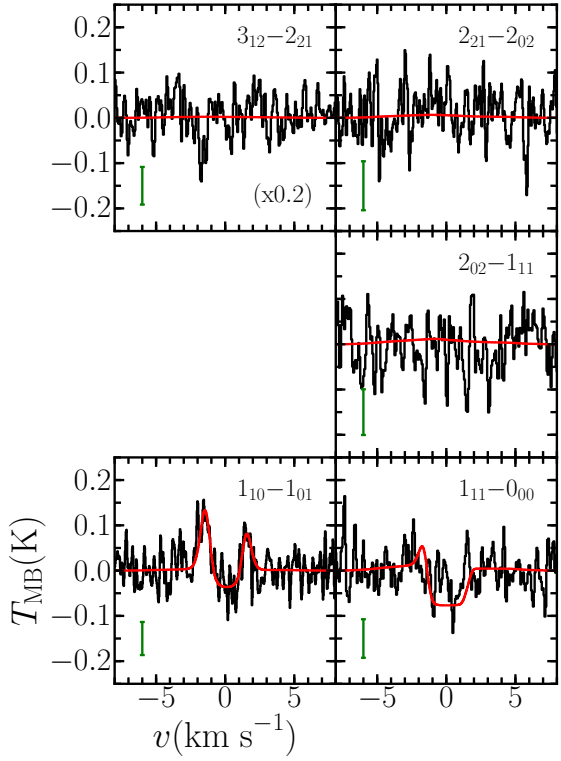}
\caption{Comparison of outflow and baseline subtracted observations for L1527 (black) with the best-fit model (red). The green error bar and scaling factors are the same as in Figure~\ref{F:infall_iras4a_fit}.}
\label{F:infall_sample_l1527}
\end{center}
\end{figure}

\begin{figure}
\begin{center}
\includegraphics[width=0.4\textwidth]{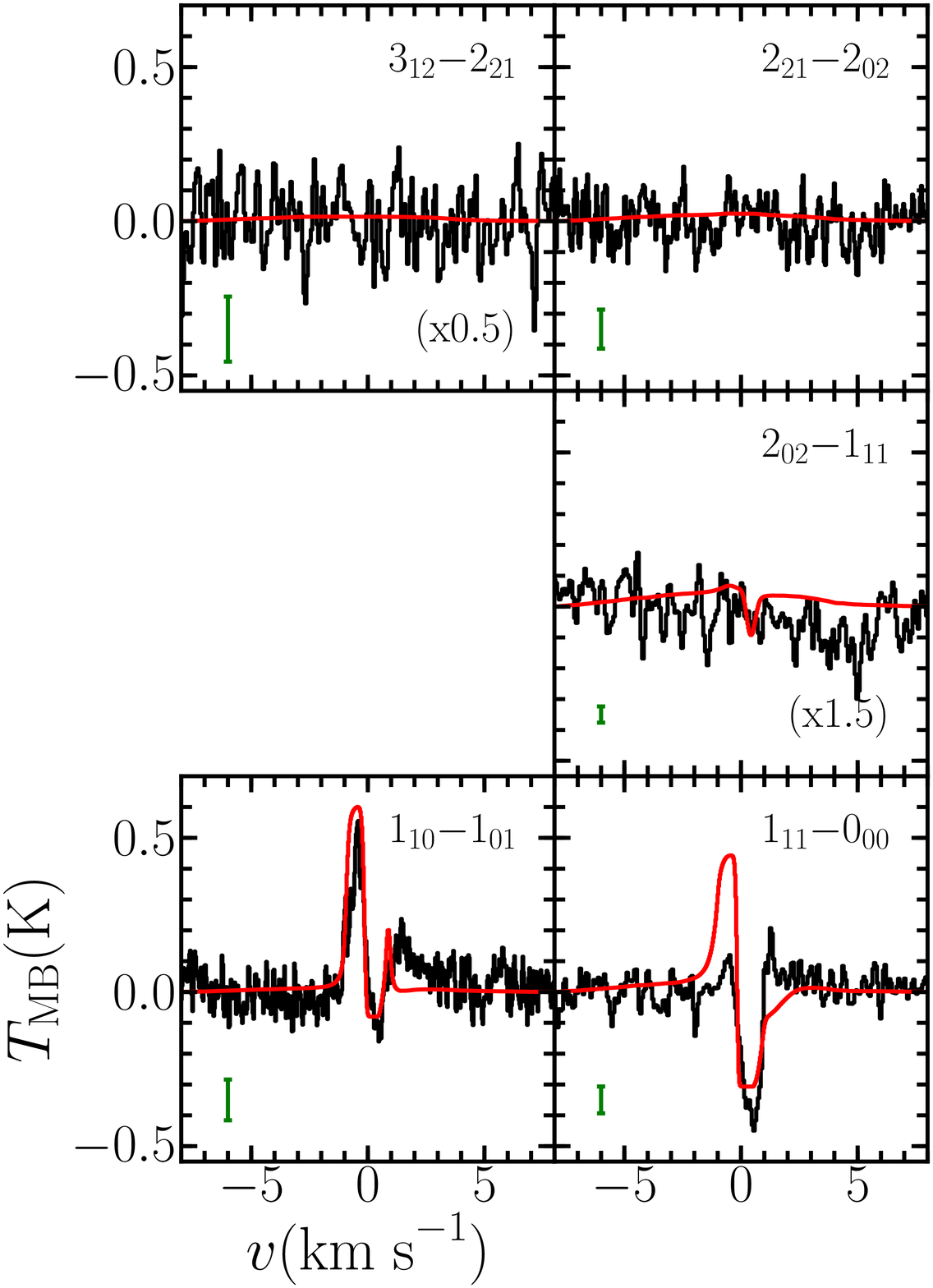}
\caption{As in Figure~\ref{F:infall_sample_l1527} but for BHR71}
\label{F:infall_sample_bhr71}
\end{center}
\end{figure}

\begin{figure}
\begin{center}
\includegraphics[width=0.4\textwidth]{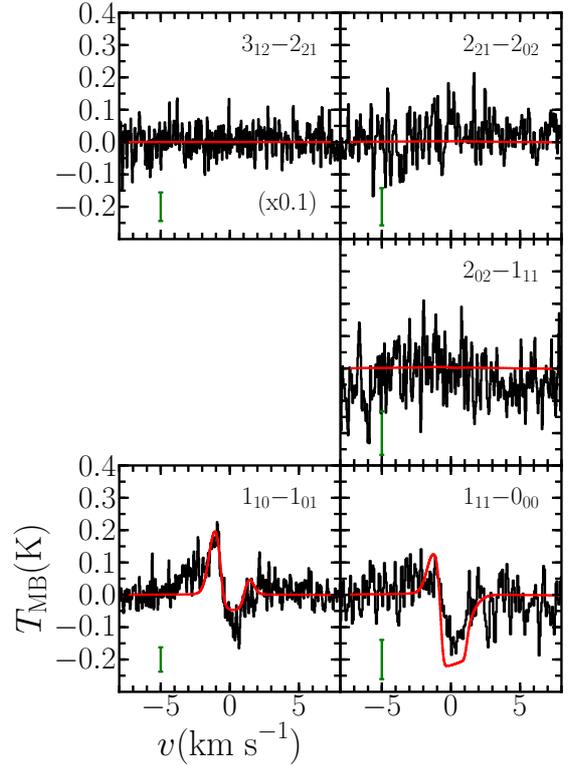}
\caption{As in Figure~\ref{F:infall_sample_l1527} but for IRAS15398}
\label{F:infall_sample_iras15398}
\end{center}
\end{figure}

\begin{figure}
\begin{center}
\includegraphics[width=0.4\textwidth]{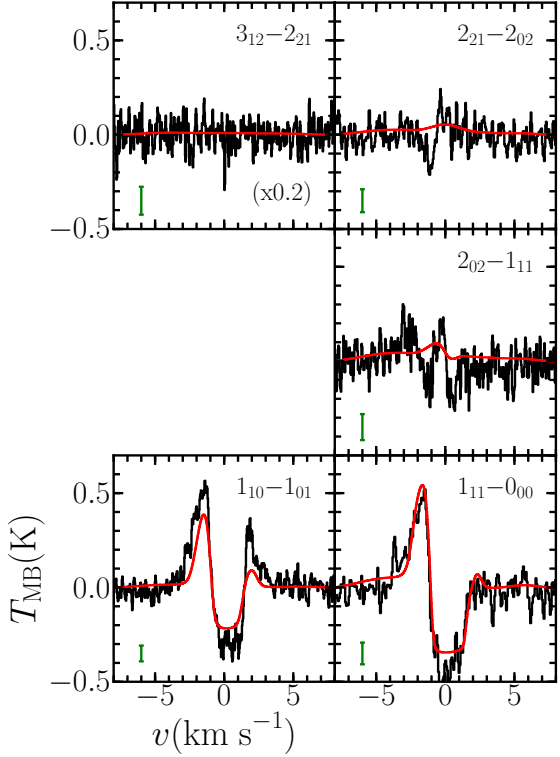}
\caption{As in Figure~\ref{F:infall_sample_l1527} but for GSS30}
\label{F:infall_sample_gss30}
\end{center}
\end{figure}

\begin{figure}
\begin{center}
\includegraphics[width=0.4\textwidth]{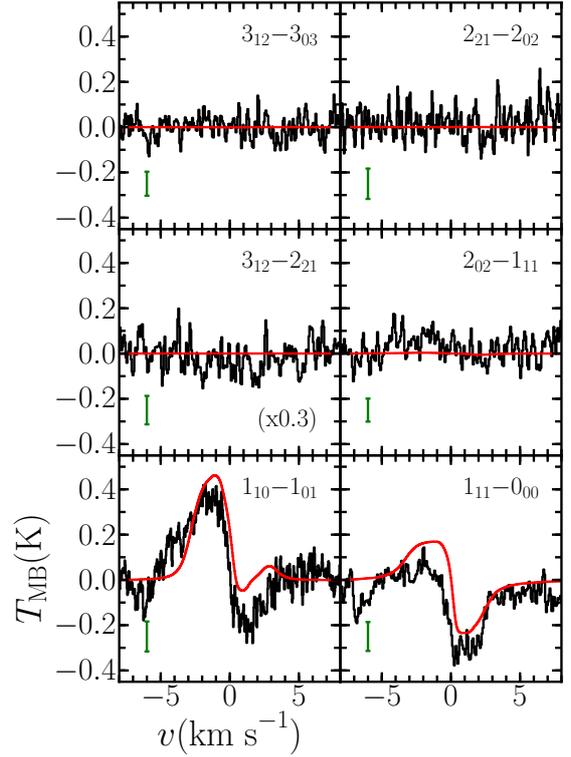}
\caption{As in Figure~\ref{F:infall_sample_l1527} but for Ser-SMM4}
\label{F:infall_sample_smm4}
\end{center}
\end{figure}

\begin{figure}
\begin{center}
\includegraphics[width=0.4\textwidth]{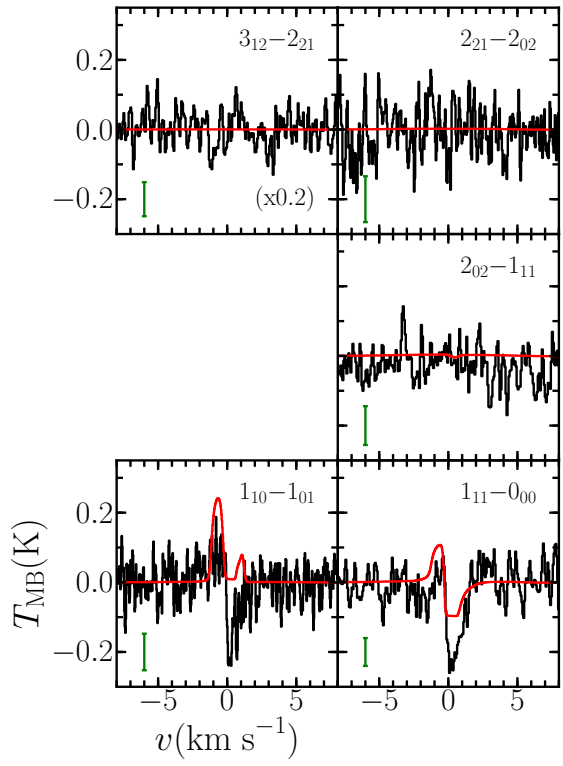}
\caption{As in Figure~\ref{F:infall_sample_l1527} but for L1157}
\label{F:infall_sample_l1157}
\end{center}
\end{figure}

Figures~\ref{F:infall_sample_l1527} to \ref{F:infall_sample_l1157} show comparisons between the observations and the best-fit models for the larger sample of sources. Table~\ref{T:obsids} provides the observation identification numbers for all \textit{Herschel} observations included in this paper.

\end{document}